\documentclass[twocolumn]{aastex63} 

\usepackage{savesym}
\savesymbol{tablenum}
\usepackage{siunitx}
\restoresymbol{SIX}{tablenum}

\usepackage{gensymb}
\usepackage[super]{nth}
\usepackage{graphicx}

\usepackage[caption=false]{subfig}

\DeclareFixedFont{\ttb}{T1}{txtt}{bx}{n}{8} 
\DeclareFixedFont{\ttm}{T1}{txtt}{m}{n}{8}  

\usepackage{color}
\definecolor{deepblue}{rgb}{0,0,0.5}
\definecolor{deepred}{rgb}{0.6,0,0}
\definecolor{deepgreen}{rgb}{0,0.5,0}

\usepackage{listings}
\newcommand\pythonstyle{\lstset{
language=Python,
basicstyle=\ttm,
morekeywords={self},              
keywordstyle=\ttb\color{deepblue},
emph={MyClass,__init__},          
emphstyle=\ttb\color{deepred},    
stringstyle=\color{deepgreen},
frame=tb,                         
showstringspaces=false
}}

\lstnewenvironment{python}[1][]
{
\pythonstyle
\lstset{#1}
}
{}

\newcommand\pythoninline[1]{{\pythonstyle\lstinline!#1!}}

\DeclareSIUnit \parsec {pc}
\DeclareSIUnit \msun {\mbox{$M_\odot$}}
\DeclareSIUnit \year {yr}
\DeclareSIUnit \radian {rad}
\DeclareSIUnit \jansky {Jy}

\revised{\today}
\shorttitle{The young stars in the Galactic Center}
\shortauthors{von Fellenberg et al.}
\graphicspath{{./}{figures/}}
\usepackage{amsmath}
\usepackage{comment}

\begin{document}

\title{The young stars in the Galactic Center 
}

\correspondingauthor{Sebastiano von Fellenberg}
\email{sfellenberg@mpifr-bonn.mpg.de}

\author[0000-0002-9156-2249]{Sebastiano von Fellenberg}
\affiliation{Max-Planck-Institute for Radio Astronomy,
auf dem H{\"u}gel 69, D-53121 Bonn, Germany}
\affiliation{Max-Planck-Institute for Extraterrestrial Physics,
Gie{\ss}enbachstr. 1, D-85748 Garching, Germany}

\author[0000-0002-5708-0481]{Stefan Gillessen}
\affiliation{Max-Planck-Institute for Extraterrestrial Physics,
Gie{\ss}enbachstr. 1, D-85748 Garching, Germany}

\author{Julia Stadler}
\affiliation{Max-Planck-Institute for Astrophysics,
Karl-Schwarzschild-Straße 1, D-85748 Garching, Germany}

\author{Michi Baub{\"o}ck}
\affiliation{University of Illinois Urbana-Champaign,
1002 W. Green Street, Urbana, IL 61801}

\author[0000-0002-2767-9653]{Reinhard Genzel}
\affiliation{Max-Planck-Institute for Extraterrestrial Physics,
Gie{\ss}enbachstr. 1, D-85748 Garching, Germany}
\affiliation{Departments of Physics and Astronomy, Le Conte Hall, University of California, Berkeley, CA 94720, USA}

\author{Tim de Zeeuw}
\affiliation{Sterrewacht Leiden, Leiden University, 
Postbus 9513, 2300 RA Leiden, The Netherlands}
\affiliation{Max-Planck-Institute for Extraterrestrial Physics,
Gie{\ss}enbachstr. 1, D-85748 Garching, Germany}

\author{Oliver Pfuhl}
\affiliation{European Southern Observatory, 
Karl-Schwarzschild-Straße 2, D-85748 Garching, Germany}

\author{Pau Amaro Seoane}
\affiliation{Institute for Multidisciplinary Mathematics, UPV, Spain}
\affiliation{Max-Planck-Institute for Extraterrestrial Physics,
Gie{\ss}enbachstr. 1, D-85748 Garching, Germany}
\affiliation{Kavli Institute for Astronomy and Astrophysics, Peking University, China}
\affiliation{AMSS, CAS, Beijing, China}

\author[0000-0002-4494-5503]{Antonia Drescher}
\affiliation{Max-Planck-Institute for Extraterrestrial Physics,
Gie{\ss}enbachstr. 1, D-85748 Garching, Germany}

\author{Frank Eisenhauer}
\affiliation{Max-Planck-Institute for Extraterrestrial Physics,
Gie{\ss}enbachstr. 1, D-85748 Garching, Germany}


\author[0000-0003-1366-1695]{Maryam Habibi}
\affiliation{Max-Planck-Institute for Extraterrestrial Physics,
Gie{\ss}enbachstr. 1, D-85748 Garching, Germany}

\author[0000-0003-1572-0396]{Thomas Ott}
\affiliation{Max-Planck-Institute for Extraterrestrial Physics,
Gie{\ss}enbachstr. 1, D-85748 Garching, Germany}


\author[0000-0002-0327-6585]{Felix Widmann}
\affiliation{Max-Planck-Institute for Extraterrestrial Physics,
Gie{\ss}enbachstr. 1, D-85748 Garching, Germany}

\author{Alice Young}
\affiliation{Max-Planck-Institute for Extraterrestrial Physics,
Gie{\ss}enbachstr. 1, D-85748 Garching, Germany}


\begin{abstract}
\noindent We present a large ${\sim 30'' \times 30''}$ spectroscopic survey of the Galactic Center using the SINFONI IFU at the VLT. Combining observations of the last two decades we compile spectra of over $2800$ stars. Using the Bracket-$\gamma$ absorption lines we identify $195$ young stars, extending the list of known young stars by $79$. In order to explore the angular momentum distribution of the young stars, we introduce an isotropic cluster prior. This prior reproduces an isotropic cluster in a mathematically exact way, which we test through numerical simulations.
We calculate the posterior angular momentum space as function of projected separation from Sgr~A*. We find that the observed young star distribution is substantially different from an isotropic cluster. We identify the previously reported feature of the clockwise disk and find that its angular momentum changes as function of separation from the black hole, and thus confirm a warp of the clockwise disk ($p \sim 99.2\%$). At large separations, we discover three prominent overdensities of angular momentum. One overdensity has been reported previously, the counter-clockwise disk. The other two are new.
Determining the likely members of these structures, we find that as many as $75\%$ of stars can be associated with one of these features.
Stars belonging to the warped clockwise-disk show a top heavy K-band luminosity function, while stars belonging to the larger separation features do not.
Our observations are in good agreement with the predictions of simulations of in-situ star formation, and argue for common formation of these structures.
\end{abstract}

\keywords{editorials, notices --- 
miscellaneous --- catalogs --- surveys}


\section{Introduction} \label{sec:intro}
The first infrared observations of the Galactic Center (GC) revealed that the central region of the Milky Way is surprisingly bright \citep[][]{Becklin1968, Becklin1975}. Due to the advent of ever higher resolution observations we now know that this light originates from a cluster of young, massive stars, many of O-type or Wolf-Rayet stars, residing in the central parsec  \citep[][]{Krabbe1991, Genzel1994, Simons1996, Blum1996}. The presence of young stars close to the massive black hole is puzzling, since star formation should be suppressed in the tidal field of the large mass. At the same time, the lifetimes of such stars is so short that they cannot have migrated from far.

\noindent The most important clue to solving this puzzle came from resolved stellar kinematics \citep[][]{Genzel2000, Ghez2000, Genzel2003_stellarcusp, Levin2003, Beloborodov2006, Lu2006, Paumard2006, Lu2009, Bartko2009}. The young stars to a large extent reside in two counter-rotating disks. This is now understood as a result of their formation from massive (\SI{\sim e5}{\msun}) gaseous disks a few Myr ago \citep[][]{Bonnell2008, HobbsNayakshin2009}. This picture is supported by the fact that the observed (and hence also initial) mass function is very top-heavy \citep[][]{Bartko2010} and by the radial surface density profile $\sim r^{-2}$. Further, the dynamical structure shows a warp for the clockwise disk \citep[][]{Bartko2009}, which might be a natural consequence of resonant relaxation \citep[][]{Kocsis2011}.

\noindent While the basic findings are agreed upon, there are several details which are not fully settled: \citet{Do2013} and \citet{Lu2013} find a less top-heavy mass function than \citet{Bartko2010}. The statistical significance of the presence of the counter-clockwise disk is low and has been disputed in \citet{Yelda2014}. The same authors also do not find the clockwise disks' warp.

\noindent Since these studies, the underlying data base has grown further. More stars in the GC field have been observed spectroscopically (which is the key for spectral typing and being able to include them into the kinematic analysis). Further, the number of stars for which we can report full orbital solutions has increased due to the longer time coverage. Given these advances and the open questions, we here present a re-analysis of the dynamics of the young stars in the GC.

\section{Data set}

\subsection{Observations}
We compile spectroscopic GC observations spanning almost two decades. Our observations consist of AO-assisted SINFONI observations, of which most are obtained with the combined H+K band, with a pixel scale of $100~\mathrm{mas}$. We re-reduce and analyse all GC SINFONI pointings. A considerable fraction of these data were analysed in previous publications, e.g. \cite{Paumard2006, Bartko2009, Bartko2010, Pfuhl2011, Pfuhl2014}. Furthermore, we analyse previously unpublished observations of the GC, which were obtained as back-up observations of the continuous monitoring of the motions of the stars in the GC \citep[][]{Gillessen2009, Gillessen2017, GravityCollaboration2021_abberations}. 

\noindent For stars closer than $2~\mathrm{''}$ to the black hole, our astrometry is determined from the continuous monitoring program, while for stars at larger projected distances, we rely on the astrometry presented by \cite{Trippe2008} or \cite{Fritz2011} if available. \autoref{fig:completeness} shows an overview of the SINFONI exposures. Our spectroscopic coverage increased substantially compared to previous studies (e.g. \citealt{Pfuhl2011} and \citealt{Yelda2014}). While we have reduced the gaps in our coverage, it is biased towards the natural guide star used by the SINFONI AO system in the North-East. Our observations cover a square spanning ${\sim}20''$ to ${\sim}-10''$ offset from Sgr~A* in right ascension and ${\sim}-10''$ to ${\sim}+20''$ in declination. The integration depth across this square is however not homogeneous, with some patches suffering from poor quality. Further, bright sources outshine near-by fainter stars in some patches. Only few Southern or North-Western exposures exist which rely on the laser guide star system \citep[][]{Bonnet2004}. We stack exposures from different epochs if multiple exposures exist. We account for Earth's motion around the Sun by shifting the wavelength axis of each exposure to the local standard of rest before combination. We try to classify all stars photometrically discernible in the exposures into either young or old type by extracting spectra and identifying the emission and absorption lines.
We classify stars as old if the CO band heads around ${\sim}2.3~\mathrm{\mu m}$ are detected. Young stars are classified by the detection of the Bracket $\gamma$ (Br$\gamma$) line at $2.166~\mathrm{\mu m}$ (and other lines). 
Because our observations seldomly allow for the identification of stars fainter than $K_{mag}=15$, such a simple classification scheme suffices to determine the age of the stars \citep[][]{Do2013}. The classification of young stars is complicated by the background Br$\gamma$ emission of the ionized gas in the Galactic Center, which can mimic young star features in the spectra if the background subtraction is poor. Thus we allow for stars to remain un-classified, if the stars do not show CO-band heads, and the Br$\gamma$ line is not credibly detected. 

\noindent We classified over $2800$ stars into un-classifiable, old, or young star. For all classifiable stars, we determine the radial velocity. In this work, we only investigate the credibly detected young stars and no old stars.

\subsection{Radial velocity measurement of young stars}
Measuring radial velocities of young stars is complicated by the presence of multiple gas emission clouds that contaminate the Br$\gamma$ absorption line of the stars \citep[e.g.][]{Paumard2006}. By selecting a suitable background, we minimize the effect of ionized gas emission. However, this approach is limited. For stars with difficult background subtraction, we revert to the Helium absorption line at \SI{2.113}{\micro \meter}, which is less affected by background emission but is significantly harder to detect. This leads to low SNR in the line detections for many stars. We use line-maps to visually confirm stars with faint lines, as well as the consistency of Br$\gamma$ and Helium velocities. We account for this difficulty by conservatively estimating the uncertainty of the radial velocity. Explicitly, for each star we obtain three different spectra using three different background aperture masks and determine the radial velocity for each. We determine the radial velocity from the RMS of the three spectra, but further increase the derived value if the line is poorly measured. For lines with discernible Helium absorption we fit a template young star spectrum, otherwise we revert to double and single Gauss fits. We checked for consistency of the methods if more than one method was applicable. Further we check that the derived velocity is different from that of the local ionized gas emission. The mean radial velocity uncertainty is \SI{58}{km/s}, with many stars having radial velocity uncertainties larger than \SI{100}{km/s}. 

\subsection{Spectroscopic completeness}
\begin{figure}
    \centering
    \includegraphics[width=0.485\textwidth, trim={12cm 0 4cm 0}, clip]{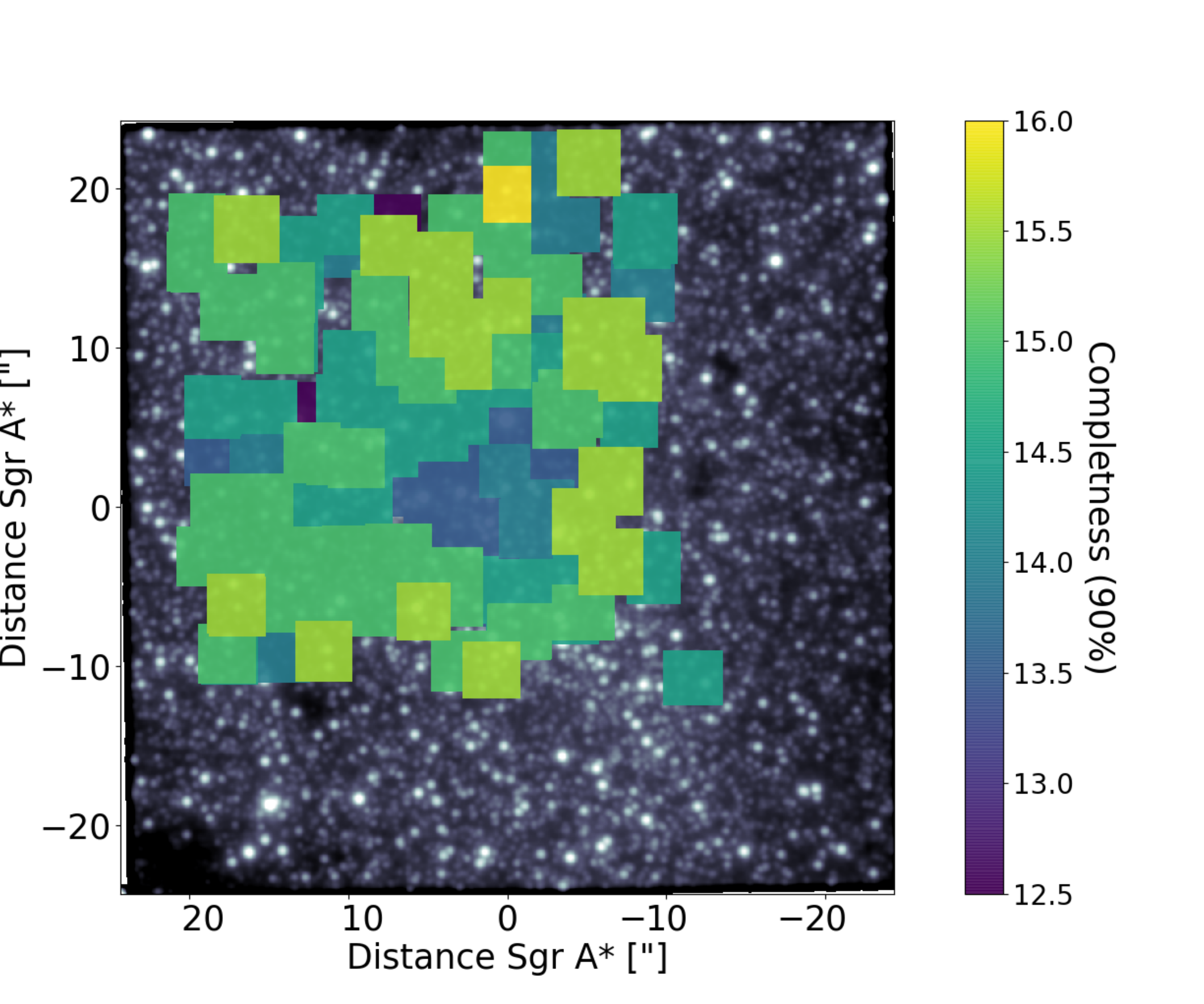}
    \caption[SINFONI completeness map of the Galactic Center]{$90\%$ Completeness estimate based on cross-referencing spectropically identified stars with the catalogue by \cite{Trippe2008}.}
    \label{fig:completeness}
\end{figure}

We estimate the fraction of stars we are able to classify. This can be done by planting point sources of different brightness in the images and estimating their detectablilty. Because our coverage is very patchy, and the integration depth is different for each pointing, we revert to a simpler technique: we assume that the catalogue by \cite{Trippe2008} is photometrically complete up to $K_{mag}=16$. Under this assumption, we cross-referenced all stars which we were able to classify as either young or old to this catalogue. By binning the catalog stars in steps of 0.5 magnitudes, we can count the fraction of stars we were able to classify spectroscopically. \autoref{fig:completeness} shows the resulting completeness maps. We identify stars up to $K_{mag}=14.5$ for most of our covered area. The central region, which has been observed the most frequently, is less complete than the outer regions. This is a consequence of the increasing number of (bright) stars towards the center. The decreasing number of stars at larger projected distances thus compensates for the shorter integration time. Other than the decreasing completeness in the central region there is no substantial bias in neither Northern, North-Eastern nor Eastern direction and our completeness estimate is comparable with that found in \cite{Pfuhl2011}.

\subsection{Stars with full orbits}
\begin{figure*}
    \centering
    \includegraphics[width=0.985\textwidth, trim={0cm 0 0cm 0}]{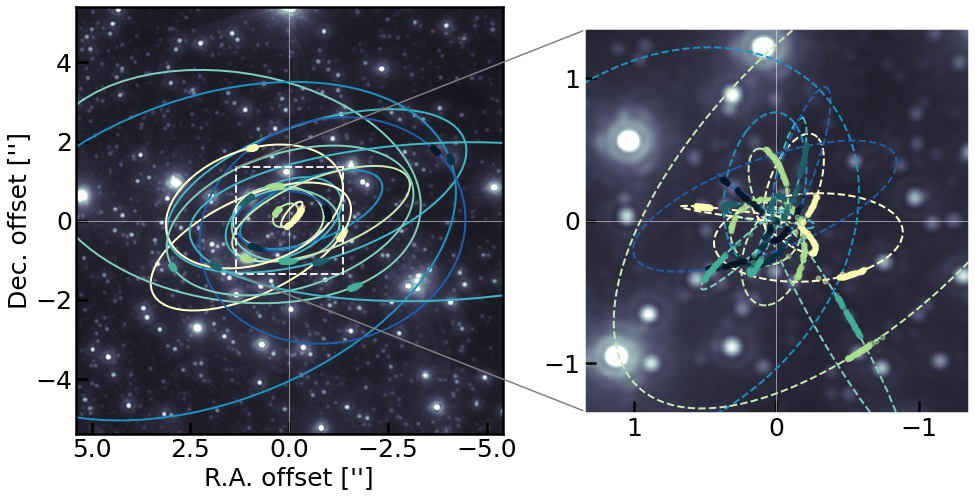}
    \caption[Orbits and astrometric measurements of the 36 young stars]{Orbits and astrometric measurements of the $36$ young stars in the Galactic Center. The track of astrometric measurements is over-plotted as darkened points. The orbits of the young stars belonging to the clockwise disk are shown. The inset shows the ``Sgr~A* star cluster'' of the innermost young stars which are on preferentially eccentric and randomly distributed orbits. }
    \label{fig:orbitPlot}
\end{figure*}

Overall 36 young stars have known full orbital solutions. We give their orbital solutions, and uncertainties in \autoref{tab:orbit_stars} in \autoref{app:youngst_stars_with_orbits}. \autoref{fig:orbitPlot} shows the inferred orbits. Of these orbits, 31 were presented in previous studies \citep[][]{Gillessen2009, Gillessen2017}, but we update their orbital estimates. The procedure of the orbit determination is described in detail in these references. Five stars have newly found orbital solutions that we add here. 

\subsection{Stars without orbits}
For 159 young stars we are able to determine five of the six phase space coordinates ($x$, $y$, $v_x$, $v_y$, $v_z$), which are given in \autoref{tab:youngStasPhaseSpace} and plotted in \autoref{fig:lookup} in \autoref{app:youngStasPhaseSpace}. The total number of known young stars therefore has increased by 79 with respect to \cite{Yelda2014}. For several of the known young stars we updated the radial velocity using the combination of available observations which maximize the signal to noise in the spectra. For many stars, this means discarding observations because a given star is affected by hot pixels or other data glitches.
In the following we will describe our star list and compare it with previously published star lists.

\subsection{Stars with radial velocity from the literature}
For several stars we resort to previously published values. For instance, we do not reanalyse the radial velocities of any of the Wolf-Rayet stars. For those, the radial velocities need to be derived from a stellar atmosphere model, and the uncertainty is dominated by the modeling and not the SNR of the spectra. Thus, re-derived spectra would not improve the radial velocity (F. Martins, private comm.). Furthermore, we adopt the radial velocities reported in \cite{Yelda2014} for stars which we either did not observe, or when our spectrum is too poor to derive a radial velocity, but does not contradict the classification. For stars which are continuously monitored in the central arcsecond we either use the radial velocity published in \cite{Gillessen2009}, \cite{Gillessen2017} or an updated value. 

\subsection{Stars in the literature removed from star list}
For five stars reported as young in \cite{Yelda2014} we identified CO-bandheads in the spectra. This may be the result of confusion. For instance, our spatial resolution may not suffice to disentangle a young star next to a bright old one. Nevertheless, we remove these stars from our list. We removed the stars: S1-19, S4-287, S7-36, S10-34, and S13-3. 

\section{Analysis: Theory and numerical experiments}\label{sec:analysis}
The initial conditions of a test particle in a fixed gravitational potential have six degrees of freedom, corresponding to the initial position and velocity of the particle. It is standard practice to express those in terms of the classical orbital elements: the semi-major axis $a$, the eccentricity $e$, the inclination $i$, the longitude of the pericenter $\omega$, the position angle of the ascending node $\Omega$, and the epoch of pericenter passage $t_P$. 

\noindent In order to determine these six numbers, one needs to measure six dynamical quantities. From multi-epoch astrometry in the GC, one can determine the on-sky position $(x,y)$ and proper motion $(v_x,v_y)$ of the object. Thus, one needs two more dynamical quantities in order to determine an orbit. From spectroscopy one can get the radial velocity of a star $(v_z)$. The missing $z$-coordinate is not accessible directly at the GC distance of \SI{\sim8.28}{\kilo \parsec} \citep[][]{GravityCollaboration2022_gheziswrong}, but its absolute value can be determined by measuring an acceleration, either by detecting curvature in the on-sky orbital trace, or by a change in radial velocity.

\noindent For stars with (at least) six dynamical quantities measured, standard fitting techniques uniquely determine the orbital elements, see for example \cite{Gillessen2009, Gillessen2017}. If only five dynamical quantities are known, in almost all cases one lacks an acceleration measurement, i.e. information on $z$. Yet, some constraints on the orbital parameters can be constructed. For example the angular momentum vector direction can be limited to lie within a one-dimensional half large-circle across the sphere of possible orientations \citep[][]{Paumard2006, Lu2009}. We call such stars ``5D-constrained''.

\noindent The key finding of earlier works was that when one compares the angular momentum distribution of the 5D-constrained stars one finds an overdensity in angular momentum in specific direction. The simplest explanation for that finding is that these stars rotate in a common disk. This interpretation was independently confirmed by stars for which full orbits have been determined \citep[][]{Gillessen2009, Yelda2014, Boehle2016, Gillessen2017}. 

\noindent The probability distribution of the orbital angular momentum vector for a given star depends on the assumptions one makes on the missing information, i.e. the $z$-coordinate. Hence the exact dynamical structure (and thus the significance of certain features such as disks and warps) of the young star sample depends on the choice of the $z$-prior.

\subsection{Determining the distribution of angular momentum vectors of the 5D-constrained stars}\label{subsection:determination_of_angular_momentum}
In order to estimate the smoothed distribution of angular momentum vectors of the 5D-constrained stars, we use the following procedure:

\begin{enumerate}
    \item Generate 10000 realizations of each star, where the $x$, $y$, $v_x$, $v_y$ and $v_z$ coordinates are sampled from the respective measured values and errors, assuming Gaussian distributions. The $z$ coordinate is sampled from the chosen $z$-prior distribution, see \autoref{sec:z-priors}.
    \item Compute the orbital elements corresponding to the phase space coordinates for each of the 10000 realizations of each observed star. 
    \item Like in \cite{Yelda2014}, we compute the \nth{3}-neighbour density of angular momentum vector directions at the desired grid points over the unit sphere spanned by $(i,\Omega)$ for each of the 10000 realizations of the sample stars. The neighbour density allows to obtain smooth maps from the discrete distributions.
\end{enumerate}

\subsection{Constraining the $z$-values}
\label{sec:z-priors}
In the above analysis, one needs to choose what to assume for the distribution of $z$-values. A natural upper limit on $|z|$ is obtained by imposing that the orbits need to be bound. This yields a maximum $z$-value as a function of projected 2D distance from the massive black hole:
\begin{equation}
   |z_\mathrm{max}| = \sqrt{\left(\frac{2 G M_\bullet}{v_x^2 + v_y^2 + v_z^2}\right)^2 - R^2},
\end{equation}
where $R=\sqrt{x^2+y^2}$ is the 2D projected radius and $M_\bullet$ the mass of the MBH. We use this upper limit when sampling the $z$ coordinate and draw a new $z$-coordinate in case $z$ was sampled outside the allowed bounds. 

\noindent Further, 5D-constrained stars yield an upper limit on the acceleration $a_\mathrm{max}$. This corresponds to a minimum $|z|$ value
\begin{equation}
   |z_\mathrm{min}| = \sqrt{\frac{G M_\bullet}{a_\mathrm{max}} - R^2}.
\end{equation}

\noindent For the distribution of $z$-values between the extreme values two choices have been made in the past: 
\begin{itemize}
    \item The so-called ``stellar cusp prior" \citep[][]{Bartko2009}, which assumes a power-law distribution of $z$-values, based on the observed power-law density profile of the stellar cusp in the GC \citep[][]{Genzel2003_stellarcusp, Gallego-Cano2018, Schoedel2018}. %
    \item A ``uniform acceleration prior" \citep[][]{Lu2009, Yelda2014}, where the z-values are computed from a uniform distribution of accelerations.
\end{itemize}

\noindent In \autoref{appendix:priors} we show that both these priors are not ideal in that they do not recover the isotropic structure in the simulated mock data.
In order to overcome this problem, we propose a new prior that mitigates the problems of the other two:
\begin{itemize}
    \item The ``isotropic cluster prior'' which evaluates the probability distribution function of an isotropic cluster for each star. Since the functions for an isotropic cluster are analytically defined, they can be evaluated using the available observational data of each star. The difficulty lies in expressing the distribution function given in orbital elements, in phase space coordinates, in which our observations are obtained. We derive the procedure in \autoref{subsec:isotropy}.
\end{itemize}

\noindent The choice of a certain prior implies choosing the corresponding null hypothesis against which the dynamical structure can be tested. Thus, the prior choice necessarily introduces some prejudice on what one believes the dynamical structure in GC is.

\subsection{The isotropic cluster prior}\label{subsec:isotropy}
\begin{figure}
    \centering
    \includegraphics[width=0.485\textwidth]{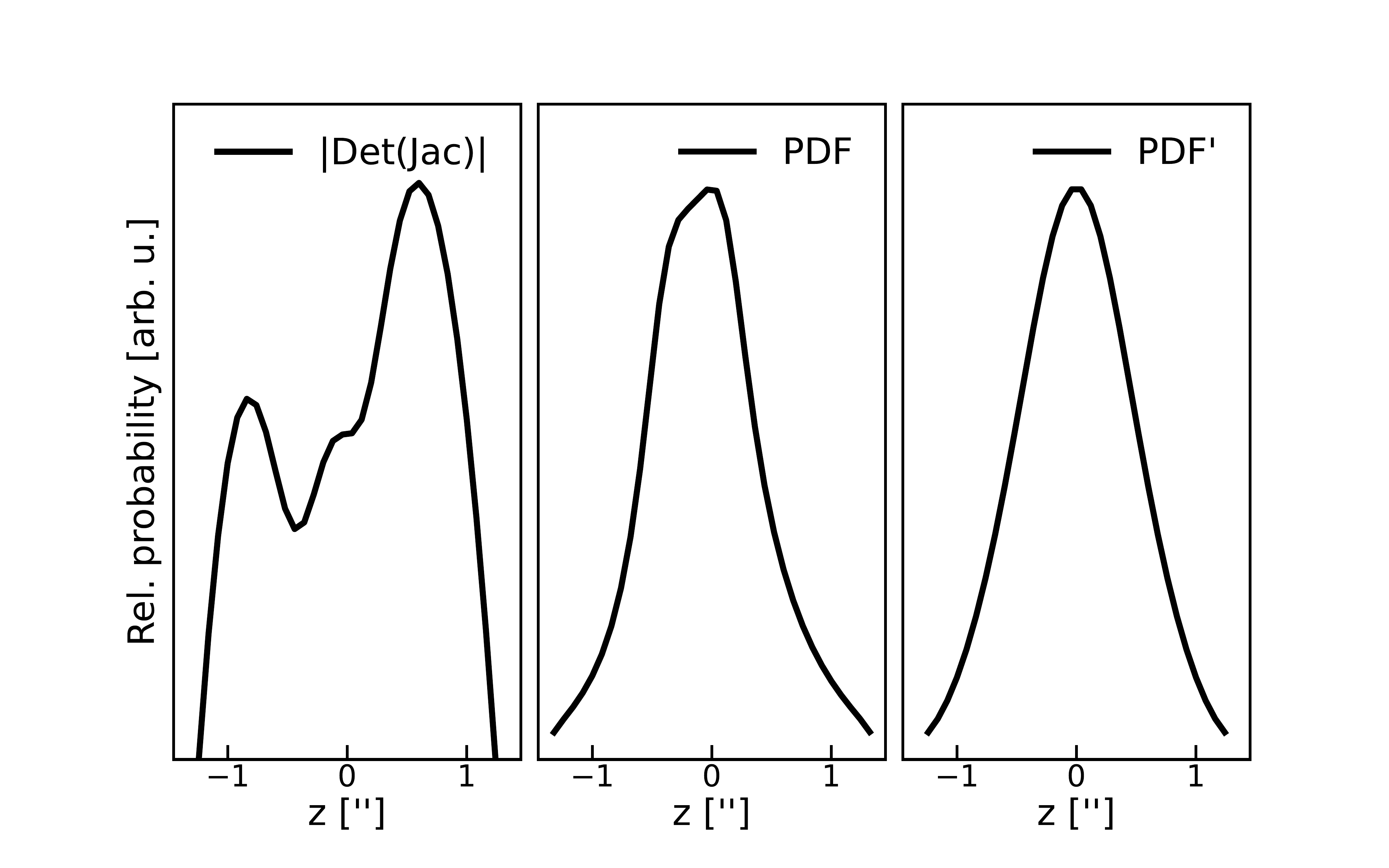}
    \caption[Isotropic cluster prior PDF]{Left to right: Determinant of Jacobian ($|\rm{det}(\rm{Jac})|$), orbital element probability distribution function ($PDF_{\rm{iso-clus.}}$) and coordinate transformed probability distribution function ($PDF'_{\rm{iso-clus.}}$) of the star IRS34W as function of z distance. The transformed $PDF'_{\rm{iso-clus.}}$ is the product of determinant and the isotropic cluster PDF. Note that the y-axis for each plot is different.}
    \label{fig:clusterPrior}
\end{figure}

In \autoref{appendix:isotropic_cluster}, we described the numerical recipe to sample orbital elements for an isotropic cluster from the respective probability distribution functions (PDFs). Since the PDFs are independent, the combined PDF describing an isotropic cluster is simply the product of the individual distributions:
\begin{equation}
\begin{aligned}
    \rm{PDF_{iso-clus.}} =& \rm{PDF}(a) \cdot \rm{PDF}(e) \cdot \rm{PDF}(i) \cdot\\
    &\rm{PDF}(\Omega) \cdot \rm{PDF}(\omega) \cdot \rm{PDF}(\rm{M})\\
    =&\dfrac{p-1}{a_{min}}\left(\dfrac{a}{a_{min}}\right)^{-p}\cdot 2\sin(i)\cdot\\
    &\mathcal{U}_{[0, 2\pi[}(\Omega)\cdot\mathcal{U}_{[0, 2\pi[}(\omega)\cdot\mathcal{U}_{[0, 2\pi[}(M),
    \label{eqn:isotropicClusterPDF}
\end{aligned}
\end{equation}
where $\mathcal{U}$ stands for the uniform distribution on the respective interval.  We need to express the known distributions of orbital elements as distributions of phase space coordinates. 
Effectively, we are thus interested in the correct coordinate transformation of the orbital element distributions to phase space coordinate distributions. 
A probability distribution can be transformed to a different coordinate system by accounting for the volume filling factor:

\begin{align}
    \rm{PDF_{sys1}} = |det(\rm{Jac})| \cdot \rm{PDF_{sys2}},
\end{align}
where $\det(\mathrm{Jac})$ stands for the determinant of the Jacobian matrix of the coordinate transformation. In the case of the orbital element coordinate transformation, the Jacobian matrix consists of the $36$ partial derivatives of the coordinate transforms. Once the analytical form of the determinant has been determined, we obtain the analytical expression for the $z$ distribution of stars in an isotropic cluster:
\begin{equation}
\begin{aligned}
    \rm{PDF'_{\rm{iso-clus.}}}(z | \rm{pc_{obs}}) = & |\rm{det}(\rm{Jac}(a(z| \rm{pc_{obs}}), \dots)|\cdot \\
    &\rm{PDF_{\rm{iso-clus.}}}(a(z| \rm{pc_{obs}}), \dots),
    \label{eqn:isotropicCluster}
\end{aligned}
\end{equation}
where $\rm{pc_{obs}}$ stands for the observed phase space coordinates $x_{obs}$, $y_{obs}$, $vx_{obs}$, $vy_{obs}$, and $vz_{obs}$. We implement the determinant in compiled C, which allows very fast evaluation.
\autoref{fig:clusterPrior} shows an example of the probability distribution of the z values for the star IRS34W \citep[][]{Paumard2006}. Without the determinant the function is asymmetric, only after multiplication with the determinant the PDF is symmetric around $z=0$. 

\subsection{Significance of observations}\label{subsec:isotropicCluster_generation}
The null hypothesis we test is an isotropic cluster, since an old, relaxed distribution should reach asymptotically that state \citep[][]{Bahcall1976, Pfuhl2011}. The procedure to generate an isotropic cluster is described in \citet{Schodel2003}, and in \autoref{appendix:isotropic_cluster}. Once a mock observation of an isotropic cluster has been generated, we determine its angular momentum distribution the same way as for the real observations outlined in the last section. We use $10000$ mock observations to calculate the mean and standard deviation in each pixel of the $(i, \Omega)$ map and define the pixel significance $\sigma_{\rm{pixel}}$ as:

\begin{equation}
    \sigma_{\rm{pixel}} = \dfrac{s_{\rm{pixel, obs}} - <s_{\rm{pixel, sim}}>}{RMS(s_{\rm{pixel, sim}})}
\end{equation}

\noindent where $s_{\rm{pixel, obs}}$ stands for the observed pixel value in the $(i, \Omega)$ map, and $s_{\rm{pixel, sim}}$ stands for the simulated pixel values. This approach is based on the method described in \cite{Li1983}. In \autoref{appendix:bartko_v_yelda} we show that the derived $\sigma_{\rm{pixel}}$ do not correspond to Gaussian confidence intervals. We use $100000$ mock observations to convert the derived $\sigma_{\rm{pixel}}$ to confidence values and in the following state both values. Note that \cite{Yelda2014} use a different definition significance based on peak values observed in the $(i, \Omega)$ maps. We give conversion estimates between these differently derived significance estimates and discuss the consequences of these different definitions in \autoref{appendix:bartko_v_yelda}. 

\section{Results: Application to data}
\subsection{Angular momentum distribution of the young stars in the Galactic Center}\label{sec:angular_momentum_distribution_gc}
\begin{figure*}
    \centering
    \includegraphics[width=0.95\textwidth, trim={7cm 0cm 7cm 0cm}]{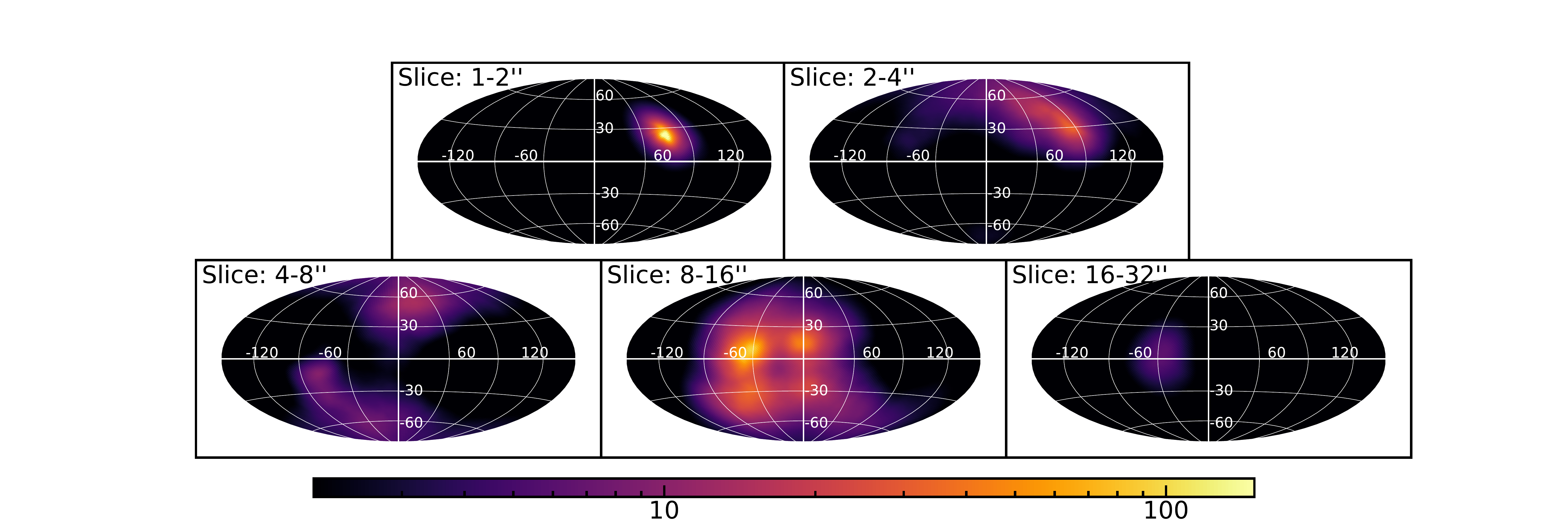}
    \caption[Significance of the overdensity in logarithmic scale of the angular momentum distribution]{Significance ($\sigma_{\rm{pixel}}$) of the overdensity of the angular momentum distribution in logarithmic scale as function of projected radius slice, computed using the isotropic cluster prior. See text for details.}
    \label{fig:significance}
\end{figure*}

In the following section we discuss the angular momentum distribution of young stars in the GC. \autoref{fig:significance} shows the overdensity of angular momentum for six projected radius slices. The density maps are computed using a $3$-nearest neighbour smoothing (\autoref{subsection:determination_of_angular_momentum}). For stars with determined orbits we sample 100 realizations of the angular momentum vector from the respective uncertainty estimates. For stars without determined orbits, we sample 100 z values from the isotropic cluster prior, and 100 realizations of the measured phasespace vector from the respective uncertainties. We bin our young star samples in six bins with increasing projected distance from the black hole. \autoref{tab:overdensities} summarizes the structures discussed in the following subsections and states the significance of the features expressed in $\sigma_{\rm{pixel}}$ as well as a confidence limit (see \autoref{appendix:bartko_v_yelda} for details of the confidence calculation).

\subsubsection{The inner region}
Our updated star sample confirms the presence of a warped clockwise disk for stars in a region ranging from $\sim 1\mathrm{''}$ to $\sim 4\mathrm{''}$ (middle and right plot of top panel in \autoref{fig:significance}). In this inner most region, most stars are aligned coherently. We call this the inner part of the clockwise disk, abbreviated CW1. For the radial bin ranging from $2\mathrm{''}$ to $4\mathrm{''}$, CW1 is less dominant and starts to change smoothly to the outer part of the clockwise disk (CW2).

\begin{figure}
    \centering
    \includegraphics[width=0.48\textwidth, trim={5cm 0 5cm 0},clip]{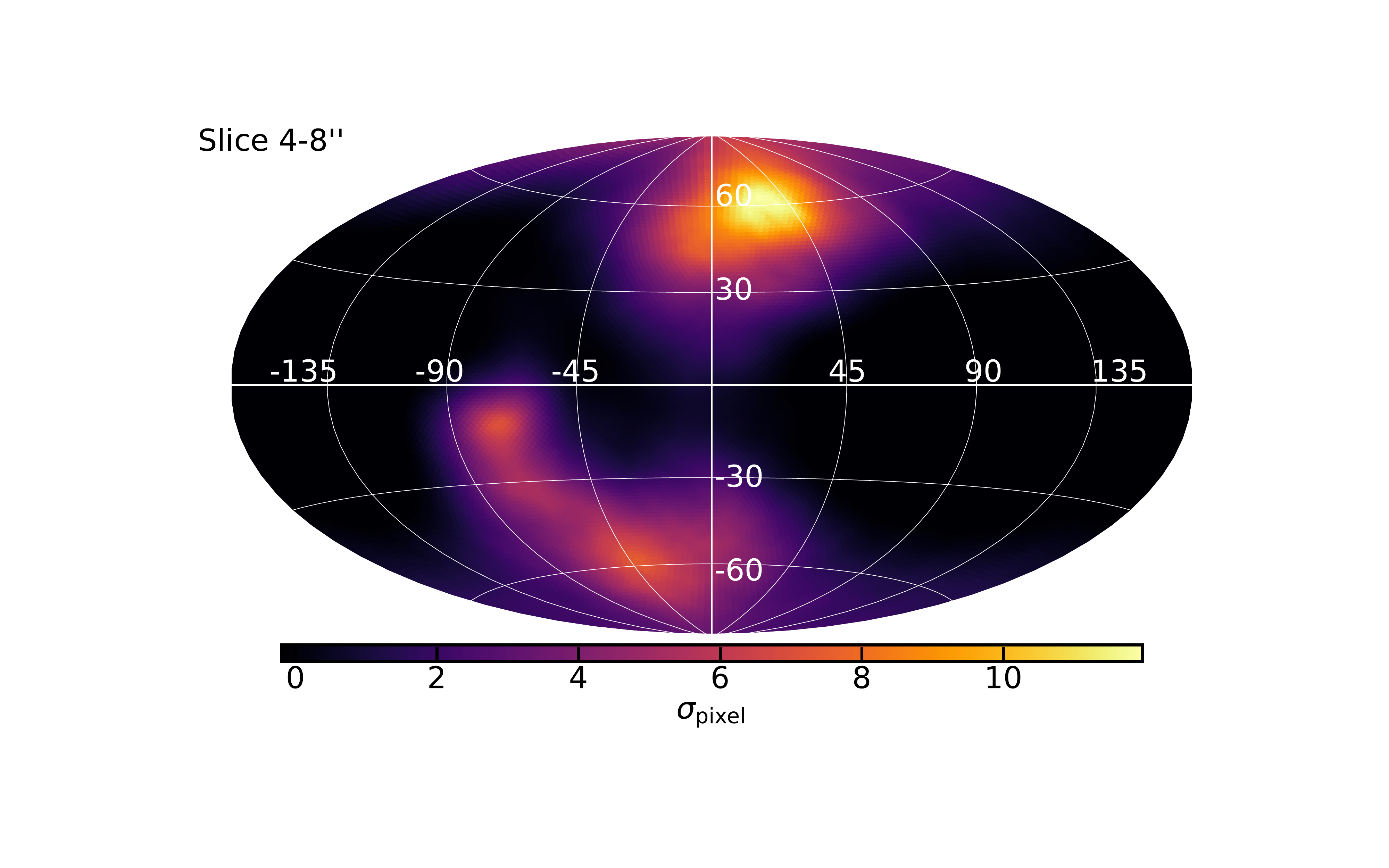}
    \caption[Significance ($\sigma_{\rm{pixel}}$) of the over-density of angular momentum for stars from $4\mathrm{''}$ to $8\mathrm{''}$.]{Significance ($\sigma_{\rm{pixel}}$) of the over-density of angular momentum for stars at a projected distance from $4\mathrm{''}$ to $8\mathrm{''}$. The figure is identical to the bottom-left most panel of \autoref{fig:significance}, however the color scaling is adapted and in linear scale.}
    \label{fig:significance48}
\end{figure}

\subsubsection{The intermediate region}
For the radial bin ranging from $4\mathrm{''}$ to $8\mathrm{''}$ (bottom left panel in \autoref{fig:significance} and \autoref{fig:significance48}), no single disk structure dominates the density map. \cite{Bartko2009} and \cite{Bartko2010} found an overdensity for their sample of 30 stars in the range $3.5\mathrm{''}$ to $7\mathrm{''}$. They interpret this as a warped extension of the clockwise disk -- here called CW2. The significance of the outer part of the warped clockwise disk was estimated at $\sim 6\sigma$ using the stellar cusp prior \citep[][]{Bartko2009}. However, this outer part was disputed by \cite{Yelda2014}. We confirm the CW2 disk at a significance of $\sim 12\sigma_{\rm{pixel}}$ corresponding to a p-value of $p\sim99.2$ (\autoref{fig:significance48}). Further, we find the onset of the counter-clockwise feature (CCW/F1) at a significance of $\sim 10\sigma_{\rm{pixel}}$ reported by \cite{Genzel2003_stars, Paumard2006, Bartko2009} in this intermediate region. 

\subsubsection{The outer region}
\begin{figure}
    \centering
    \includegraphics[width=0.48\textwidth, trim={5cm 0 5cm 0},clip]{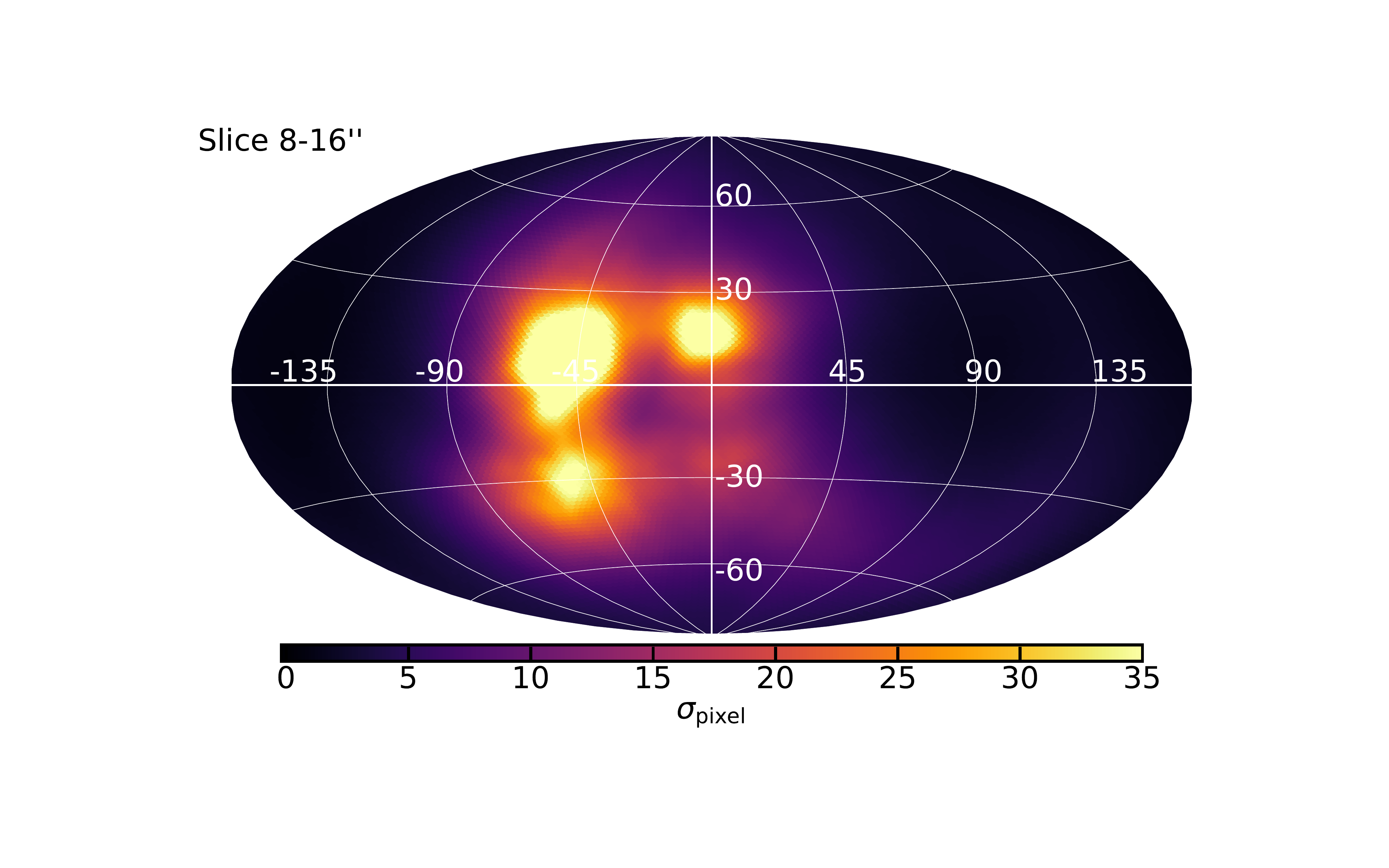}
    \caption[Same as \autoref{fig:significance48} from $8\mathrm{''}$ to $16\mathrm{''}$.]{Same as \autoref{fig:significance48} projected distance from $8\mathrm{''}$ to $16\mathrm{''}$, again with adapted color scaling and in linear scale.}
    \label{fig:significance816}
\end{figure}

For projected distances larger than $8\mathrm{''}$ we find three prominent features (bottom middle panel in \autoref{fig:significance} and \autoref{fig:significance816}). First, we confirm an overdensity of angular momenta at $(\phi,\theta) = \sim (0\degree, 16\degree)$ significant at $\sim 35\sigma_{\rm{pixel}}$. This feature was first reported in the outer most radial bin studied by \cite{Bartko2009} and attributed to the clockwise disk. We call this feature the outer filament 2 (F2). Second, we find the outer continuation of the CCW/F1 feature with similar significance ($\sim 35 \sigma_{\rm{pixel}}$). Most prominently, we find a previously un-reported feature at $(\phi, \theta)=\sim (-44\degree, 12\degree)$ at very high significance of $>100\sigma_{\rm{pixel}}$. We call this feature the second outer filament -- F3. 

The location and significance of all these overdensity features is tabulated in \autoref{tab:overdensities}.
\begin{table*}
    \centering
    \begin{tabular}{l|c|c|c|c|c}
         Name & Abb. & ($\theta, \phi)$ & $\sigma_{\rm{pixel}}$ & Approx. confidence  & Approx. Gaussian $\sigma_{\rm{Gauss}}$\\
         \hline
         Inner clockwise disk & CW1 & $(73.1\degree , 34.5\degree)$ & $>100~\sigma_{\rm{pixel}}$& $\gg 99.999\%$  & $ \gg 4 ~\sigma_{\rm{Gauss}}$\\
         Outer clockwise disk & CW2 & $(23.3\degree , 55.6\degree)$ &$\sim12~\sigma_{\rm{pixel}}$& $\sim99.2\%$ & $\sim 2.5~\sigma_{\rm{Gauss}}$ \\
         Counter-clockwise disk/filament & CCW/F1  & $(-47\degree , -30.0\degree)$ & $\sim35~\sigma_{\rm{pixel}}$& $\sim99.99\%$ & $\sim 4~\sigma_{\rm{Gauss}}$\\
         Outer filament 2& F2 & $(0.0\degree , 16.0\degree)$ & $\sim35~\sigma_{\rm{pixel}}$& $\sim99.99\%$ & $\sim 4~\sigma_{\rm{Gauss}}$\\
         Outer filament 3& F3 & $(-44.2\degree , 11.5\degree)$ & $>100~\sigma_{\rm{pixel}}$& $\gg99.999\%$ & $ \gg 4 ~\sigma_{\rm{Gauss}}$\\
    \end{tabular}
    \caption[Angular momenta direction, derived pixel significance (\autoref{eqn:pixel_sig}) and the corresponding confidence limits and Gaussian $\sigma_{\rm{Gauss}}$ for the different kinematic features in the Galactic Center]{Angular momenta direction, derived pixel significance (\autoref{eqn:pixel_sig}) and the corresponding confidence limits and Gaussian $\sigma_{\rm{Gauss}}$ for the different kinematic features in the Galactic Center.}
    \label{tab:overdensities}
\end{table*}

\section{Results: Estimating the disk membership fraction}\label{sec:estimating-disk-membership}
In order to assess the disk membership of each star to one of the features, we numerically integrate the star's PDF to calculate the Bayesian evidence. We do so using the statistical software package dynesty \citep[][]{Speagle2020_dynesty, Skilling2004_dynesty, Skilling2006_dynesty, Skilling2006_dynesty2}. We sample the likelihood function in phase space coordinates, which allows sampling only the allowed part of phase space and include the prior information on the feature location as additional term in the likelihood function $L_{\rm{feat}}$, which is a Gaussian prior with a specified width at the feature location. We do not impose other prior knowledge the stars other than the disk location and thus we assume a flat prior on the z coordinate of each star, constrained only by the maximum z-distance allowed for a bound orbit. Further, we assume a flat prior on the remaining phase space coordinates with width equal to four times the standard deviation of each coordinate expectation value. In contrast to the isotropic cluster prior used in \autoref{sec:z-priors}, the volume element is numerically derived by the evaluation of the likelihood function by dynesty. Explicitly, we integrate the following likelihood function:
\begin{figure}
    \centering
    \includegraphics[width=0.485\textwidth]{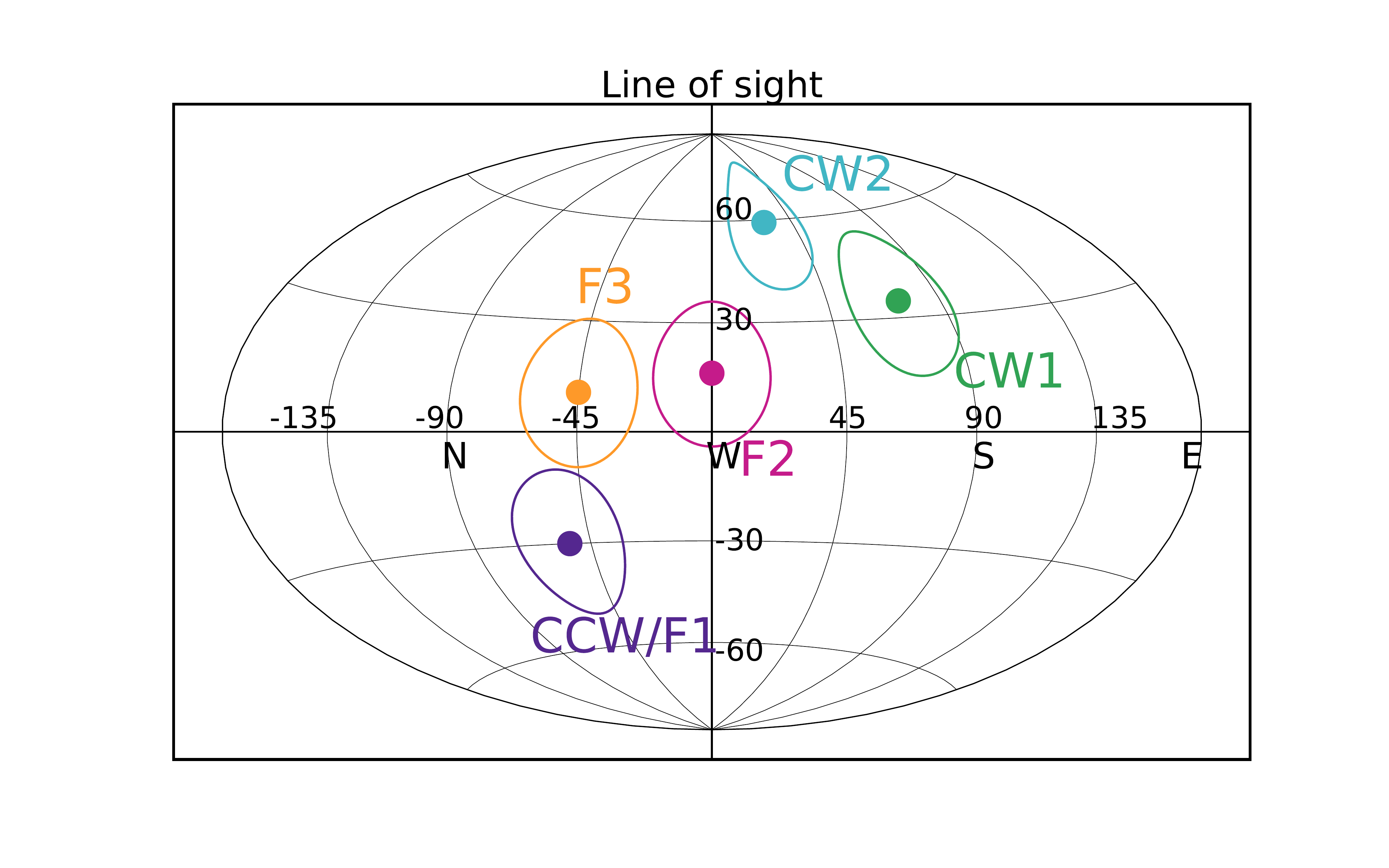}
    \caption[Prior location for angular momentum overdensity]{Illustration of the location and width of the feat priors assumed, tabulated in \autoref{tab:overdensities}.}
    \label{fig:prior_loc}
\end{figure}
\begin{equation}
    \begin{aligned}
        \log L_{\rm{model}} =& \log L_{\rm{feat}} + \log L_{\rm{star}} \\
               =& -0.5 (d(i_{\rm{feat}}, \Omega_{\rm{feat}}, i(x, \dots), \Omega(x, \dots))^2/\sigma_{\rm{feat}}^2 \\
                &+ \sum_n (x_{n, \rm{obs}}-x_i)^2/\sigma_{x_n,\rm{obs}}^2),
               \label{eqn:likelihood_stardisk}
    \end{aligned}
\end{equation}
where $d(i_{\rm{feat}}, \Omega_{\rm{feat}}, i(x, \dots), \Omega(x, \dots))$ stands for the spherical cap distance\footnote{i.e. the Haversine distance} from the feat angular location $(i_{\rm{feat}}, \Omega_{\rm{feat}})$, computed for each sample of the phase space coordinates $(x, y, z, v_x, v_y, v_z)$. $\sigma_{\rm{feat}}$ is the opening angle of the feat, which we set to $20\degree$ for all features. We illustrate the feat priors in \autoref{fig:prior_loc}. 

Similarly, we can define the likelihood of a star without the disk prior:
\begin{equation}
    \begin{aligned}
        \log L_{\rm{star}}  &= \sum_i (x_{i, obs}-x_i)^2/\sigma_{x_i}^2.
       \label{eqn:nullhypthosis_stardisk}
    \end{aligned}
\end{equation}
Because we integrate the same phase-space ($pc$) prior volume for each star, the log evidence evaluates to the same value $\int \log L_{\rm{star}}~ d\vec{pc} = -5.8$ for each star. For stars with orbital solutions it suffices to sample $i$ and $\Omega$. \autoref{eqn:likelihood_stardisk} thus can be rewritten as:
\begin{equation}
    \begin{aligned}
    \log L_{\rm{model}} =& \log L_{\rm{feat}} + \log L_{\rm{star}} \\
               =& -0.5\times (d(i_{\rm{feat}}, \Omega_{\rm{feat}}, i, \Omega))^2/\sigma_{\rm{feat}}^2 +\\
               & (i_{\rm{obs}}-i)^2/\sigma_{ i_{\rm{obs}}}^2 + (\Omega_{\rm{obs}}-\Omega)^2/\sigma_{\Omega_{\rm{obs}}}^2),
    \end{aligned}
\end{equation}

The corresponding integral to \autoref{eqn:nullhypthosis_stardisk} for the stars with determined orbits evaluates to a log evidence of $\sim-2.3$. 

\noindent By comparing the log evidence of $L_{\rm{model}}$ with the log evidence $L_{\rm{star}}$, we can compare the feature membership probability for stars with and without orbital solutions.
In order to establish feature membership, we require the difference of the log evidence $L_{\rm{model}} - L_{\rm{star}}$ to be smaller than $2$. Essentially our procedure corresponds to a log-likelihood cut. However, it can be viewed from a Bayesian model selection point of view where one compares evidence ratios: If the relative log evidence of $L_{\rm{star}}$ and $L_{\rm{model}}$ is smaller than 2, the star is consistent with belonging to the respective feature.

\autoref{tab:cw_disk} and \autoref{tab:inner_warp} report stars consistent with belonging to the CW1 and CW2, \autoref{tab:ccw_disk} reports the stars consistent with belonging to the CCW/F1, \autoref{tab:outer_warp} and \autoref{tab:new_disk} reports the stars consistent with being on the F2 and F3 features.

\subsection{Is it necessary to de-bias the disk fraction?}
\cite{Yelda2014} estimated the true disk fraction by comparing the observed distribution against their simulations of an (approximate) isotropic cluster mixed with a stellar disk. This approach is correct under the assumption that the young stars not aligned with the disk are in an isotropic distribution. If the distribution of the not-aligned stars is anisotropic, for instance if several streams of stars exist, this approach underestimates the number of disk members. We thus do not estimate the true disk fraction under the assumption of a single disk + isotropic cluster model. Consequently, we can not tell the difference of a by-chance aligned star from that of a true feature member. Our feature fraction estimate is $100\%$, which should be understood as an upper limit. Further, we impose that each star is at most member of one feature. If a star could be associated with more than one feature, we count it to the feature with the lowest $\Delta$ evidence.

\noindent Ultimately, our feature membership depends on the prior width and location of the features and the evidence cut. Optimally, these feature properties should be inferred from the data too, which however requires a hierarchical approach which is beyond the scope of this work.

\section{Results: Properties of the young stellar components}\label{sec:properties_of_young_stars}
Marginalizing the prior and likelihood functions in \autoref{eqn:likelihood_stardisk}, we obtain posterior phasespace distributions for each star. For the stars that satisfy our feature membership criterion, we compute the orbital elements and obtain the posterior density estimate of orbital elements. Because all stars are independent from one another, we can combine the posterior estimates from each sample and obtain the joint orbital element distribution. 

\noindent In the following we analyse the posterior semi-major axes and eccentricity distributions as well as their luminosity distributions. In this discussion, only the prior on the preferred direction of angular momenta and our observational data enters (\autoref{tab:summary_disk}). We do not require stars to have certain projected distances or eccentricities. However, we require that all stars belong to at most one feature. Further we remove B type stars that lie within $0.8''$ of Sgr~A*. These so called S-stars have a relaxed angular momentum distribution \citep[e.g. ][]{Gillessen2017}, thus we assume that they their alignment with the any of the features is by chance. In the following observed number counts and summary statistics will be presented, without the inclusion of a completeness correction.

\begin{table*}[]
    \centering
    \begin{tabular}{c|c|c|c|c}
         Disk Name & Number  & Brighter/Fainter  & IQR  & IQR \\
         & of Stars& than $\mathrm{K_{mag}}=14$ & semi-major axes & eccentricity \\
         \hline
         CW1   & $33$ & $24/9=2.7$ & $1.6'',~2.1'',~2.7''$  & $0.4,~0.5,~0.6$\\
         CW2   & $13$ & $12/2=6 $ & $5.8'',~7.0'',~8.5''$  & $0.2,~0.4,~0.5$\\
         CCW/F1 & $33$ & $21/12=1.8$ & $5.4'',~7.4'',~12.1''$ & $0.4,~0.5,~0.6$\\
         F2   & $37$ & $23/8=2.8$ & $2.2'',~5.6'',~9.3''$  & $0.4,~0.6,~0.9$\\
         F3  & $36$ & $20/16=1.3$ & $3.7'',~8.8'',~12.4''$ & $0.6,~0.7,~0.9$\\
    \end{tabular}
    \caption[Kinematic features of the young star population in the Galactic Center]{Significant kinematic features of the young star population in the Galactic Center. Number of stars, luminosity, semi-major axis and eccentricity distribution. The number counts of stars are the observed values, i.e. no completeness correction is applied.}
    \label{tab:summary_disk}
\end{table*}

\subsection{The warped clockwise disk and its stars}
Our updated star sample confirms the presence of a warped clockwise disk. 
We do not model the clockwise disk with a ``warp'', i.e. smooth change of angular momentum. Instead, we define two angular momentum directions motivated from the observed overdensities of angular momentum compared to an isotropic cluster (see \autoref{fig:significance} and \autoref{tab:overdensities}).
This allows to check if stars consistent with belonging to the respective features are indeed similar to one another, without imposing the ``warpedness'' already in the prior. 

\noindent We find $33$ stars which are consistent with being in the CW1 disk.
Only four stars are at a projected distance greater than $4\mathrm{''}$, with the largest projected distance being $\sim10\mathrm{''}$. The median projected distance is $2.0\mathrm{''}$, the interquartile range (IQR) of the clockwise disk feature stars is $1.0\mathrm{''}$ and $3.2\mathrm{''}$. The majority of the clockwise disk stars are bright: $24$ of the $33$ stars are brighter than $K_{mag}=14$. Nevertheless, there is no distinctive brightness cut that leads to disk membership: nine stars are fainter than $K_{mag}=14$, six of which have determined orbital solutions (\rm{R1, S5, S11, S31, S66, S87}). The  median $K_{mag}$ of the clockwise disk is $12.7$. 

\begin{figure}
    \centering
    \includegraphics[width=0.485\textwidth]{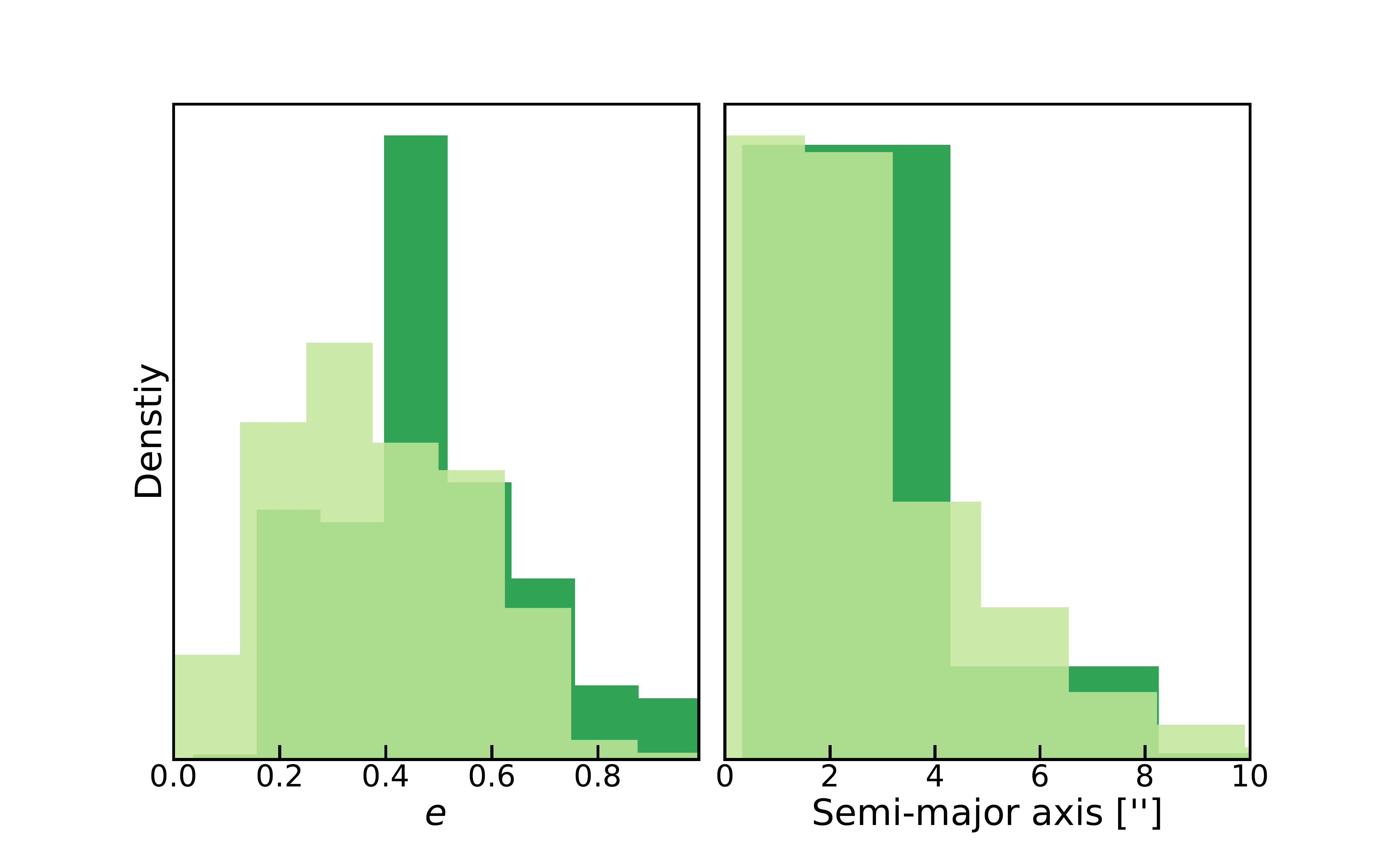}
    \caption[Posterior distribution of orbital elements of inner warped clockwise disk stars]{Distribution of the eccentricity and the semi-major axis of the stars consistent with belonging to the inner part of the warped clockwise disk. The dark green histograms show the properties of the 5D-constrained stars, the light green histogram show the distribution of stars with determined orbital solutions. The grey and black vertical lines indicate the median, and the dashed lines indicate the IQR of the stars with and without orbit.}
    \label{fig:cwdisk_elements}
\end{figure}

\noindent Combining the posteriors we obtain the joint distribution of orbital elements. For stars with determined orbital solutions, we sample orbital elements from the respective orbit posterior distributions. \autoref{fig:cwdisk_elements} plots the distribution of the eccentricities and the semi-major axes. The stars without determined orbits typically have non-zero eccentricities, with a median eccentricity $\sim 0.5$, highly eccentric orbits are however not preferred by our data. The median semi-major axis is $2.1\mathrm{''}$. The distribution of the stars with determined orbital solutions broadly agrees with those without a fully determined orbit. The stars with determined orbital solutions have however slightly lower eccentricities, and very high eccentricities are completely suppressed.

In summary, the CW1 disk is made up of predominantly, but not exclusively, O/WR type stars, on eccentric orbits in the proximity of the black hole.

$13$ stars are consistent with belonging to the CW2 disk. All but two stars are brighter than $K_{mag}=14$. One faint star is S60 and belongs to the S-star cluster, so we remove it from the sample. The median magnitude is $12.7$. All but one star have projected distances ranging between $4\mathrm{''}$ and $\sim 8.6\mathrm{''}$, the median projected distance is $\sim6.6~\mathrm{''}$. 

\noindent The eccentricity distribution is comparable to the eccentricity distribution of the CW1 disk, with median eccentricity of $\sim 0.4$, high eccentricities are not favoured by our data; the median semi major axis is $7.0\mathrm{''}$. The star R70 has a determined orbital solution with a semi-major axis of ${\sim}3.5\mathrm{''}$ and an eccentricity of $0.3$, consistent with the 5D-constrained stars.

\noindent In summary, the CW2 disk consists of eleven O/WR stars and one B stars, on slightly eccentric orbits. The stars are bright, very similar to the CW1 stars, but are found at larger radii. 

\begin{figure*}
    \centering
    \includegraphics[width=0.485\textwidth, trim={2.5cm 2.5cm 2.5cm 2.5cm},clip]{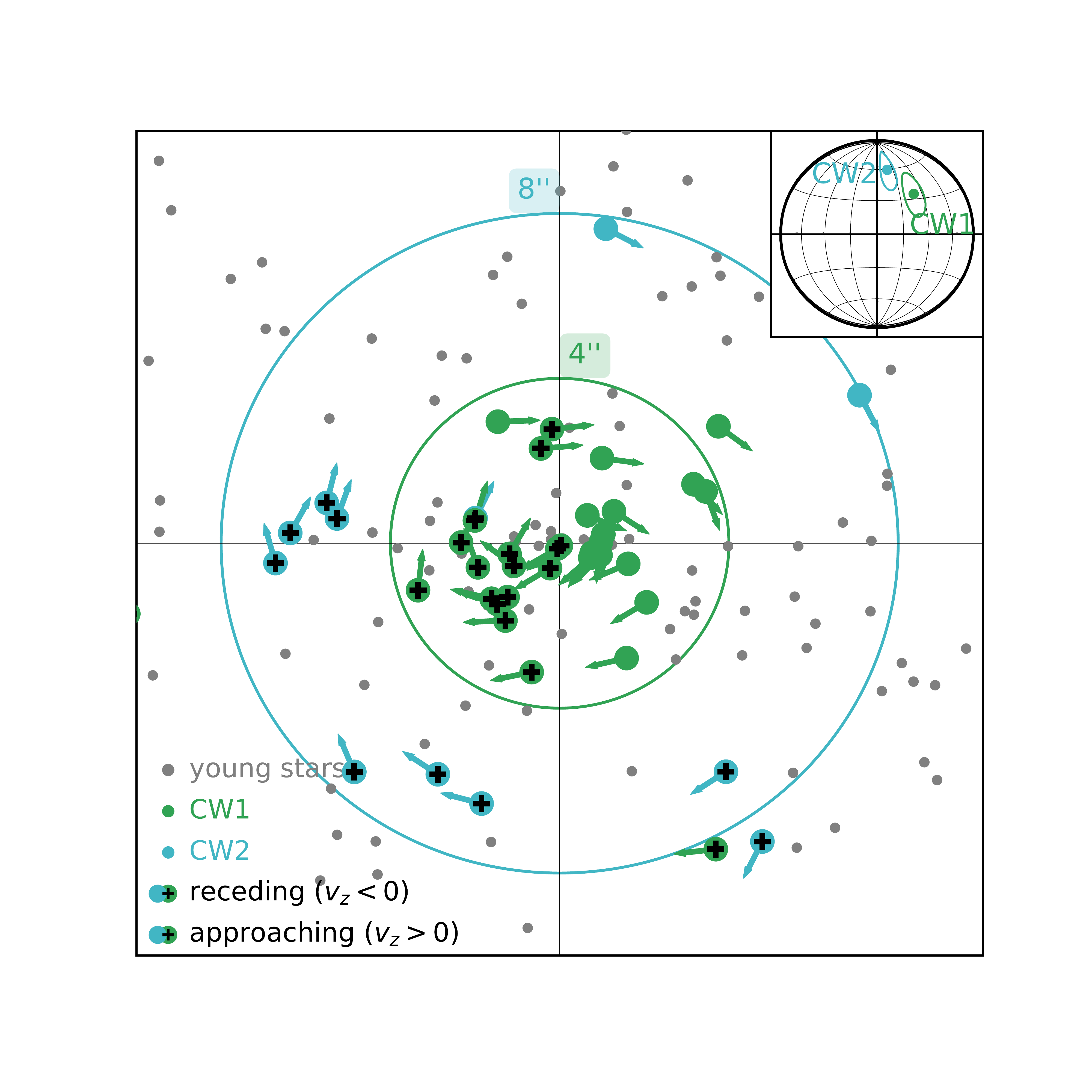}
    \caption{Comparison of stars belonging to the CW1 (green) and CW2 (blue) disk. Stars that do not belong to either feature are plotted in gray. The inset shows the prior width and direction of the feature prior $L_{\rm{feat}}$ based on which the colored stars are selected (i.e. $\Delta$ log evidence $> 2$, \autoref{sec:estimating-disk-membership}). Stars marked with a ``plus'' have positive radial velocity, unmarked stars have a negative radial velocity. The arrows indicate the direction of the projected velocity. In both cases we evaluate the disk membership probability for all young stars, i.e. no radial binning has been applied.}
    \label{fig:comparison_inner_outer_cw}
\end{figure*}

\autoref{fig:comparison_inner_outer_cw} demonstrates the morphological difference between the inner and the outer part of the warped clockwise disk. All but three stars belonging to the inner part of the disk are found centralized within $5''$. 
The stars in the outer part are thus only different by their angular momentum direction as function of radius. This is consistent with a ``warp-picture'', which is a sole result of the data and not the prior. The feature prior $L_{\rm{feat}}$ does not impose a radial dependence, or a magnitude selection.

\begin{figure*}
\centering
\subfloat[]{\includegraphics[width=0.7\columnwidth, trim={4.cm 4.cm 4.cm 4.cm},clip]{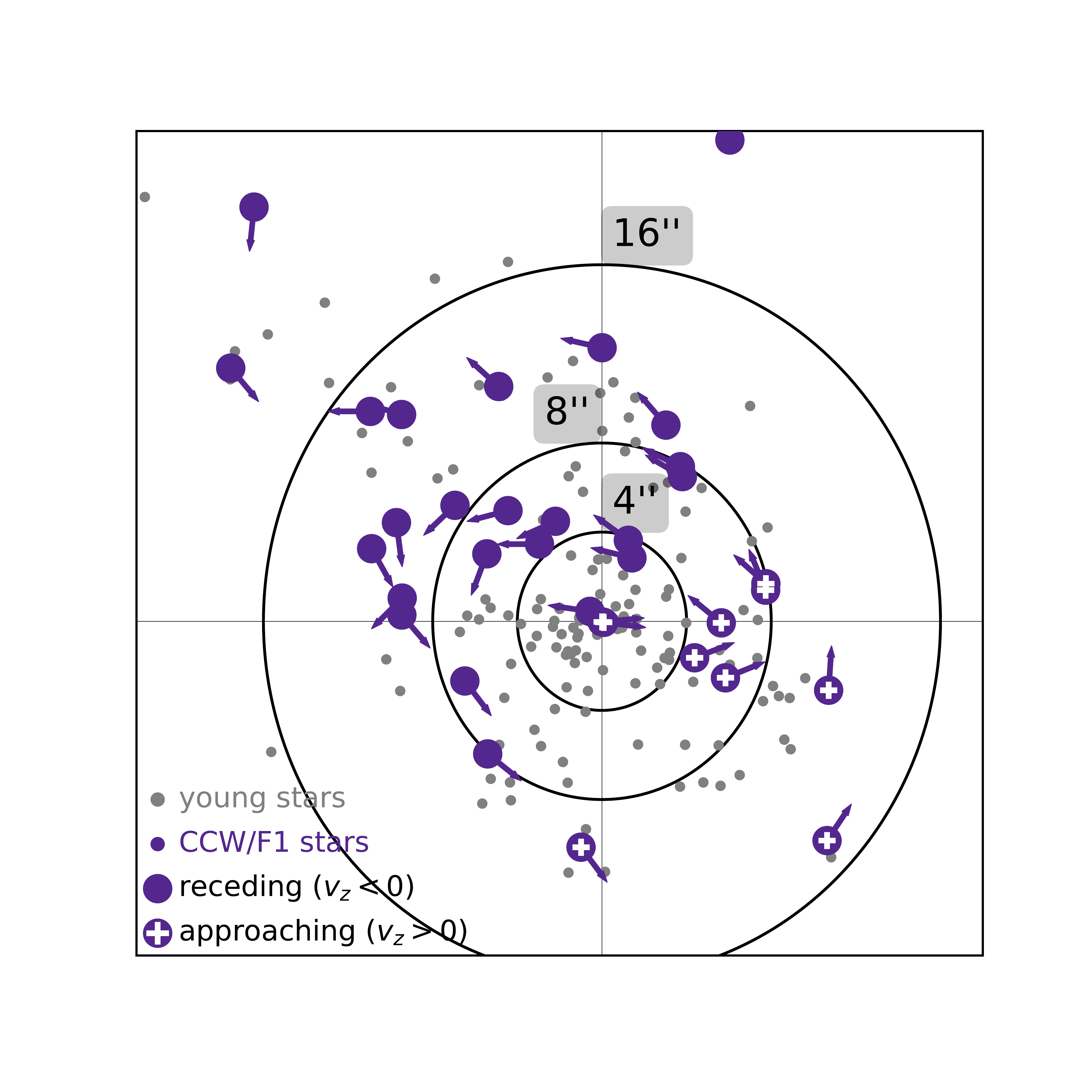}}
\hfill
\subfloat[]{\includegraphics[width=0.7\columnwidth, trim={4.cm 4.cm 4.cm 4.cm},clip]{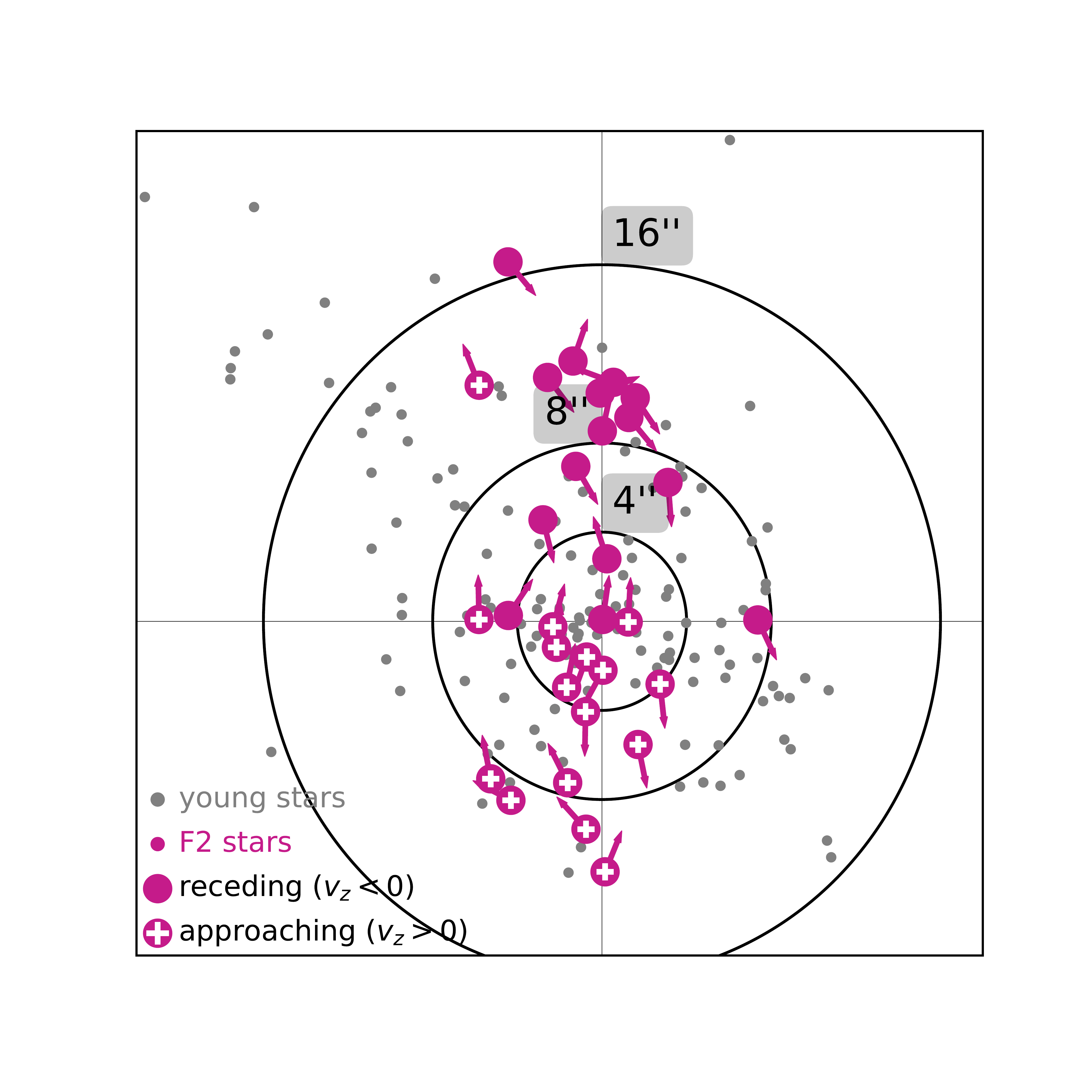}}
\hfill
\subfloat[]{\includegraphics[width=0.7\columnwidth, trim={4.cm 4.cm 4.cm 4.cm},clip]{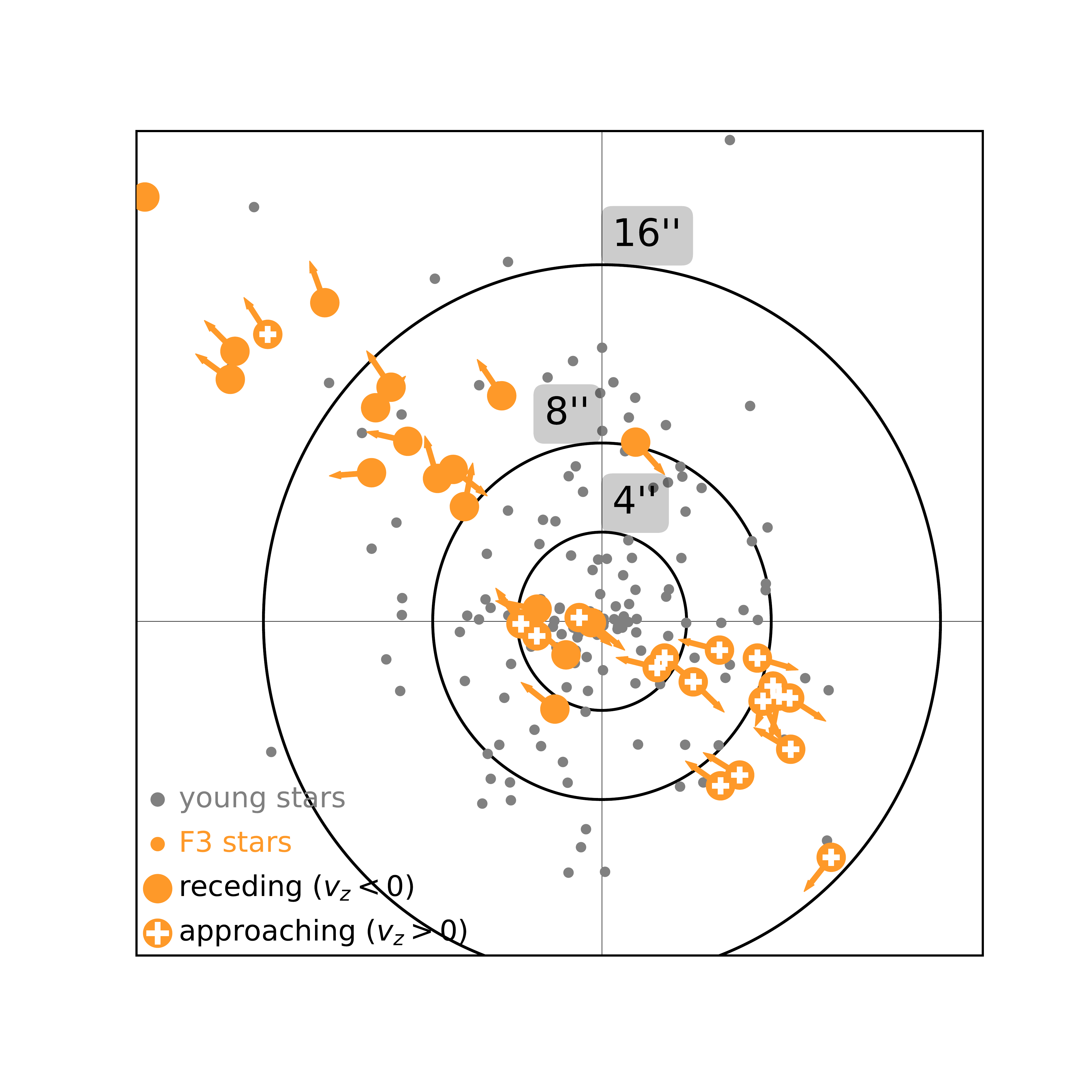}}
\caption{Same as \autoref{fig:comparison_inner_outer_cw} for CCW/F1 (purple), F2 (pink), and F3 (orange).}
\label{fig:star_postion_in_the_GC}
\end{figure*}

\subsection{The CCW/F1 feature and its stars}
$33$ stars are consistent with belonging to the CCW/F1 feature. Only two stars, S4 and S12, have full orbital solutions, both of which are S-stars which we again discard. $12$ stars are brighter than $K_{mag}=14$, $21$ are fainter, and the feature is therefore more skewed towards fainter, B-type stars, compared to the clockwise disk. The median magnitude is $13.5$. This feature contains stars mostly at large projected distance, with a median projected distance of $8.0\mathrm{''}$. 

The stars are on modestly eccentric orbits, with a median eccentricity of $0.5$. Unlike for the CW1/2 features, highly eccentric orbits are not entirely suppressed. However, the fraction of the CCW/F1 stars on highly eccentric orbits (eccentricity $> 0.9$) is small.  
The semi-major axes distribution is similar to the observed projected distances, with a median semi-major axes of $7.4\mathrm{''}$.

\subsection{The F2 feature and it's stars}
$37$ stars are consistent with belonging to the F2 feature. Similar to the CW disk, the majority of stars are brighter than $K_{mag}=14$: eight stars are fainter, 23 brighter. The median magnitude is $13.2$. The stars are at a median projected distance of $7.0\mathrm{''}$.

\noindent The median eccentricity is $0.7$, higher than for the CW disk and the median of semi-major axes is $5.6\mathrm{''}$.

\subsection{The F3 feature and it's stars}
The F3 feature consists of $36$ stars. Two stars have a full orbital solution which however belong to the S-star cluster. Two thirds of the stars belonging to this feature are brighter than $K_{mag}=14$ (16 B-type stars, 20 O/WR-type stars). Like for the CCW disk, the stars are at large projected distances: the median distance is $8.8\mathrm{''}$. 

\noindent Most of the F3  stars are preferentially on highly eccentric orbits, which differentiates the F3 feature from the other features. Indication for such almost radial orbits have already been reported in \cite{Madigan2014}. The median eccentricity is $0.7$. The distribution of semi-major axes is comparable to the observed projected distances with median $7.8$. Intriguingly, the orientation of the F3 feature is in the same plane as the Galaxy, however the sense of rotation is opposite to it \citep[][]{Paumard2006}. 

\subsection{K-band Luminosity function of the young stars}\label{sec:KLF}
Following \cite{Bartko2010}, we define the K-band luminosity function (KLF) as:

\begin{equation}
    KLF(R_1, R_2, m_K) = \dfrac{N_{\rm{stars, obs}}(R_1, R_2, m_K)}{A_{\rm{eff}}(R_1, R_2, m_K)}
\end{equation}

where $N_{\rm{stars, obs}}$ is the number of observed stars in the radial slice ($R_1$, $R_2$) and magnitude bin $m_K$, and $A_{\rm{eff}}$ is the effective area in the radial and magnitude bin.
If our observations were perfect, the effective area would simply depend on the radial slice. However, our spectroscopic integration depth varies from pointing to pointing and our classification fidelity strongly depends on the background gas emission and the presence of bright stars. Thus we have to take the completeness of our observations into account, and the effective area is computed as:

\begin{equation}
    A_{\rm{eff}}(R_1, R_2, m_K) = \int\int_{\sqrt{x^2 + y^2} \geqslant R_1}^{\sqrt{x^2 + y^2} \leqslant R_2} \epsilon(x, y, m_K) ~dx, dy
\end{equation}

\noindent We compute the completeness similar to \cite{Do2013}, comparing the number of photometrically identified stars present in a given SINFONI pointing, with the number of stars we were able to classify spectroscopically. This method allows exploiting the much better photometric information of a star. 

Explicitly, we use the star catalogue presented in \cite{Trippe2008} and assume it to be complete to $K_{mag}=16$. Than, for each pointing, we calculate the ratio of spectropically classified stars to the total number of stars.

\begin{equation}
    \epsilon(x_{\rm{pointing}}, y_{\rm{pointing}}, m_K) = 1 - \dfrac{N_{\rm{unclassified}}(m_K)}{N_{\rm{total}}(m_K)}
\end{equation}
\noindent We assess the uncertainty of the completeness correction by simulating different completeness map based on the number of classified and unclassified stars. We detail the procedure in \autoref{appendix:completeness_error}.

\noindent We compare the KLF found in our work with the ones found by \cite{Do2013} and \cite{Bartko2010} in \autoref{appendix:klf_comparison}. Overall the agreement between our values and those found in \cite{Bartko2010} is good, despite using a different method for determining the completeness correction. 
In contrast to \cite{Do2013}, we do not infer the completeness corrected KLF and completeness correction simultaneously, but calculate it from the star counts once the completeness of each pointing is determined. The most important difference to \cite{Do2013} is that we do not enforce classification into either young or old, but allow a star to remain unclassified. Consequently, we do not differentiate young and old star completeness, but simply calculate the fraction of classified and unclassified stars. This prevents that stars with no signal in the spectrum are classified, and thus conservatively estimates the completeness. The completeness correction is none-negligible for deriving the KLF. For instance the overall completeness in the $K_{\rm{mag}} = 13.5$ to $K_{\rm{mag}} = 14.5$ bin is on the order of $56\%$.

\begin{figure}
    \centering
    \includegraphics[width=0.485\textwidth, trim={0cm 0cm 2.5cm 0cm},clip]{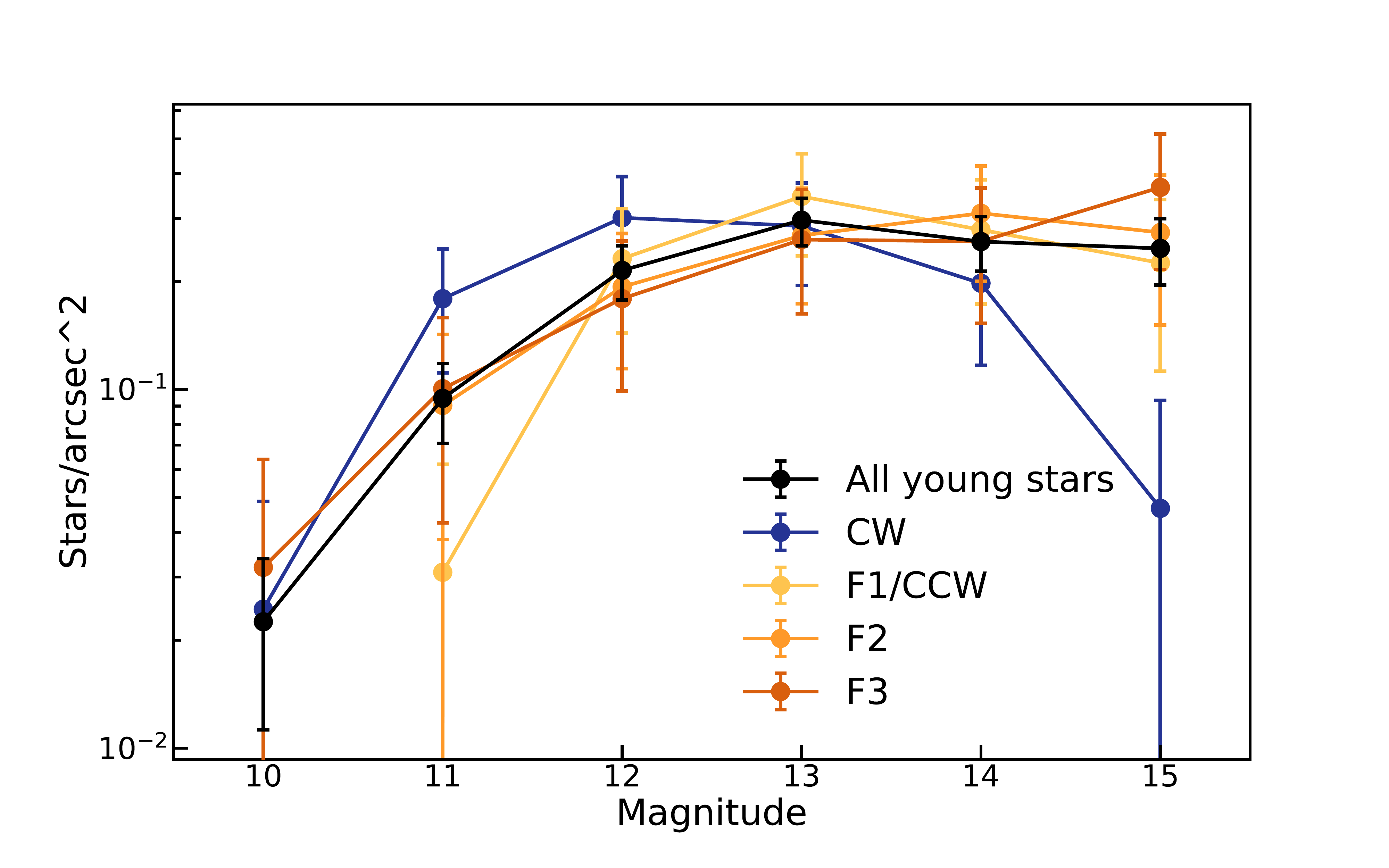}
    \caption{KLFs of stars associated with the different kinematic features and the KLF of all young stars including the central S-stars. No explicit projected radius cut is applied, but we combine the stars of clockwise disk (CW1 and CW2).}
    \label{fig:klf_disks}
\end{figure}

\noindent In order to assess whether the higher O+WR to B star ratio of the CW disk and the other kinematic features is an artifact of the incomplete observations, we plot the completeness-corrected KLF of the stars associated with the respective features as well as the KLF of the all young stars combined (including the central young S-stars) in \autoref{fig:klf_disks}. The KLFs of the outer features are essentially indistinguishable, while the KLF of warped clockwise disk peaks at $K_{mag} = 12$ and is thus top-heavy. The overall KLF follows those of the outer features, as those dominate the star count.

\subsection{Summary of the young stellar components}
This analysis has revealed that the young stars can be categorized into five different significant features. As many as $75\%$ of the young stars ($152$ of $201$) are consistent with belonging to one of these features. Our analysis has shown that these features cannot only be separated by their angular momentum but also by their distance from Sgr~A*. 
The CW disk forms a coherent structure ranging from $\sim1''$ to $\sim 8''$.
The F2 feature has a large rangs from $\sim3''$ to $\sim 10''$.
The CCW/F1, and F2 features extend the furthest from Sgr~A*. We compare the different eccentricity distributions, semi-major axis distributions and the $K_{mag}$ distributions in the appendix (\autoref{fig:comparison_distributions}). 

\section{Discussion}
We carried out a spectroscopic survey of the central $(+20, -10)$, $(-20, +10)$ arcseconds of the GC. We reanalyzed, combined and updated the spectra derived for all GC stars observed with ESO's SINFONI instrument taken in AO mode. This lead to spectra for over $2800$ stars. We classified the stars into old, if CO-band heads are discernible, young, if the Br$\gamma$ line (and other young star lines) are discernible, or unclassifiable, if no line is discernible. This led to the identification of a total of $201$ young stars. For $35$ young stars full orbital solutions can be derived. Three stars have too high radial velocities to be on bound orbits\footnote{This is likely a consequence of a poorly determined radial velocity or a confusion event.}. For the remaining $158$ stars, only radial velocities could be determined. 
We extend previous Monte-Carlo studies presented in \cite{Lu2009, Bartko2009, Bartko2010, Yelda2014} by introducing a new prior. The proposed prior maps an isotropic cluster onto itself without bias in angular momentum. It is not ``better'' for answering the question of the true angular momentum distribution of the young stars. However, it allows for a clean definition of a null-hypothesis: How different is the observed distribution from an isotropic cluster. In particular, we ask how different the observed angular momentum distribution is from that of the old star cluster present in the GC, which is an isotropic cluster to good approximation \citep[e.g.][]{Pfuhl2014}. 

\subsection{Distribution of angular momentum of young stars in the Galactic Center}
We find five significant different kinematic features compared to an isotropic cluster. Further, we have found that the vast majority ($75\%$) of stars can be attributed to one of these five features. 
The angular momentum distribution in the GC is therefore very rich, and significantly different from the old star population. 
We demonstrate that the young stars reside in a warped-disk and several outer filaments. 
Such a rich structure has been proposed by several simulations of in-situ star formation in an in-falling gas cloud scenario.
\cite{Bonnell2008} demonstrated that stars can form in massive gas clouds around a massive black hole like Sgr~A* and speculated that multiple young star rings may be present in the GC. \cite{LockmannBaumgard2009} have demonstrated that in the presence of two separate disk systems (like the clockwise and the counter-clockwise system), the disks tidally interact with one another causing a warping of the disks. Further, \cite{Kocsis2011} show a warping of the disk naturally arises from the interaction of the disk with the potential of the embedding old star cluster. Our observations are fully consistent with the results of these simulations and theoretical arguments.

\subsection{K-band luminosity function}\label{sec:klf_difference}
\autoref{fig:klf_1_12} and \autoref{fig:klf_disks} show the KLF of the young stars within $1$ to $12$ arcseconds and the KLFs of the kinematic components. 
The KLF derived from our revised sample of young stars and revised completeness correction is consistent with the KLF reported in \cite{Bartko2010}. Compared to \cite{Bartko2009} who used source implanting to derive the completeness, and following \cite{Do2013}, we use the information available in photometric observations of the Galactic Center to improve the completeness correction. Nevertheless, the KLF derived here is more top-heavy than the one in \cite{Do2013}.

\noindent Comparing the KLFs of the different kinematic features we find that the combined KLF of the CW1 and CW2 features is more top heavy than that of the other features. The KLFs of the outer features are comparable to the KLF of the inner S-star cluster and the young stars at large separations reported in \cite{Bartko2009} and are thus consistent with a normal Salpeter/Koupra IMF \citep[][]{Salpeter1955, Kroupa2001} of $dN/dm \propto m^{-2.15}$. 
Nevertheless, the statistical significance of this result is low. For instance, the p-value of the K-S-test between the CW disk KLF and the CCW/F1 feature is $0.23$, yielding only marginal evidence that these distributions are different. If one combines the three outer features, the significance is increased (p-value $0.03$). Thus we caution that this result remains tentative, and that the decisive magnitude bin to confirm the down tick of the warped clockwise disk KLF are stars fainter than $K_{mag}=15$ and thus beyond the fidelity of this work. 

Assuming the difference is not an artifact of the completeness correction, two possibilities emerge. If star formation is universally top heavy in the Galactic Center, the absence of brighter stars may be result of a main sequence cut off. This would imply that the outer structures are older and bear witness of star formation event prior to the formation of the clockwise disk stars. Alternatively, the IMF in an accreting gas cloud may be radially dependent and the observed KLF could thus be explained in a singular star formation event. 

\subsection{A warped disk and several filaments of young stars}
In section \ref{sec:properties_of_young_stars} we show that the CW disk forms a coherent structure ranging from $1''$ to $8''$. The stars share very similar eccentricities, have similar angular momentum directions and are predominantly made up of O and WR type stars. They are mainly different by the angular momentum as function of separation from the black hole, consistent with the warp-picture. The other features are harder to explain as an extension or warp of the clockwise disk. 

\noindent The CCW/F1 feature possesses a similar eccentricity distribution as the two inner features but shows a drastically different angular momentum direction.

\noindent The F2 feature shares some similarities with the CW disk and could thus be its outer most extension. For instance, some of the F2 stars have low eccentricities. However, many of the F2 stars have higher eccentricities atypical for the CW disk stars.

\noindent The F3 feature has a different angular momentum direction than the warped clockwise disk, which is not too different from that of the CCW/F1 feature. It is the most eccentric feature. 

\noindent Ultimately, all three outer features share a very similar K-band luminosity function, which appears to be different from the one of the CW disk. 

Several simulations of gas accretion disk produced such rich features \citep[][]{Nayakshin2006, Nayakshin2007, Bonnell2008, LockmannBaumgard2009}.
Of particular interest is the scenario which was studied in \cite{HobbsNayakshin2009}, in which two Giant Molecular Clouds collide. After the initial collision the two clouds are sent on a plunging orbit and accrete onto Sgr~A*. A central accretion disk forms and, depending on the initial conditions, several gas streamers. In both the central disk, as well as the gas-streamers stars subsequently form. 

\noindent Several of the predictions made in this scenario are consistent with our observations. In the simulations with large impact parameter, the inner-most accretion disk stays in a rather compact region around the black hole, consistent with the inner region of the CW disk in the GC. The remnants of the colliding gas clouds form filaments at larger separations, which do not share the same angular momentum direction as the central disk. This could correspond to the stellar populations found in the CCW/F1, F2, and F3 features. Furthermore, the disks found in the simulations show large scale warps, perfectly consistent with the observed change in angular momentum direction of the CW disk. In these simulations, the central disk circularizes after an initial period of highly eccentric orbits, while the stars further out remain on more eccentric orbits. We find a similar behavior, but note that our values are overall more eccentric than found in this set of simulations. 

\noindent In the simulations the star formation is different in the inner and outer regions. While \cite{HobbsNayakshin2009} caution that their star-formation prescription may be oversimplified, in their simulations, mostly heavy stars form in the central disk, and the IMF is substantially less top-heavy in the outer filaments. This is consistent with our observations too: The ratio of observed O+WR to B stars is much higher in the warped clockwise disk than in the outer structures. Our data confirms: while the KLFs of the outer structures are indistinguishable, the KLF of the warped clockwise disks peaks at $K_{mag}=12$, and stars fainter than $K_{mag}=14$ are rare.

While matching the observations well, the effect of the embedding old nuclear star cluster is typically ignored in such simulations of star formation. However, through the process of vector resonant relaxation, the embedding old star cluster facilitates a fast reorientation of the angular momentum direction. This process can lead to a wrapping of initially coherent disk-like structures in time scales comparable to the age of the young stars \citep{Kocsis2011}, and can lead clustering in angular momentum space for stars of different mass \citep{Hungarians2018}. For an isotropic background potential, the timescale at which an initially coherent structure is dissolved increases with separation from the center \citep{Kocsis2015} and the existence of the clockwise disk structure constrains the efficiency of vector resonant relaxation processes in the Galactic Center \citep{Martinez2020}, it is thus not clear if the observed structure can be explained solemnly by the effect of the embedding cluster. 

Another intriguing aspect of young star population is the central S-star cluster: the central $0.8''$ of the Galactic Center is populated by B  type stars, with an approximately isotropic angular momentum distribution, and a super-thermal eccentricity distribution \citep{Gillessen2009, Boehle2016, Gillessen2017}. The very different angular momentum distribution and the evident absence of any WR- and O-type stars on tight orbits poses a significant complication of the physical picture for the formation of these young stars. One solution is presented in \cite{Chen2014}, who propose a so-called rapidly evolving region (RER) in the inner region of clockwise disk. If the initial disk is sufficiently heavy, stars within the RER rapidly exchange eccentricity and semi-major axis through Kozai-Lidov like resonances. This reorients the orbits into the observed super-thermal, isotropic distribution. The presence of the outer filaments may be indicative that enough initial mass was present for such a RER to form. Further, this scenario can also explain the absence of WR- and O-type stars in the central region, as those stars migrate close enough towards the black hole to be tidally disrupted. If the central S-star cluster would belong to the same dynamical component as the CW disk stars, the number of fainter B-type stars is increased, altering KLF of this feature. We discuss this possibility in \autoref{appendix:sstars_klf}.

Nevertheless, the strict isotropy of the angular momentum has been challenged by \cite{Basel2020} who report two planes of resonances\footnote{Note that \cite{Basel2020} refer to the plane of orbits as `disk'. However, the stars in their `disks' show both clock and counter-clockwise rotation, which is inconsistent with our nomenclature.}, with an apparent overdensity in the orbital inclination of both the late and early type stars in the central arcsecond. Our analysis provides an appropriate tool to measure the anisotropy of an observed stellar distribution by comparing it against an isotropic cluster. We could not confirm a significant anisotropy in the central region other than in the direction of the clockwise disk (see \autoref{appendix:anisotropy} for details). However, we did not include late type stars which contribute to their findings. In addition, our test is more constraining as it requires isotropy in all orbital elements, rather than just in the orbital inclination. Thus we cannot rule out that such planes exist.

Lastly, the observed young stars might not have formed together at all. The main sequence life time of B-stars is much longer than that of the O and WR stars, further star formation in the Galactic Center has been shown to to be episodic with evidence of star formation in the recent history ($\sim100$ Myr) \citep[][]{Pfuhl2014, Schoedel2020}. The study of the age distribution of the bright young stars in the CW and CCW features has, however, not revealed a secondary star formation event \citep{Bartko2010, Lu2013}. \cite{Madigan2014} discuss a scenario in which older B stars ($60$ to $100$ Myr) interact with the potential of the black hole, the isotropic old star cluster and the CW disk. In this scenario binary stars are placed on near radial orbits with very high eccentricities due to the interaction with massive perturbers \citep{Perets2009, Perets2010}. After some time the binary is disrupted and one partner is captured on a tight orbit around the black hole. This scenario could thus explain the high eccentricity orbits observed for the F3 feature, implying that F3 stars correspond to the stars found in the magnitude bin $K_{\rm{mag}}>15$ of \cite{Madigan2014}. Conveniently, this scenario explains the B-stars found in the inner most region around Sgr~A*, which however seem to have similar ages as the CW disk stars \citep{Habibi2017}.

\section{Conclusion}
\noindent We confirm the presence of the warped clockwise disk \citep[][]{Bartko2009, Bartko2010}: it exhibits a smooth change in angular momentum as function of radius. 
We confirm the presence of an outer kinematic feature (F2), which \cite{Bartko2009} attributed to the CW disk. The feature shares similarities with the CW disk, but also with the features at larger separations. We associate it to the other outer structures, but note that it remains possible that it is part of the CW disk. 
Further, we confirm the presence of a counter-clockwise feature at large separations reported in \cite{Genzel2003_stars, Paumard2006, Bartko2009}. This feature was deemed insignificant by other work \cite{Lu2006, Lu2009, Yelda2014}. We find that this feature consists mostly of stars at large projected separations, explaining the difficulty of establishing significance in past studies which had smaller spatial coverage. 
In addition to the features which have been discussed in the literature before, we identify a new feature F3 which, like the CCW/F1 and F2 features, is at large projected distances from the black hole. The F3 feature is, however, substantially more eccentric and we thus argue that the systems are distinct.  

\noindent This rich structure in kinematic features has been suggested in different simulations of star formation in an accretion disk around Sgr~A*. 
The set of simulations by \cite{HobbsNayakshin2009} in which two giant molecular clouds collide and subsequently accrete show intriguingly comparable features to the ones observed: A small, medium eccentric disk in close proximity of Sgr~A*; several remnant star streamers at larger separation which have substantially different angular momenta directions; and higher eccentricities at larger separations. Further, the simulations show differences in the distribution of O and WR type stars, with the most heavy stars found in the inner disk -- consistent with the apparent distribution of O and WR type stars in the Galactic Center. We thus argue that the simultaneous formation of all young stars in the Galactic Center remains a feasible scenario, consistent with the latest analysis of the age distribution of the S-star cluster \citep{Habibi2017}. However, the dramatically different kinematic distribution of the B-stars in the central arcsecond remains serious challenge \citep{Boehle2016, Gillessen2017} for such a common formation scenario, and more detailed analysis of the age distribution of the young stars are required to confirm or rule out a single star formation event some $\sim6~\mathrm{Myr}$ ago.

\acknowledgments
We thank the referee for a very quick, yet thorough report which helped to improve the paper. AD, SVF and FW acknowledge support by the Max Planck International Research School. 

\newpage
\bibliography{bib/sample63}{}

\begin{thebibliography}{}
\expandafter\ifx\csname natexlab\endcsname\relax\def\natexlab#1{#1}\fi
\providecommand{\url}[1]{\href{#1}{#1}}
\providecommand{\dodoi}[1]{doi:~\href{http://doi.org/#1}{\nolinkurl{#1}}}
\providecommand{\doeprint}[1]{\href{http://ascl.net/#1}{\nolinkurl{http://ascl.net/#1}}}
\providecommand{\doarXiv}[1]{\href{https://arxiv.org/abs/#1}{\nolinkurl{https://arxiv.org/abs/#1}}}

\bibitem[{{Ali} {et~al.}(2020){Ali}, {Paul}, {Eckart}, {Parsa}, {Zajacek},
  {Pei{\ss}ker}, {Subroweit}, {Valencia-S.}, {Thomkins}, \&
  {Witzel}}]{Basel2020}
{Ali}, B., {Paul}, D., {Eckart}, A., {et~al.} 2020, \apj, 896, 100,
  \dodoi{10.3847/1538-4357/ab93ae}

\bibitem[{{Bahcall} \& {Wolf}(1976)}]{Bahcall1976}
{Bahcall}, J.~N., \& {Wolf}, R.~A. 1976, \apj, 209, 214, \dodoi{10.1086/154711}

\bibitem[{{Bartko} {et~al.}(2009){Bartko}, {Martins}, {Fritz}, {Genzel},
  {Levin}, {Perets}, {Paumard}, {Nayakshin}, {Gerhard}, {Alexander},
  {Dodds-Eden}, {Eisenhauer}, {Gillessen}, {Mascetti}, {Ott}, {Perrin},
  {Pfuhl}, {Reid}, {Rouan}, {Sternberg}, \& {Trippe}}]{Bartko2009}
{Bartko}, H., {Martins}, F., {Fritz}, T.~K., {et~al.} 2009, \apj, 697, 1741,
  \dodoi{10.1088/0004-637X/697/2/1741}

\bibitem[{{Bartko} {et~al.}(2010){Bartko}, {Martins}, {Trippe}, {Fritz},
  {Genzel}, {Ott}, {Eisenhauer}, {Gillessen}, {Paumard}, {Alexander},
  {Dodds-Eden}, {Gerhard}, {Levin}, {Mascetti}, {Nayakshin}, {Perets},
  {Perrin}, {Pfuhl}, {Reid}, {Rouan}, {Zilka}, \& {Sternberg}}]{Bartko2010}
{Bartko}, H., {Martins}, F., {Trippe}, S., {et~al.} 2010, \apj, 708, 834,
  \dodoi{10.1088/0004-637X/708/1/834}

\bibitem[{{Becklin} \& {Neugebauer}(1968)}]{Becklin1968}
{Becklin}, E.~E., \& {Neugebauer}, G. 1968, \apj, 151, 145,
  \dodoi{10.1086/149425}

\bibitem[{{Becklin} \& {Neugebauer}(1975)}]{Becklin1975}
---. 1975, \apjl, 200, L71, \dodoi{10.1086/181899}

\bibitem[{{Beloborodov} {et~al.}(2006){Beloborodov}, {Levin}, {Eisenhauer},
  {Genzel}, {Paumard}, {Gillessen}, \& {Ott}}]{Beloborodov2006}
{Beloborodov}, A.~M., {Levin}, Y., {Eisenhauer}, F., {et~al.} 2006, \apj, 648,
  405, \dodoi{10.1086/504279}

\bibitem[{{Blum} {et~al.}(1996){Blum}, {Sellgren}, \& {Depoy}}]{Blum1996}
{Blum}, R.~D., {Sellgren}, K., \& {Depoy}, D.~L. 1996, \apj, 470, 864,
  \dodoi{10.1086/177917}

\bibitem[{{Boehle} {et~al.}(2016){Boehle}, {Ghez}, {Sch{\"o}del}, {Meyer},
  {Yelda}, {Albers}, {Martinez}, {Becklin}, {Do}, {Lu}, {Matthews}, {Morris},
  {Sitarski}, \& {Witzel}}]{Boehle2016}
{Boehle}, A., {Ghez}, A.~M., {Sch{\"o}del}, R., {et~al.} 2016, \apj, 830, 17,
  \dodoi{10.3847/0004-637X/830/1/17}

\bibitem[{{Bonnell} \& {Rice}(2008)}]{Bonnell2008}
{Bonnell}, I.~A., \& {Rice}, W.~K.~M. 2008, Science, 321, 1060,
  \dodoi{10.1126/science.1160653}

\bibitem[{{Bonnet} {et~al.}(2004){Bonnet}, {Conzelmann}, {Delabre},
  {Donaldson}, {Fedrigo}, {Hubin}, {Kissler-Patig}, {Lizon}, {Paufique},
  {Rossi}, {Stroebele}, \& {Tordo}}]{Bonnet2004}
{Bonnet}, H., {Conzelmann}, R., {Delabre}, B., {et~al.} 2004, in Society of
  Photo-Optical Instrumentation Engineers (SPIE) Conference Series, Vol. 5490,
  Advancements in Adaptive Optics, ed. D.~{Bonaccini Calia}, B.~L.
  {Ellerbroek}, \& R.~{Ragazzoni}, 130--138, \dodoi{10.1117/12.551187}

\bibitem[{{Chen} \& {Amaro-Seoane}(2014)}]{Chen2014}
{Chen}, X., \& {Amaro-Seoane}, P. 2014, \apjl, 786, L14,
  \dodoi{10.1088/2041-8205/786/2/L14}

\bibitem[{{Do} {et~al.}(2013){Do}, {Lu}, {Ghez}, {Morris}, {Yelda}, {Martinez},
  {Wright}, \& {Matthews}}]{Do2013}
{Do}, T., {Lu}, J.~R., {Ghez}, A.~M., {et~al.} 2013, \apj, 764, 154,
  \dodoi{10.1088/0004-637X/764/2/154}

\bibitem[{{Fritz} {et~al.}(2011){Fritz}, {Gillessen}, {Dodds-Eden}, {Lutz},
  {Genzel}, {Raab}, {Ott}, {Pfuhl}, {Eisenhauer}, \& {Yusef-Zadeh}}]{Fritz2011}
{Fritz}, T.~K., {Gillessen}, S., {Dodds-Eden}, K., {et~al.} 2011, \apj, 737,
  73, \dodoi{10.1088/0004-637X/737/2/73}

\bibitem[{{Gallego-Cano} {et~al.}(2018){Gallego-Cano}, {Sch{\"o}del}, {Dong},
  {Nogueras-Lara}, {Gallego-Calvente}, {Amaro-Seoane}, \&
  {Baumgardt}}]{Gallego-Cano2018}
{Gallego-Cano}, E., {Sch{\"o}del}, R., {Dong}, H., {et~al.} 2018, \aap, 609,
  A26, \dodoi{10.1051/0004-6361/201730451}

\bibitem[{{Genzel} {et~al.}(1994){Genzel}, {Hollenbach}, \&
  {Townes}}]{Genzel1994}
{Genzel}, R., {Hollenbach}, D., \& {Townes}, C.~H. 1994, Reports on Progress in
  Physics, 57, 417, \dodoi{10.1088/0034-4885/57/5/001}

\bibitem[{{Genzel} {et~al.}(2000){Genzel}, {Pichon}, {Eckart}, {Gerhard}, \&
  {Ott}}]{Genzel2000}
{Genzel}, R., {Pichon}, C., {Eckart}, A., {Gerhard}, O.~E., \& {Ott}, T. 2000,
  \mnras, 317, 348, \dodoi{10.1046/j.1365-8711.2000.03582.x}

\bibitem[{{Genzel} {et~al.}(2003){Genzel}, {Sch{\"o}del}, {Ott}, {Eisenhauer},
  {Hofmann}, {Lehnert}, {Eckart}, {Alexander}, {Sternberg}, {Lenzen},
  {Cl{\'e}net}, {Lacombe}, {Rouan}, {Renzini}, \&
  {Tacconi-Garman}}]{Genzel2003_stellarcusp}
{Genzel}, R., {Sch{\"o}del}, R., {Ott}, T., {et~al.} 2003, \apj, 594, 812,
  \dodoi{10.1086/377127}

\bibitem[{Genzel {et~al.}(2003)Genzel, Schoedel, Alexander, Sternberg, Lenzen,
  Clenet, Lacombe, Rouan, Renzini, \& Tacconi-Garman}]{Genzel2003_stars}
Genzel, R., Schoedel, R., Alexander, T., {et~al.} 2003, The Astrophysical
  Journal, 812.
\newblock \url{https://iopscience.iop.org/article/10.1086/377127/pdf}

\bibitem[{{Ghez} {et~al.}(2000){Ghez}, {Morris}, {Becklin}, {Tanner}, \&
  {Kremenek}}]{Ghez2000}
{Ghez}, A.~M., {Morris}, M., {Becklin}, E.~E., {Tanner}, A., \& {Kremenek}, T.
  2000, \nat, 407, 349, \dodoi{10.1038/35030032}

\bibitem[{{Gillessen} {et~al.}(2009){Gillessen}, {Eisenhauer}, {Trippe},
  {Alexand er}, {Genzel}, {Martins}, \& {Ott}}]{Gillessen2009}
{Gillessen}, S., {Eisenhauer}, F., {Trippe}, S., {et~al.} 2009, \apj, 692,
  1075, \dodoi{10.1088/0004-637X/692/2/1075}

\bibitem[{{Gillessen} {et~al.}(2017){Gillessen}, {Plewa}, {Eisenhauer}, {Sari},
  {Waisberg}, {Habibi}, {Pfuhl}, {George}, {Dexter}, {von Fellenberg}, {Ott},
  \& {Genzel}}]{Gillessen2017}
{Gillessen}, S., {Plewa}, P.~M., {Eisenhauer}, F., {et~al.} 2017, \apj, 837,
  30, \dodoi{10.3847/1538-4357/aa5c41}

\bibitem[{{Giral Mart{\'\i}nez} {et~al.}(2020){Giral Mart{\'\i}nez}, {Fouvry},
  \& {Pichon}}]{Martinez2020}
{Giral Mart{\'\i}nez}, J., {Fouvry}, J.-B., \& {Pichon}, C. 2020, \mnras, 499,
  2714, \dodoi{10.1093/mnras/staa2722}

\bibitem[{{Gravity Collaboration} {et~al.}(2021){Gravity Collaboration},
  {Abuter}, {Amorim}, {Baub{\"o}ck}, {Berger}, {Bonnet}, {Brandner},
  {Cl{\'e}net}, {Davies}, {de Zeeuw}, {Dexter}, {Dallilar}, {Drescher},
  {Eckart}, {Eisenhauer}, {F{\"o}rster Schreiber}, {Garcia}, {Gao}, {Gendron},
  {Genzel}, {Gillessen}, {Habibi}, {Haubois}, {Hei{\ss}el}, {Henning},
  {Hippler}, {Horrobin}, {Jim{\'e}nez-Rosales}, {Jochum}, {Jocou}, {Kaufer},
  {Kervella}, {Lacour}, {Lapeyr{\`e}re}, {Le Bouquin}, {L{\'e}na}, {Lutz},
  {Nowak}, {Ott}, {Paumard}, {Perraut}, {Perrin}, {Pfuhl}, {Rabien},
  {Rodr{\'\i}guez-Coira}, {Shangguan}, {Shimizu}, {Scheithauer}, {Stadler},
  {Straub}, {Straubmeier}, {Sturm}, {Tacconi}, {Vincent}, {von Fellenberg},
  {Waisberg}, {Widmann}, {Wieprecht}, {Wiezorrek}, {Woillez}, {Yazici},
  {Young}, \& {Zins}}]{GravityCollaboration2021_abberations}
{Gravity Collaboration}, {Abuter}, R., {Amorim}, A., {et~al.} 2021, \aap, 647,
  A59, \dodoi{10.1051/0004-6361/202040208}

\bibitem[{{Gravity Collaboration} {et~al.}(2022){Gravity Collaboration},
  {Abuter}, {Aimar}, {Amorim}, {Ball}, {Baub{\"o}ck}, {Berger}, {Bonnet},
  {Bourdarot}, {Brandner}, {Cardoso}, {Cl{\'e}net}, {Dallilar}, {Davies}, {de
  Zeeuw}, {Dexter}, {Drescher}, {Eisenhauer}, {F{\"o}rster Schreiber},
  {Foschi}, {Garcia}, {Gao}, {Gendron}, {Genzel}, {Gillessen}, {Habibi},
  {Haubois}, {Hei{\ss}el}, {Henning}, {Hippler}, {Horrobin}, {Jochum}, {Jocou},
  {Kaufer}, {Kervella}, {Lacour}, {Lapeyr{\`e}re}, {Le Bouquin}, {L{\'e}na},
  {Lutz}, {Ott}, {Paumard}, {Perraut}, {Perrin}, {Pfuhl}, {Rabien},
  {Shangguan}, {Shimizu}, {Scheithauer}, {Stadler}, {Stephens}, {Straub},
  {Straubmeier}, {Sturm}, {Tacconi}, {Tristram}, {Vincent}, {von Fellenberg},
  {Widmann}, {Wieprecht}, {Wiezorrek}, {Woillez}, {Yazici}, \&
  {Young}}]{GravityCollaboration2022_gheziswrong}
{Gravity Collaboration}, {Abuter}, R., {Aimar}, N., {et~al.} 2022, \aap, 657,
  L12, \dodoi{10.1051/0004-6361/202142465}

\bibitem[{Habibi {et~al.}(2017)Habibi, Gillessen, Martins, Eisenhauer, Plewa,
  Pfuhl, George, Dexter, Waisberg, Ott, von Fellenberg, Baub{\"{o}}ck,
  Jimenez-Rosales, \& Genzel}]{Habibi2017}
Habibi, M., Gillessen, S., Martins, F., {et~al.} 2017, \apj, 847, 120,
  \dodoi{10.3847/1538-4357/aa876f}

\bibitem[{{H.E.S.S. Collaboration} {et~al.}(2018){H.E.S.S. Collaboration},
  {Abdalla}, {Abramowski}, {Aharonian}, {Ait Benkhali}, {Ang{\"u}ner},
  {Arakawa}, {Arrieta}, {Aubert}, {Backes}, {Balzer}, {Barnard}, {Becherini},
  {Becker Tjus}, {Berge}, {Bernhard}, {Bernl{\"o}hr}, {Blackwell},
  {B{\"o}ttcher}, {Boisson}, {Bolmont}, {Bonnefoy}, {Bordas}, {Bregeon},
  {Brun}, {Brun}, {Bryan}, {B{\"u}chele}, {Bulik}, {Capasso}, {Carrigan},
  {Caroff}, {Carosi}, {Casanova}, {Cerruti}, {Chakraborty}, {Chaves}, {Chen},
  {Chevalier}, {Colafrancesco}, {Condon}, {Conrad}, {Davids}, {Decock}, {Deil},
  {Devin}, {deWilt}, {Dirson}, {Djannati-Ata{\"\i}}, {Domainko}, {Donath},
  {Drury}, {Dutson}, {Dyks}, {Edwards}, {Egberts}, {Eger}, {Emery},
  {Ernenwein}, {Eschbach}, {Farnier}, {Fegan}, {Fernandes}, {Fiasson},
  {Fontaine}, {F{\"o}rster}, {Funk}, {F{\"u}{\ss}ling}, {Gabici}, {Gallant},
  {Garrigoux}, {Gast}, {Gat{\'e}}, {Giavitto}, {Giebels}, {Glawion},
  {Glicenstein}, {Gottschall}, {Grondin}, {Hahn}, {Haupt}, {Hawkes},
  {Heinzelmann}, {Henri}, {Hermann}, {Hinton}, {Hofmann}, {Hoischen}, {Holch},
  {Holler}, {Horns}, {Ivascenko}, {Iwasaki}, {Jacholkowska}, {Jamrozy},
  {Jankowsky}, {Jankowsky}, {Jingo}, {Jouvin}, {Jung-Richardt}, {Kastendieck},
  {Katarzy{\'n}ski}, {Katsuragawa}, {Katz}, {Kerszberg}, {Khangulyan},
  {Kh{\'e}lifi}, {King}, {Klepser}, {Klochkov}, {Klu{\'z}niak}, {Komin},
  {Kosack}, {Krakau}, {Kraus}, {Kr{\"u}ger}, {Laffon}, {Lamanna}, {Lau},
  {Lees}, {Lefaucheur}, {Lemi{\`e}re}, {Lemoine-Goumard}, {Lenain}, {Leser},
  {Lohse}, {Lorentz}, {Liu}, {L{\'o}pez-Coto}, {Lypova}, {Marandon},
  {Malyshev}, {Marcowith}, {Mariaud}, {Marx}, {Maurin}, {Maxted}, {Mayer},
  {Meintjes}, {Meyer}, {Mitchell}, {Moderski}, {Mohamed}, {Mohrmann},
  {Mor{\r{a}}}, {Moulin}, {Murach}, {Nakashima}, {de Naurois}, {Ndiyavala},
  {Niederwanger}, {Niemiec}, {Oakes}, {O'Brien}, {Odaka}, {Ohm}, {Ostrowski},
  {Oya}, {Padovani}, {Panter}, {Parsons}, {Paz Arribas}, {Pekeur}, {Pelletier},
  {Perennes}, {Petrucci}, {Peyaud}, {Piel}, {Pita}, {Poireau}, {Poon},
  {Prokhorov}, {Prokoph}, {P{\"u}hlhofer}, {Punch}, {Quirrenbach}, {Raab},
  {Rauth}, {Reimer}, {Reimer}, {Renaud}, {de los Reyes}, {Rieger}, {Rinchiuso},
  {Romoli}, {Rowell}, {Rudak}, {Rulten}, {Safi-Harb}, {Sahakian}, {Saito},
  {Sanchez}, {Santangelo}, {Sasaki}, {Schandri}, {Schlickeiser},
  {Sch{\"u}ssler}, {Schulz}, {Schwanke}, {Schwemmer}, {Seglar-Arroyo},
  {Settimo}, {Seyffert}, {Shafi}, {Shilon}, {Shiningayamwe}, {Simoni}, {Sol},
  {Spanier}, {Spir-Jacob}, {Stawarz}, {Steenkamp}, {Stegmann}, {Steppa},
  {Sushch}, {Takahashi}, {Tavernet}, {Tavernier}, {Taylor}, {Terrier},
  {Tibaldo}, {Tiziani}, {Tluczykont}, {Trichard}, {Tsirou}, {Tsuji}, {Tuffs},
  {Uchiyama}, {van der Walt}, {van Eldik}, {van Rensburg}, {van Soelen},
  {Vasileiadis}, {Veh}, {Venter}, {Viana}, {Vincent}, {Vink}, {Voisin},
  {V{\"o}lk}, {Vuillaume}, {Wadiasingh}, {Wagner}, {Wagner}, {Wagner}, {White},
  {Wierzcholska}, {Willmann}, {W{\"o}rnlein}, {Wouters}, {Yang}, {Zaborov},
  {Zacharias}, {Zanin}, {Zdziarski}, {Zech}, {Zefi}, {Ziegler}, {Zorn}, \&
  {{\.Z}ywucka}}]{HessCollab2018}
{H.E.S.S. Collaboration}, {Abdalla}, H., {Abramowski}, A., {et~al.} 2018, \aap,
  612, A1, \dodoi{10.1051/0004-6361/201732098}

\bibitem[{{Hobbs} \& {Nayakshin}(2009)}]{HobbsNayakshin2009}
{Hobbs}, A., \& {Nayakshin}, S. 2009, \mnras, 394, 191,
  \dodoi{10.1111/j.1365-2966.2008.14359.x}

\bibitem[{{Kocsis} \& {Tremaine}(2011)}]{Kocsis2011}
{Kocsis}, B., \& {Tremaine}, S. 2011, \mnras, 412, 187,
  \dodoi{10.1111/j.1365-2966.2010.17897.x}

\bibitem[{{Kocsis} \& {Tremaine}(2015)}]{Kocsis2015}
---. 2015, \mnras, 448, 3265, \dodoi{10.1093/mnras/stv057}

\bibitem[{{Krabbe} {et~al.}(1991){Krabbe}, {Genzel}, {Drapatz}, \&
  {Rotaciuc}}]{Krabbe1991}
{Krabbe}, A., {Genzel}, R., {Drapatz}, S., \& {Rotaciuc}, V. 1991, \apjl, 382,
  L19, \dodoi{10.1086/186204}

\bibitem[{Kroupa(2001)}]{Kroupa2001}
Kroupa, P. 2001, Monthly Notices of the Royal Astronomical Society, 322, 231,
  \dodoi{10.1046/j.1365-8711.2001.04022.x}

\bibitem[{{Levin} \& {Beloborodov}(2003)}]{Levin2003}
{Levin}, Y., \& {Beloborodov}, A.~M. 2003, \apjl, 590, L33,
  \dodoi{10.1086/376675}

\bibitem[{{Li} \& {Ma}(1983)}]{Li1983}
{Li}, T.~P., \& {Ma}, Y.~Q. 1983, \apj, 272, 317, \dodoi{10.1086/161295}

\bibitem[{{L{\"o}ckmann} \& {Baumgardt}(2009)}]{LockmannBaumgard2009}
{L{\"o}ckmann}, U., \& {Baumgardt}, H. 2009, \mnras, 394, 1841,
  \dodoi{10.1111/j.1365-2966.2009.14466.x}

\bibitem[{{Lu} {et~al.}(2013){Lu}, {Do}, {Ghez}, {Morris}, {Yelda}, \&
  {Matthews}}]{Lu2013}
{Lu}, J.~R., {Do}, T., {Ghez}, A.~M., {et~al.} 2013, \apj, 764, 155,
  \dodoi{10.1088/0004-637X/764/2/155}

\bibitem[{{Lu} {et~al.}(2006){Lu}, {Ghez}, {Hornstein}, {Morris}, {Matthews},
  {Thompson}, \& {Becklin}}]{Lu2006}
{Lu}, J.~R., {Ghez}, A.~M., {Hornstein}, S.~D., {et~al.} 2006, in Journal of
  Physics Conference Series, Vol.~54, Journal of Physics Conference Series,
  279--287, \dodoi{10.1088/1742-6596/54/1/044}

\bibitem[{{Lu} {et~al.}(2009){Lu}, {Ghez}, {Hornstein}, {Morris}, {Becklin}, \&
  {Matthews}}]{Lu2009}
{Lu}, J.~R., {Ghez}, A.~M., {Hornstein}, S.~D., {et~al.} 2009, \apj, 690, 1463,
  \dodoi{10.1088/0004-637X/690/2/1463}

\bibitem[{{Madigan} {et~al.}(2014){Madigan}, {Pfuhl}, {Levin}, {Gillessen},
  {Genzel}, \& {Perets}}]{Madigan2014}
{Madigan}, A.-M., {Pfuhl}, O., {Levin}, Y., {et~al.} 2014, \apj, 784, 23,
  \dodoi{10.1088/0004-637X/784/1/23}

\bibitem[{{Nayakshin}(2006)}]{Nayakshin2006}
{Nayakshin}, S. 2006, \mnras, 372, 143,
  \dodoi{10.1111/j.1365-2966.2006.10772.x}

\bibitem[{{Nayakshin} {et~al.}(2007){Nayakshin}, {Cuadra}, \&
  {Springel}}]{Nayakshin2007}
{Nayakshin}, S., {Cuadra}, J., \& {Springel}, V. 2007, \mnras, 379, 21,
  \dodoi{10.1111/j.1365-2966.2007.11938.x}

\bibitem[{{Paumard} {et~al.}(2006){Paumard}, {Genzel}, {Martins}, {Nayakshin},
  {Beloborodov}, {Levin}, {Trippe}, {Eisenhauer}, {Ott}, {Gillessen}, {Abuter},
  {Cuadra}, {Alexander}, \& {Sternberg}}]{Paumard2006}
{Paumard}, T., {Genzel}, R., {Martins}, F., {et~al.} 2006, \apj, 643, 1011,
  \dodoi{10.1086/503273}

\bibitem[{{Perets} \& {Gualandris}(2010)}]{Perets2010}
{Perets}, H.~B., \& {Gualandris}, A. 2010, \apj, 719, 220,
  \dodoi{10.1088/0004-637X/719/1/220}

\bibitem[{{Perets} {et~al.}(2009){Perets}, {Gualandris}, {Kupi}, {Merritt}, \&
  {Alexander}}]{Perets2009}
{Perets}, H.~B., {Gualandris}, A., {Kupi}, G., {Merritt}, D., \& {Alexander},
  T. 2009, \apj, 702, 884, \dodoi{10.1088/0004-637X/702/2/884}

\bibitem[{{Pfuhl} {et~al.}(2014){Pfuhl}, {Alexander}, {Gillessen}, {Martins},
  {Genzel}, {Eisenhauer}, {Fritz}, \& {Ott}}]{Pfuhl2014}
{Pfuhl}, O., {Alexander}, T., {Gillessen}, S., {et~al.} 2014, \apj, 782, 101,
  \dodoi{10.1088/0004-637X/782/2/101}

\bibitem[{{Pfuhl} {et~al.}(2011){Pfuhl}, {Fritz}, {Zilka}, {Maness},
  {Eisenhauer}, {Genzel}, {Gillessen}, {Ott}, {Dodds-Eden}, \&
  {Sternberg}}]{Pfuhl2011}
{Pfuhl}, O., {Fritz}, T.~K., {Zilka}, M., {et~al.} 2011, \apj, 741, 108,
  \dodoi{10.1088/0004-637X/741/2/108}

\bibitem[{{Salpeter}(1955)}]{Salpeter1955}
{Salpeter}, E.~E. 1955, \apj, 121, 161, \dodoi{10.1086/145971}

\bibitem[{{Sch{\"o}del} {et~al.}(2018){Sch{\"o}del}, {Gallego-Cano}, {Dong},
  {Nogueras-Lara}, {Gallego-Calvente}, {Amaro-Seoane}, \&
  {Baumgardt}}]{Schoedel2018}
{Sch{\"o}del}, R., {Gallego-Cano}, E., {Dong}, H., {et~al.} 2018, \aap, 609,
  A27, \dodoi{10.1051/0004-6361/201730452}

\bibitem[{{Sch{\"o}del} {et~al.}(2020){Sch{\"o}del}, {Nogueras-Lara},
  {Gallego-Cano}, {Shahzamanian}, {Gallego-Calvente}, \&
  {Gardini}}]{Schoedel2020}
{Sch{\"o}del}, R., {Nogueras-Lara}, F., {Gallego-Cano}, E., {et~al.} 2020,
  \aap, 641, A102, \dodoi{10.1051/0004-6361/201936688}

\bibitem[{{Sch{\"o}del} {et~al.}(2003){Sch{\"o}del}, {Ott}, {Genzel}, {Eckart},
  {Mouawad}, \& {Alexander}}]{Schodel2003}
{Sch{\"o}del}, R., {Ott}, T., {Genzel}, R., {et~al.} 2003, \apj, 596, 1015,
  \dodoi{10.1086/378122}

\bibitem[{{Simons} \& {Becklin}(1996)}]{Simons1996}
{Simons}, D.~A., \& {Becklin}, E.~E. 1996, \aj, 111, 1908,
  \dodoi{10.1086/117929}

\bibitem[{{Skilling}(2004)}]{Skilling2004_dynesty}
{Skilling}, J. 2004, in American Institute of Physics Conference Series, Vol.
  735, Bayesian Inference and Maximum Entropy Methods in Science and
  Engineering: 24th International Workshop on Bayesian Inference and Maximum
  Entropy Methods in Science and Engineering, ed. R.~{Fischer}, R.~{Preuss}, \&
  U.~V. {Toussaint}, 395--405, \dodoi{10.1063/1.1835238}

\bibitem[{Skilling(2006{\natexlab{a}})}]{Skilling2006_dynesty}
Skilling, J. 2006{\natexlab{a}}, Bayesian Analysis, 1, 833 ,
  \dodoi{10.1214/06-BA127}

\bibitem[{Skilling(2006{\natexlab{b}})}]{Skilling2006_dynesty2}
---. 2006{\natexlab{b}}, Bayesian Analysis, 1, 833 , \dodoi{10.1214/06-BA127}

\bibitem[{{Speagle}(2020)}]{Speagle2020_dynesty}
{Speagle}, J.~S. 2020, \mnras, 493, 3132, \dodoi{10.1093/mnras/staa278}

\bibitem[{{Stewart}(2009)}]{Stewart2009}
{Stewart}, I.~M. 2009, \aap, 495, 989, \dodoi{10.1051/0004-6361:200811311}

\bibitem[{{Sz{\"o}lgy{\'e}n} \& {Kocsis}(2018)}]{Hungarians2018}
{Sz{\"o}lgy{\'e}n}, {\'A}., \& {Kocsis}, B. 2018, \prl, 121, 101101,
  \dodoi{10.1103/PhysRevLett.121.101101}

\bibitem[{{Trippe} {et~al.}(2008){Trippe}, {Gillessen}, {Gerhard}, {Bartko},
  {Fritz}, {Maness}, {Eisenhauer}, {Martins}, {Ott}, {Dodds-Eden}, \&
  {Genzel}}]{Trippe2008}
{Trippe}, S., {Gillessen}, S., {Gerhard}, O.~E., {et~al.} 2008, \aap, 492, 419,
  \dodoi{10.1051/0004-6361:200810191}

\bibitem[{{Yelda} {et~al.}(2014){Yelda}, {Ghez}, {Lu}, {Do}, {Meyer}, {Morris},
  \& {Matthews}}]{Yelda2014}
{Yelda}, S., {Ghez}, A.~M., {Lu}, J.~R., {et~al.} 2014, \apj, 783, 131,
  \dodoi{10.1088/0004-637X/783/2/131}

\end{thebibliography}
\bibliographystyle{aasjournal}

\appendix

\section{Generation of mock isotropic cluster}\label{appendix:isotropic_cluster}
The procedure to generate mock observations of an isotropic cluster is adapted from \cite{Schodel2003}:
\begin{enumerate}
    \item Sample the inclination $i$ and the longitude of the ascending node $\Omega$ isotropically on a sphere.
    \item Sample the argument of pericenter uniformly from $\omega \in [0\degree, 360\degree[$.
    \item Draw the semi-major axis from a power law distribution $\frac{dN}{da} \propto a^{-\beta}$. We choose $\beta=2$ to resemble the observed distribution of stars. We sample $a$ from $0.2 ~\mathrm{arcseconds}$ to $40 ~\mathrm{arcseconds}$, in order to match the observed scales.
    \item Sample the eccentricity such that $\frac{dN}{de} \propto e$, i.e. a thermal distribution of eccentricities.
    \item Compute the true anomaly by assuming a uniform distribution of time points along the orbit: $t_{orbit} \in [0, P_{orbit}[$. This corresponds to a uniform mean anomaly distribution.
\end{enumerate}

\noindent With this recipe, we generate a cluster containing $100'000$ stars and calculate the phase space coordinates. We then discard the $z$ coordinates and draw a new $z$ coordinate from the $z$-prior distribution. From this cluster we choose $N$ stars,
as many as our data sample contains, taking into account the observational biases from the fields covered. This yields a mock data set that we analyze the same way as the real data in \autoref{subsection:determination_of_angular_momentum}. This procedure is repeated $10000$ times, creating $10000$ mock data sets from which we calculate the mean and standard deviation in each pixel.

\section{Z-priors used in previous studies and their biases}\label{appendix:priors}
In the following we describe z-priors used in previous studies, the stellar cusp prior and the uniform acceleration prior. We show that both priors generate biased posterior distributions, in which the angular momentum distribution is not isotropically distributed on a sphere.
\subsection{The stellar cusp prior}\label{subsec:stellar_cusp}
\cite{Bartko2009} introduced the stellar cusp prior. The distribution of $z$ coordinates is given by:

\begin{equation}
\begin{aligned}
    P(z | x_{obs}, y_{obs}) \propto {(x_{obs}^2 + y_{obs}^2 + z^2)^{-\frac{\beta+1}{2}}}
\label{eqn:stellarCusp}
\end{aligned}
\end{equation}
\noindent where $\beta=2$ the power-law index of intrinsic density profile of the cusp in the GC $dN/dR \propto R^{-\beta}$ \citep[][]{Genzel2003_stellarcusp} and $x_{obs}/y_{obs}$ stand for the observed star positions.  

\noindent The simulated cluster analysis with the stellar cusp prior correctly captures the input distribution of eccentricities and 
also the distribution of semi-major axes is reproduced closely. However, the argument of pericenter $\omega$ does not follow a uniform input distribution, and high inclinations are favored. This results in a boxy distribution of stars, with an  over-density towards the center (see \autoref{fig:youngstars_oe_distribution}).

\noindent The reason for this behaviour lies in the distribution of eccentricities. While the positions of the stars geometrically follow a power law slope, the probability for $z$ given $R_{obs}$ depends on how much the individual star plunges towards the black hole. When one knows the distribution of eccentricities and the distribution of semi-major axes, this ``plunging in'' is given by the observed velocity vector $\vec{v}$, a piece of knowledge that is neglected in this prior.

\subsection{The uniform acceleration prior}\label{subsec:uniform_acceleration}
The uniform acceleration prior has first been used by \cite{Lu2009} and is constructed by drawing the acceleration $a_S(R)$ uniformly in the possible range, i.e. $a_S(R) \in \left[G M_\bullet/(z_{\rm{max}}^2 +R^2)^{3/2} , \frac{G M_\bullet}{R^2}\right]$. The maximum value for $a_S(R)$ is reached for $z=0$ and its minimum value is obtained from the maximum allowed $z_{\rm{max}} = \sqrt{2GM_\bullet/|\vec{v}|^2 -R^2}$. The $z-$coordinate is then obtained from 
\begin{equation}
\begin{aligned}
    z = \sqrt{\left(\frac{G M_\bullet R_{obs}}{a_S(R)}\right)^{2/3} - R^2} \,\,.
\label{eqn:uniform_acceleration}
\end{aligned}
\end{equation}
The analysis with the uniform acceleration prior does not reproduce the linear distribution of eccentricities and produces a tail of high semi-major axes orbits. Similar to the stellar cusp prior, the argument of periapsis $\omega$ is not uniformly sampled, but shows an angular dependence. Furthermore, the orbital nodes are biased towards high values, but without the clear concentration towards $\cos(\phi) =\pm 1$ of the stellar cusp prior. Inspecting the distribution of $z$ values, one can make out a zone of avoidance close to zero. Nevertheless, at least perceptually, the uniform acceleration prior seems to fare slightly better as the distribution of $z$ values is more symmetrical than that for the stellar cusp prior. 

\noindent The reason for the mismatch from an isotropic cluster lies again in the eccentricity distribution $\frac{dN}{d\epsilon} \propto \epsilon$. Since most stars have eccentric orbits there is a high chance of observing a star far away from the black hole and close to apocenter, where the acceleration is low. Therefore, the distribution of acceleration is not uniform but varies radially and depends on the velocity $\vec{v}$ of the star. 

\noindent For both the stellar cusp and the uniform acceleration prior, the reason that the prior clusters deviate from that of an isotropic cluster is that the priors only depend on the projected radius $R_{\rm{projected}}$ and do not take the velocity of the star into account. 

\subsection{Comparison of the different priors}\label{sec:prior_comparison}
\begin{figure}
    \centering
    \includegraphics[width=0.9\textwidth]{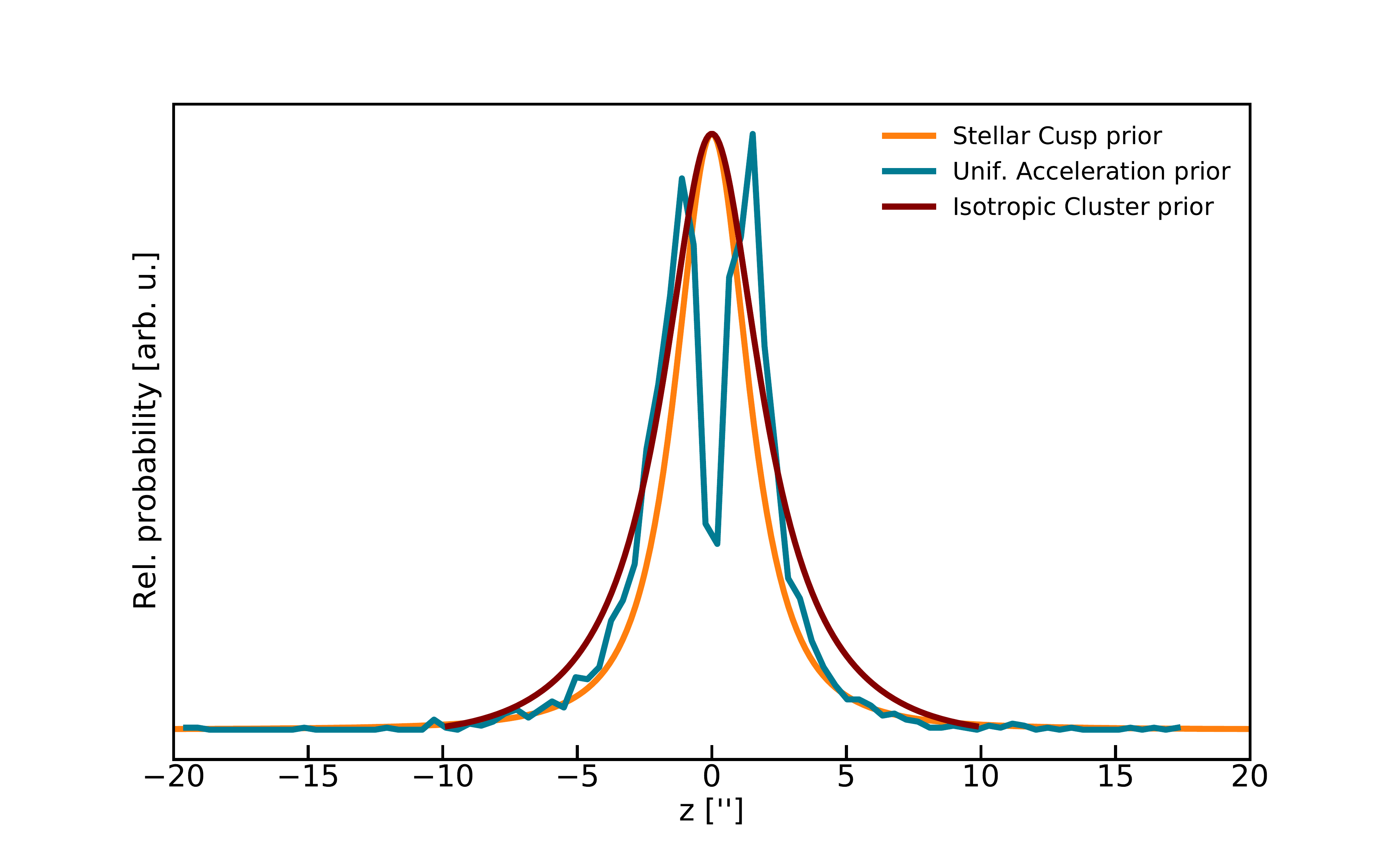}
    \caption[Comparison of the probability distribution function of the z values]{Comparison of the probability distribution function of the z values for the star E29: The orange line shows the PDF of the stellar cusp prior (\autoref{eqn:stellarCusp}), the blue line shows the histogram of z-values sampled according to the uniform acceleration prior (\autoref{eqn:uniform_acceleration}) and the dark red line shows the isotropic cluster prior (\autoref{eqn:isotropicCluster}). For better comparison we have normalized the mode of the respective distributions to one.}
    \label{fig:compare_PDFs}
\end{figure}

In this section we compare the two different priors that have been used in the past with the isotropic cluster prior.
\autoref{fig:compare_PDFs} compares the z-probability distribution functions for the three priors and the example star E29. For the stellar cusp prior and the isotropic cluster prior the analytic expression are given in \autoref{eqn:stellarCusp} and \autoref{eqn:isotropicCluster}. For the uniform acceleration prior we estimate the PDF by making a histogram of the z-values derived from the prior (see \autoref{eqn:uniform_acceleration}). 

\begin{figure*}
    \centering
    \includegraphics[width=\textwidth, trim={5cm 5cm 5cm 5cm},clip]{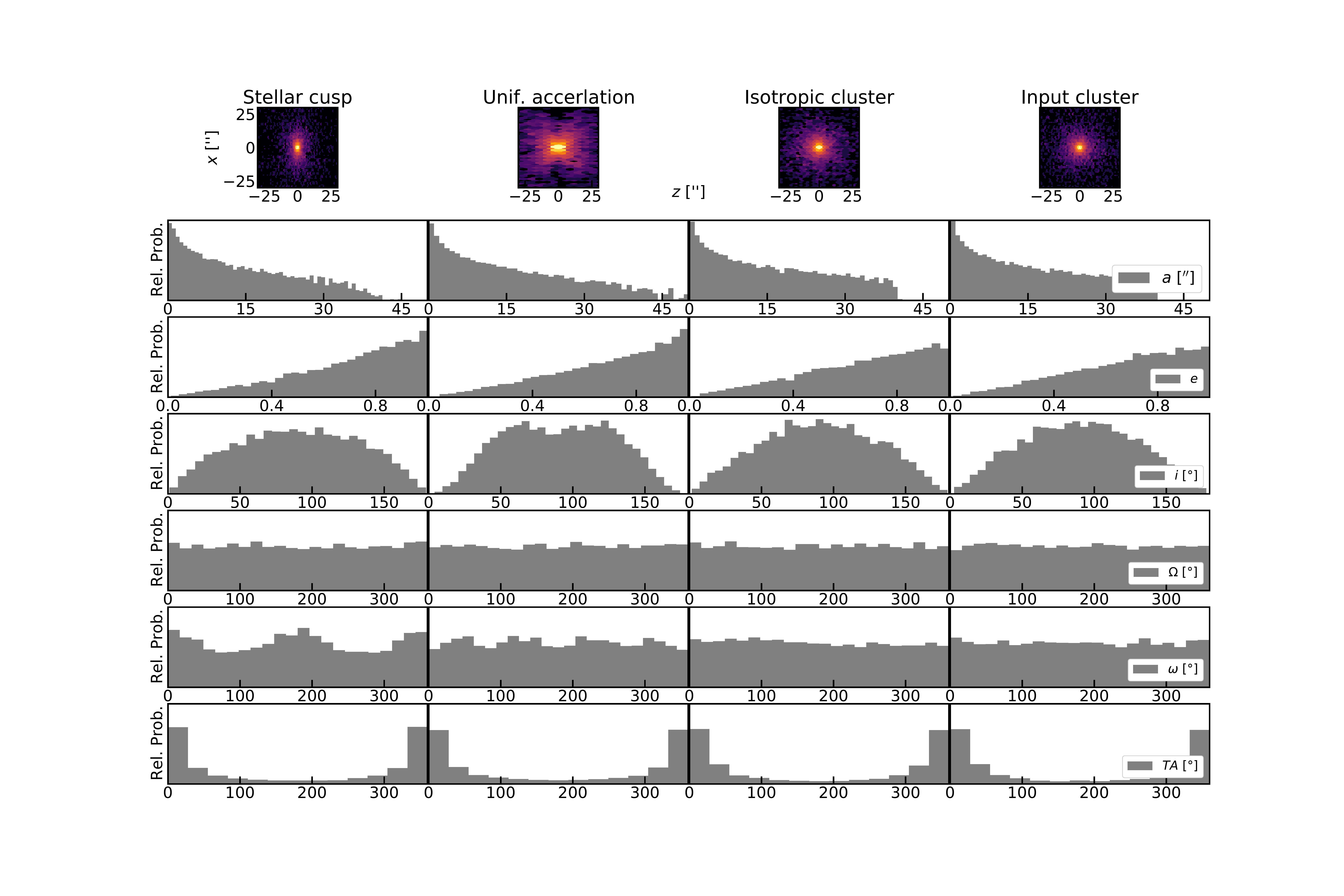}
    \caption[Orbital element distribution of different z-priors]{Distribution of orbital elements of different \textit{prior-clusters}, drawn from the stellar cusp prior, the uniform acceleration prior, the isotropic cluster prior, and the input isotropic cluster. The top panel shows the $z$ vs. $x$ distribution of the different prior clusters. See \autoref{sec:prior_comparison} for details.}
    \label{fig:youngstars_oe_distribution}
\end{figure*}

\noindent Because the $z$-PDFs are different, they will lead to different z distributions. To illustrate the effect on our null hypothesis of an isotropic cluster we conduct the following numerical experiment:
\begin{enumerate}
    \item Draw 10000 stars from an isotropic distribution according to the recipe detailed in \autoref{subsec:isotropicCluster_generation}.
    \item Discard the z-coordinate of each star.
    \item Re-draw a z-coordinate for each star from its respective prior distribution function.
\end{enumerate}
We call the resulting cluster the ``prior cluster'', in which each star has new z position. We plot the resulting $z$ vs. $x$ position in the top row of \autoref{fig:youngstars_oe_distribution} for each of the three priors as well as the input isotropic cluster. Plotting the $z$ vs $y$ coordinate yields a qualitatively identical plot. It is evident that the isotropic cluster prior best reproduces the input cluster.

To better compare the respective prior clusters with the input orbital elements we recompute the orbital elements of each star using the newly determined z-position. \autoref{fig:youngstars_oe_distribution} compares the histogram of the input orbital elements with those computed from the re-sampled stars. 

The isotropic cluster prior best reproduces the input orbital element distribution, and specifically it does not yield biased distributions of $i$ and $\Omega$ which are the parameters we are interested in. This is in contrast to the stellar cusp prior and the uniform acceleration prior. This behavior of the priors was discovered in previous studies \citep[][]{Bartko2010, Yelda2014}, and both studies tried to de-bias their study by subtracting the mean bias from the density histogram. Our unbiased prior makes this de-biasing step unnecessary. 
We conclude that the isotropic cluster prior correctly maps the input isotropic cluster on a self-similar realization of itself. We therefore achieve a meaningful null-hypothesis: How different is the observed angular momentum distribution to that of an isotropic cluster. 

We note that the isotropic cluster is not the ``best'' prior to determine the angular momentum distribution of the young stars in the GC. Given that the presence of at least one star disk is undisputed, a ``stellar disk prior'' would be more suited to determine the presence of the disk. Such a prior would however change the null-hypothesis to ``How different is the observed star distribution to the assumed stellar disk?'' and is therefore not suited to find new kinematic features. Once the determinant of the volume filling factor is determined the construction of such a ``disk prior'' is trivial, and follows the method in \autoref{subsec:isotropy}, with the suitable changes to \autoref{eqn:isotropicClusterPDF}. 

\section{Discrepancy between the Bartko et al. 2009 and Yelda et al. 2014 works}\label{appendix:bartko_v_yelda}
Both \cite{Bartko2009} (abb. Bartko09) and \cite{Yelda2014} (abb. Yelda14) agree on the presence and orientation of the clockwise disk, but the significance of the warp is disputed. This is surprising: while Yelda14 presented an improvement in all relevant numbers (number of stars, number of constrained stars, number of determined orbits) compared to the Bartko09 sample, the improvement is gradual. For example, the total number of young stars in the sample increased by 18 (from 98 to 116), but the sample includes all 98 young stars from Bartko09. 

\noindent In the following we try to explain the apparent discrepancy between the works. For this, we use the data set published in Yelda14, and use exclusively the uniform acceleration prior. We will show that the discrepancy between the work is not due to error, but is dominated by the different definitions of the significance. 

\begin{figure}
    \centering
    \includegraphics[width=0.45\textwidth, trim={5cm 0 5cm 0},clip]{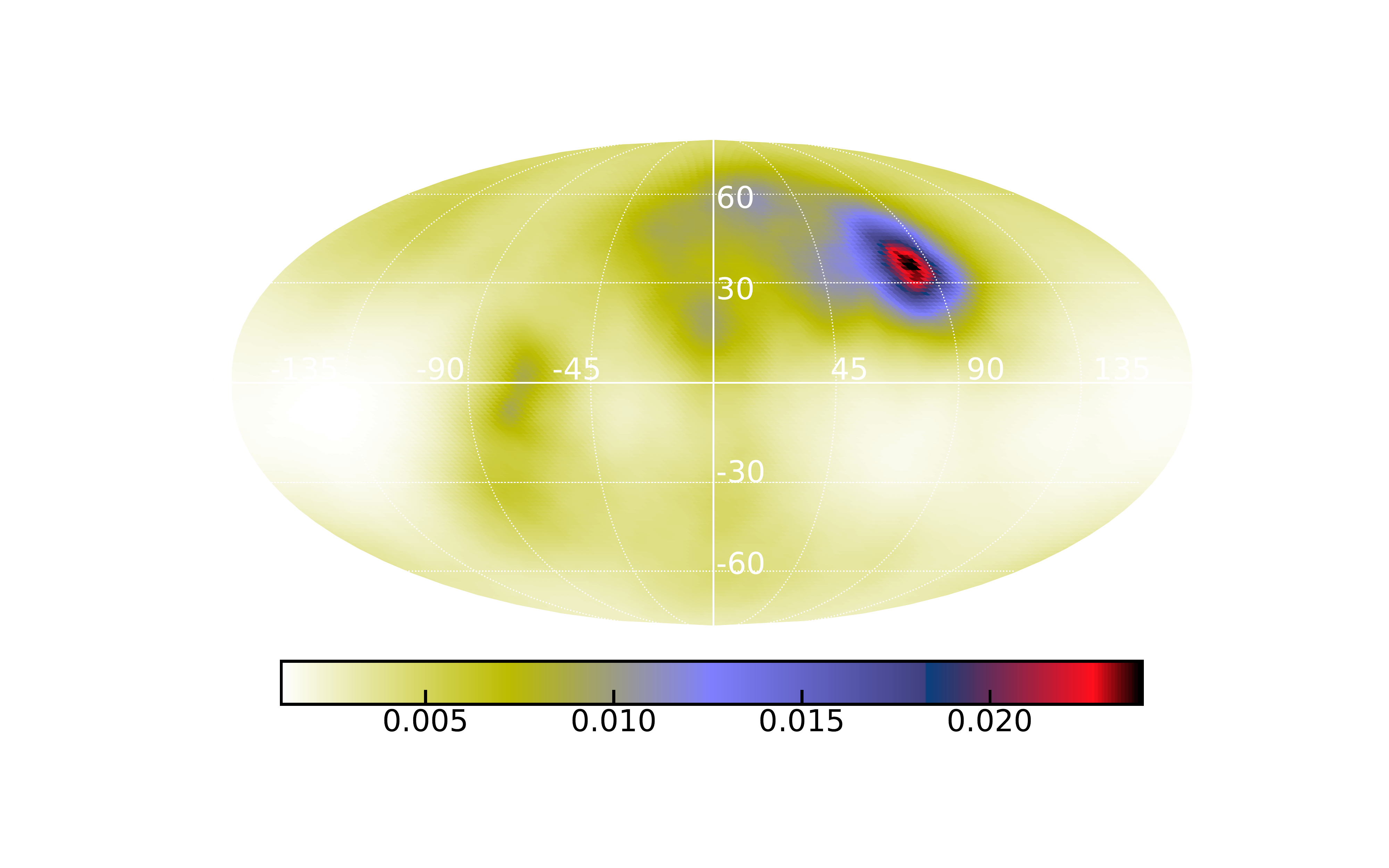}
    \caption{Comparison of the histogram of orbital nodes calculated in this work and presented in Figure 10 of \cite{Yelda2014}. We have normalized the histogram to the same minimum and maximum values, see text for details.}
    \label{fig:sebo_vs_yelda}
\end{figure}

\noindent \autoref{fig:sebo_vs_yelda} demonstrates that we can reasonably reproduce the Yelda14 results. The histogram has been normalized to match those of the bounds of the histogram of Yelda14 (Figure 10 in their work) and we use the same colormap and projection. There is broad agreement between the figure presented in Yelda14 and our reproduction\footnote{However, minor discrepancies exist. For instance, the faint feature next to the clockwise disk is more ``fuzzy'' in our reproduction. We speculate that this is most likely due to different treatment of unbound stars in the Monte-Carlo simulations, which we do not re-sample. Further, the strength of the smoothing seems decreased compared to Yelda14.}.

\begin{figure}
    \centering
    \includegraphics[width=0.45\textwidth]{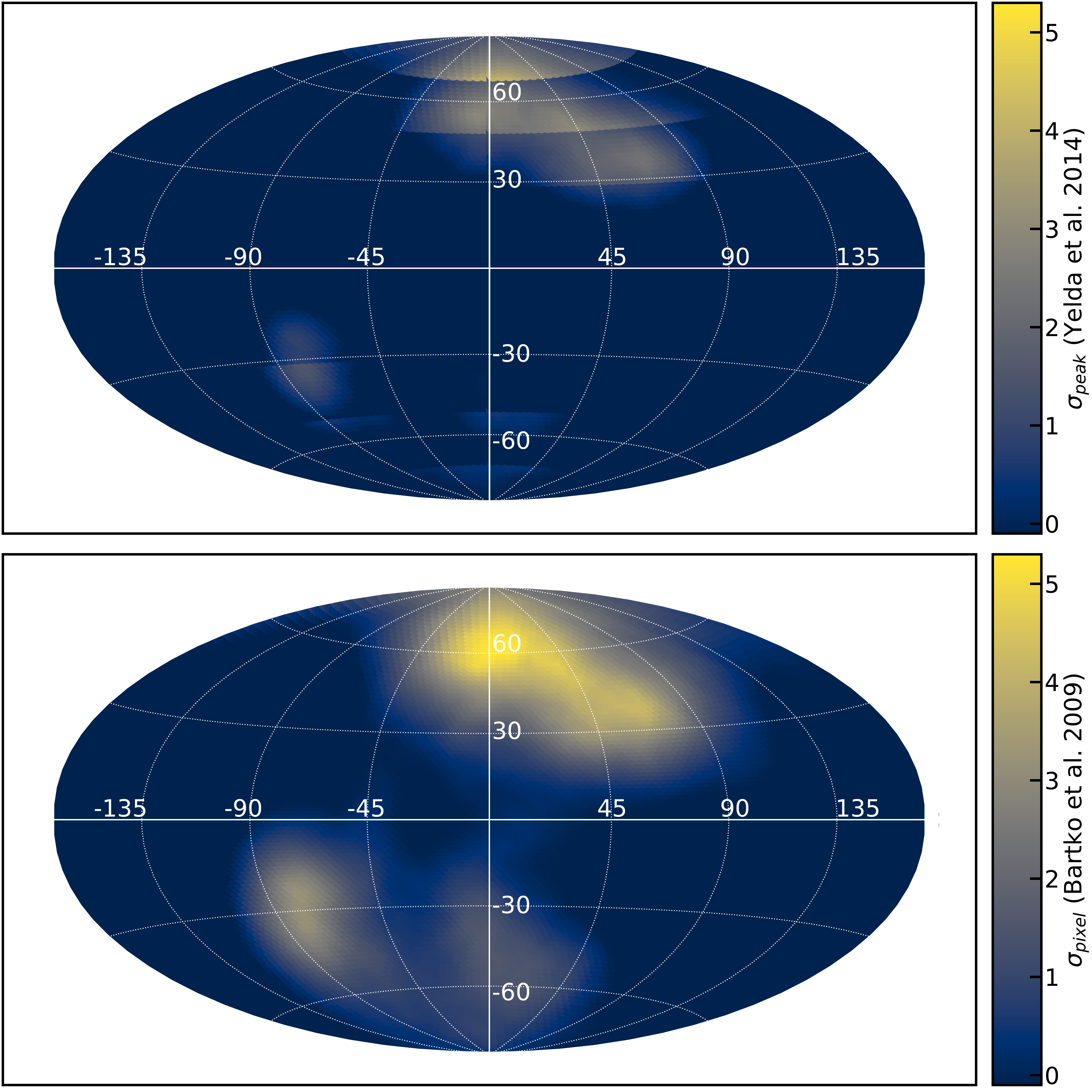}
    \caption{Difference of the between significance using the pixel and peak significance used in \cite{Bartko2009} and \cite{Yelda2014} for the radial slice ranging from $3.2''$ to $6.5''$. The data used is taken from \cite{Yelda2014}, and we use the uniform acceleration prior.}
    \label{fig:yelda_bartko_significance}
\end{figure}

\noindent Bartko14 compute the pixel-significance in the same manner as we do:
\begin{equation}
    \sigma_{\rm{pixel}} = \dfrac{s_{\rm{pixel, obs}} - <s_{\rm{pixel, sim}}>}{RMS(s_{\rm{pixel, sim}})}  
    \label{eqn:pixel_sig}
\end{equation}
where $s_{\rm{pixel, obs}}$ stands for the pixel value in the observed histogram and $s_{\rm{pixel, sim}}$ stands for the simulated pixels of the mock observations. This is based on the standard approach described in \cite{Li1983}. In contrast, Yelda14 use a peak-significance instead of a pixel-significance:
\begin{equation}
    \sigma_{\rm{peak}} = \dfrac{s_{\rm{peak, obs}} - <s_{\rm{peak, sim}}>}{RMS(s_{\rm{peak, sim}})}  
    \label{eqn:peak_sig}
\end{equation}

\noindent where $s_{\rm{peak, obs.}}$ stands for the observed peak value of a feature, and $s_{\rm{peak, sim}}$ stands for the respective peak values in the simulations. In order to account for the biases introduced by the observations and the uniform acceleration prior, they calculate the peaks in bins of $20\degree$ in latitude. \autoref{fig:yelda_bartko_significance} shows the difference in reported significance between the two methods for the radial slice ranging from $3.2''$ to $6.5''$. $\sigma_{\rm{peak}}$ is much reduced compared to $\sigma_{\rm{pixel}}$ and we recover the two seemingly competing conclusions found in Bartko09 and Yelda14: Using the pixel significance we find a significant feature at ${\sim}6\sigma_{\rm{pixel}}$ (see Figure 11 of Bartko09). In contrast, using the peak significance is $\sigma_{\rm{peak}} \lesssim 3$.

\begin{figure}
    \centering
    \includegraphics[width=0.485\textwidth]{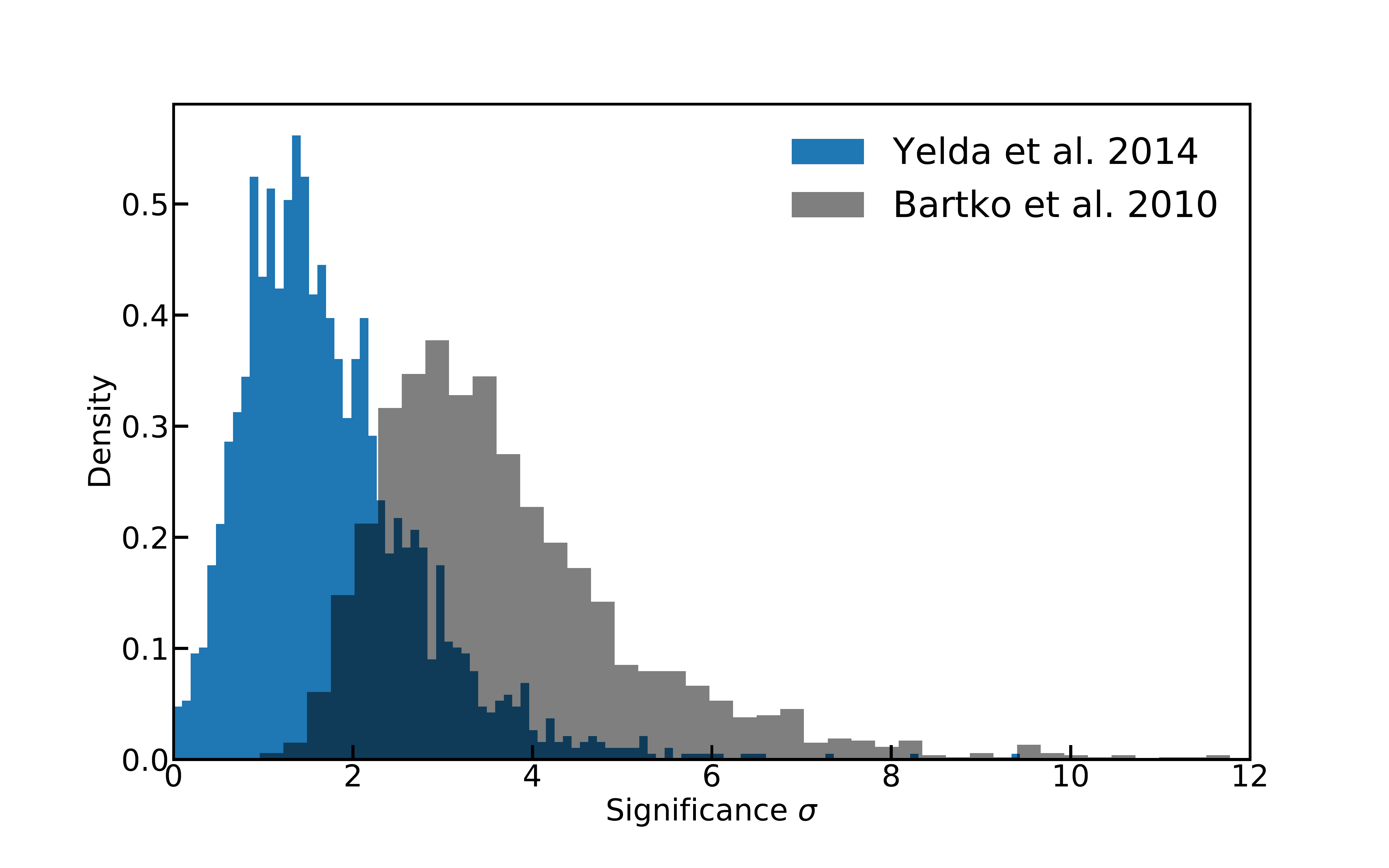}
    \caption{Histograms of the highest value of the significance calculated for each mock observation. The black histogram shows the pixel significances $\sigma_{\rm{pixel}}$ \citep[][]{Bartko2009} and blue histogram shows the peak significance $\sigma_{\rm{peak}}$ \citep[][]{Yelda2014}, respectively.}
    \label{fig:histogram_signifance}
\end{figure}

In the following, we explore the differences between the two definitions of the significance. Using a set of $2000$ mock observations of an isotropic cluster resembling our observations, we calculate the feature with the highest significance for each of the mocks. The histogram of these significances is shown in \autoref{fig:histogram_signifance}. The peak significance is more conservative, with the mode of the significance corresponding to ${\sim}2\sigma_{\rm{peak}}$, while significances of ${\sim}6\sigma_{\rm{pixel}}$ are routinely observed for the pixel significance. Despite this, the histograms are very similar, and seem to be merely shifted realizations of each other. 

\begin{figure}
    \centering
    \includegraphics[width=0.485\textwidth]{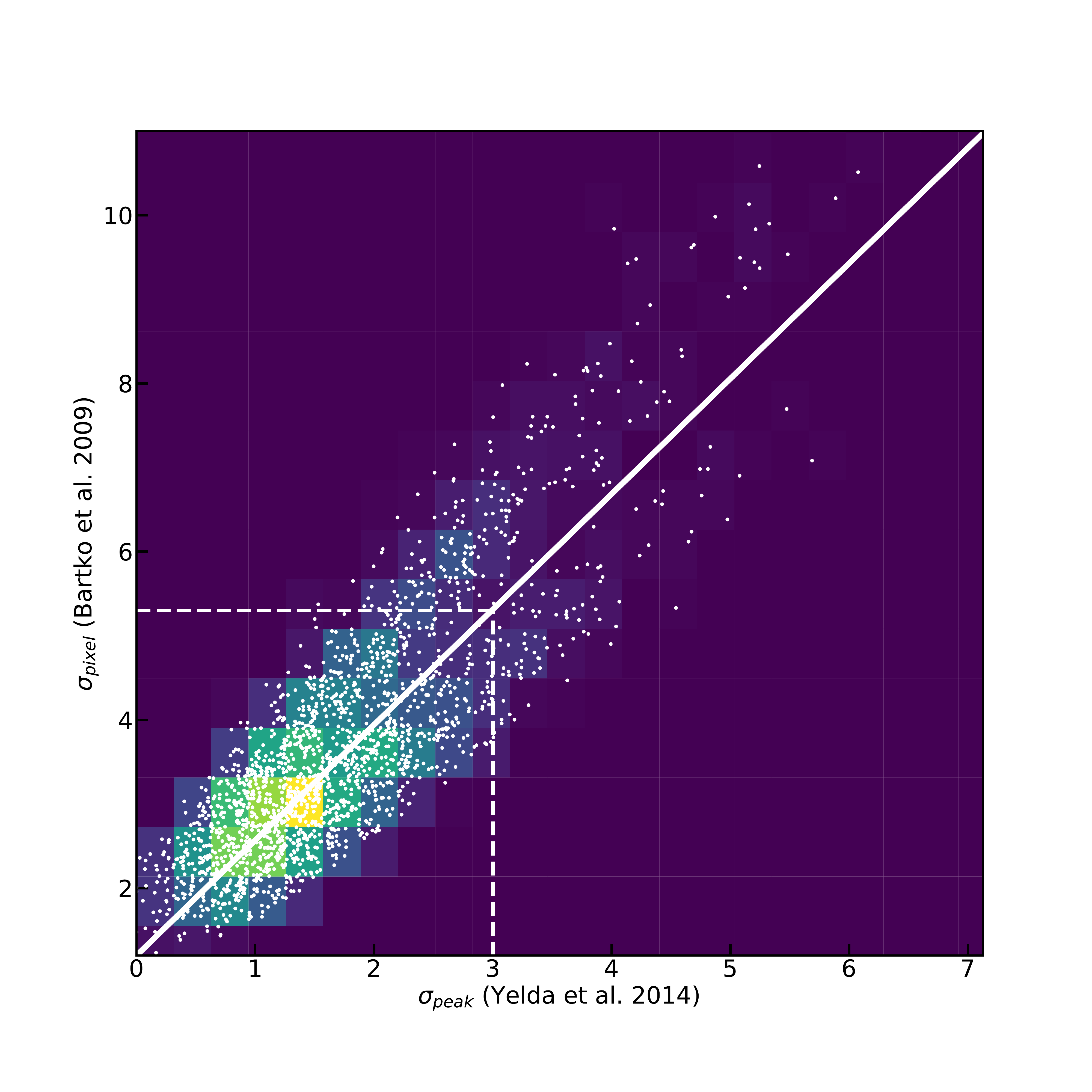}
    \caption{2D histogram of the peak and the pixel sigifcance calculated for 2000 mock observations. The white dots indicate the individual signifcance values, the thick white line indicates the trend. The horizontal dashed line indicates the peak value of $\sigma_{\rm{pixel}}$ found in \autoref{fig:yelda_bartko_significance}, and the vertical dashed line presents the projection again consistent with the respective $\sigma_{\rm{peak}}$ in \autoref{fig:yelda_bartko_significance}.}
    \label{fig:2dhistogram_signifance}
\end{figure}

\noindent This becomes even clearer when plotting the respective peak significance against the pixel significance of each mock observation (\autoref{fig:2dhistogram_signifance}). There exists a linear trend between the two definitions (indicated by the trend-line). The horizontal dashed line indicates the maximum significance $\sigma_{\rm{pixel}}$ found in the upper panel of \autoref{fig:yelda_bartko_significance} (${\sim}5.3\sigma$), and the vertical line shows the projection onto $\sigma_{\rm{peak}}=3$ which is consistent with the lower panel of \autoref{fig:yelda_bartko_significance}. 

\noindent The differences between the two methods are interesting, and it is clear from \autoref{fig:2dhistogram_signifance} that the significance $\sigma_{\rm{pixel}}$ is inconsistent with confidence ranges associated with Gaussian $\sigma$. This is not a new problem, and the difficulty of finding confidence ranges of Monte Carlo simulations is commonly discussed in other astronomical observations (see for instance section 4.4 in \cite{HessCollab2018} and \cite{Stewart2009}). In order to assess the significance in terms of confidence levels, we calculate the maximum pixel significance in $100000$ mock observations. Deriving the percentiles of this maximum pixel significance distribution thus allows to estimate probability values up to ($p\sim 99.999\%$), which we give in \autoref{tab:tablestuff}.
\begin{table}[]
    \centering
    \begin{tabular}{c|c}
    Confidence percentile & Corresponding $\sigma_{\rm{pixel}}$\\
    \hline
         68\% & $\sim 4~\sigma_{\rm{pixel}}$\\
         95\% & $\sim 7~\sigma_{\rm{pixel}}$\\
         99\% & $\sim 10~\sigma_{\rm{pixel}}$\\
         99.9\% & $\sim 18~\sigma_{\rm{pixel}}$\\
         99.99\% & $\sim 31~\sigma_{\rm{pixel}}$\\
         99.999\% & $\sim 42~\sigma_{\rm{pixel}}$\\
    \end{tabular}
    \caption{Confidence percentiles corresponding to $\sigma_{\rm{pixel}}$ rounded the integer, derived from 100000 mock simulations.}
    \label{tab:tablestuff}
\end{table}

\section{Posterior combined distributions of kinematic features}
\begin{figure*}
    \centering
    \includegraphics[width=0.985\textwidth, trim={0cm 3cm 0cm 3cm}, clip]{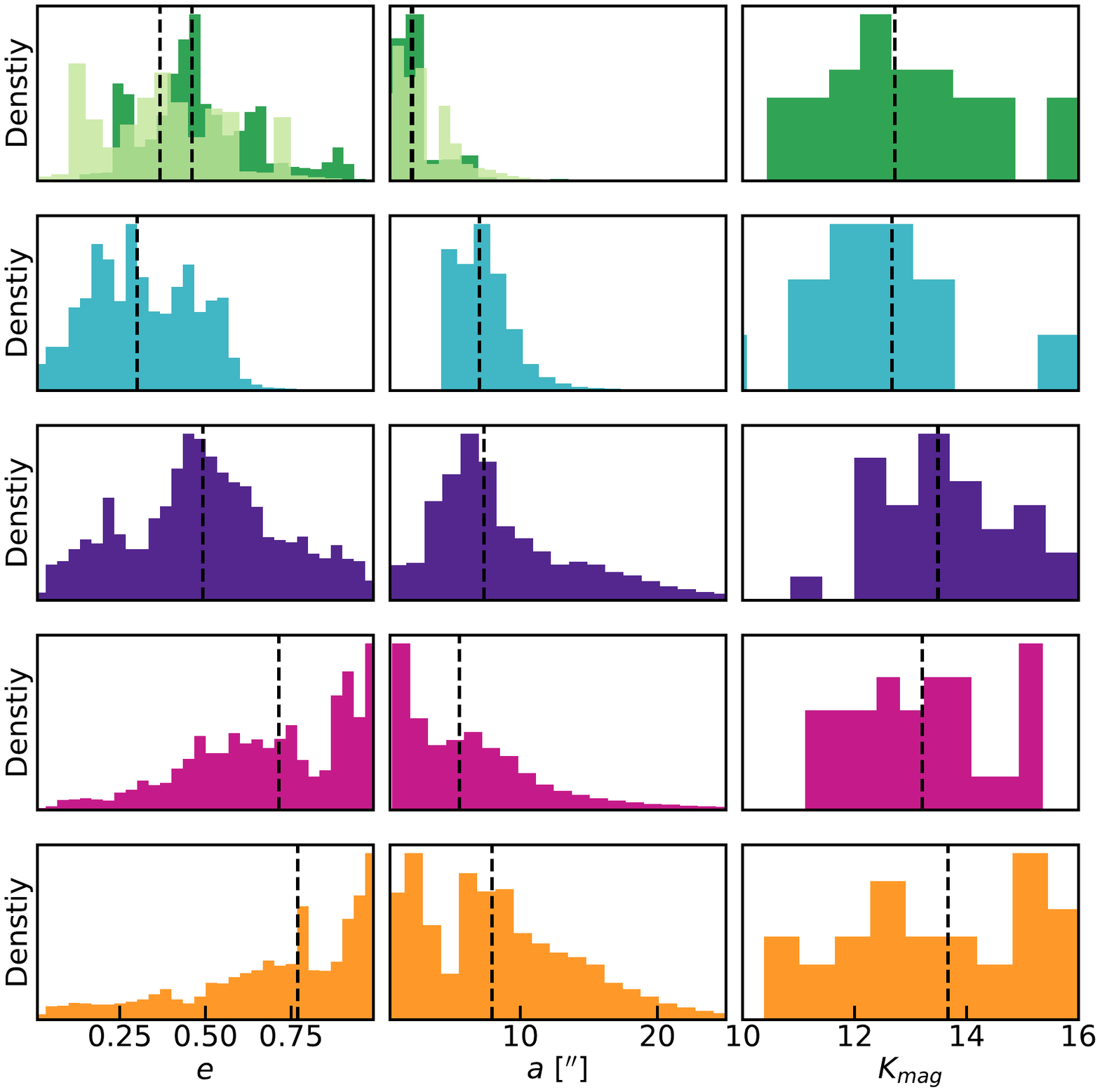}
    \caption[Comparison of the orbital element and $K_{mag}$ distributions]{Comparison of the eccentricity distributions, the semi-major axis distributions and $K_{mag}$ distributions of the different kinematic features. The top row includes both the distributions of the 5-D constrained stars (dark green) and the stars with determined orbital solutions (light green) of CW1. CW2 is shown in light blue, CCW/F1 in purple, F2 in pink and F3 in orange.}
    \label{fig:comparison_distributions}
\end{figure*}

\autoref{fig:comparison_distributions} shows the combined posterior distribution for stars beloning to the respective kinematic features (i.e. with $\Delta$ evidence $<2$). 



\section{Uncertainty estimation of the completeness correction}\label{appendix:completeness_error}

\noindent Because we do not treat the KLF as a probabilistic quantity, we need to propagate the uncertainty of the completeness correction to the KLF. To estimate the completeness uncertainty we assume that the total number of stars in a given pointing is Poissonian, and that the determined fraction is binomially distributed. Explicitly, for each pointing and magnitude bin, the Poisson rate parameter is $\lambda=N_{\rm{total}}$, and the rate parameter of the binomial distribution is $p=\dfrac{N_{\rm{unclassified}}(m_K)}{N_{\rm{total}}(m_K)}$. This allows us to draw random samples of the completeness for each pointing and magnitude bin\footnote{Code snippet in Python.}:

\begin{python}
n_samp=100
p=N_unclassified/N_total
completeness = np.empty(n_samp) #Pointing Complete.
n_total = np.empty(n_samp) #Pointing Total
Poisson = np.random.poisson 
Binomial = np.random.binomial
for n in range(n_samp):
    n_total[n] = Poisson(N_total, n_samp)
    completeness[n] = Binomial(n_total[n], p)
completeness = 1 - completeness/n_total
\end{python}
Our estimate of the completeness is derived in patches, thus there is a perfect correlation of neighbouring pixels within each patch. We account for this correlation by drawing random samples of patches rather than each pixel. With the so sampled completeness maps we can compute sampled effective area $A_{\rm{eff, sampled}}(R_1, R_2, m_K)$ for each realization. We determine the uncertainty of the effective area as the standard deviation of the sampled effective area, and calculate the uncertainty of the KLF using Gaussian error propagation which holds in the limit of many samples:
\begin{equation}
\begin{aligned}
    \sigma \mathrm{KLF} = ( & (|\partial \mathrm{KLF}/\partial A_{\rm{eff}}| \sigma A_{\rm{eff}})^2 +\\
     &(|\partial \mathrm{KLF}/ \partial N_{\rm{stars, obs}}| \sigma N_{\rm{stars, obs}})^2 ) ^{1/2},
\end{aligned}
\end{equation}
with $\sigma N_{\rm{stars, obs}} = \sqrt{N_{\rm{stars, obs}}}$.
\section{Comparison of different K-band luminosity functions reported in previous works}\label{appendix:klf_comparison}
\begin{figure}
    \centering
    \includegraphics[width=0.485\textwidth]{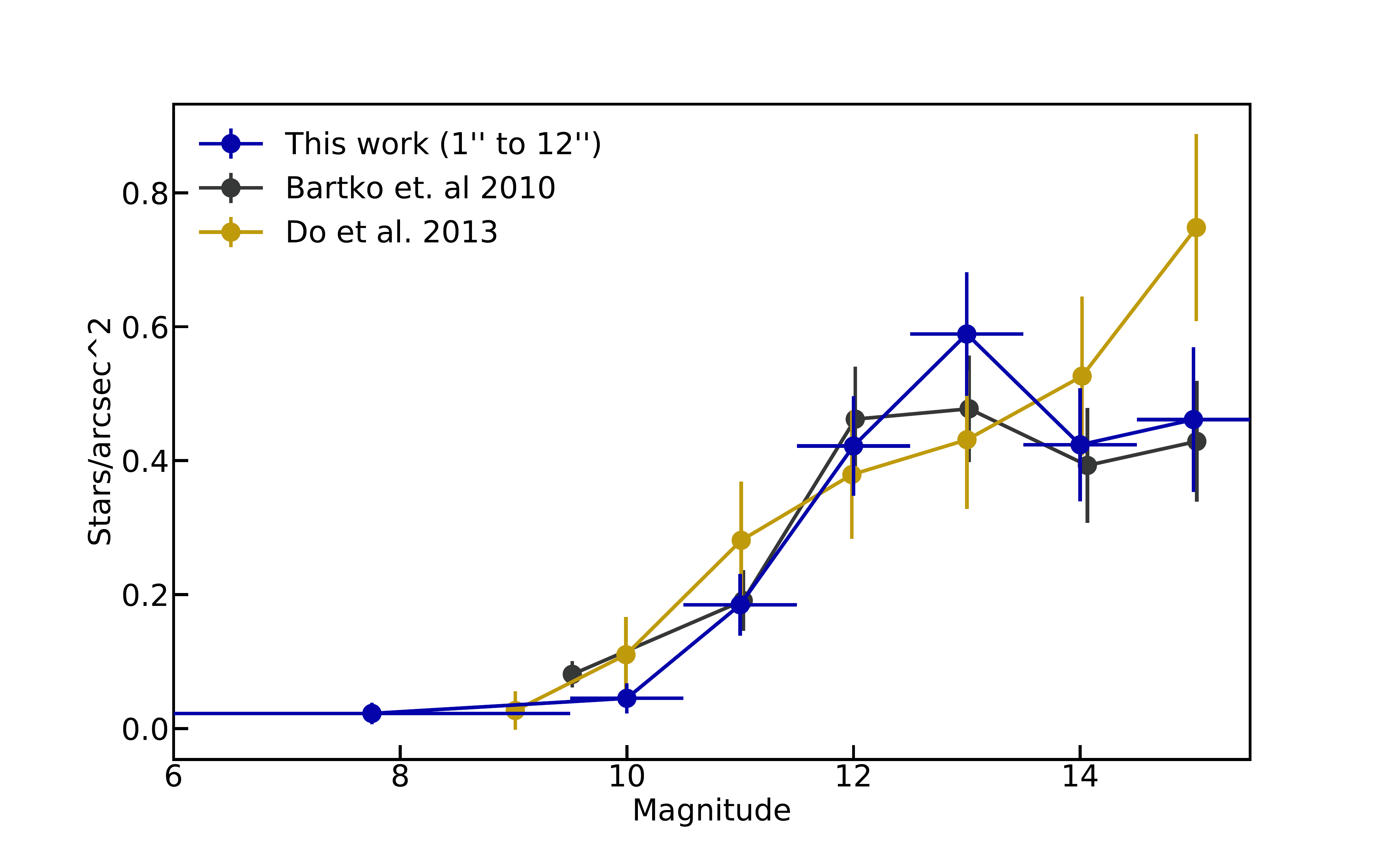}
    \caption{KLF of all young stars with projected radii ranging from $1''$ to $12''$ (black), compared to previously published KLFs by \cite{Bartko2010} and \cite{Do2013}. In order to compare the KLFs, we have normalized the distributions so that the area under the respective curves up to $K_{mag}=13$ is $1$.}
    \label{fig:klf_1_12}
\end{figure}

We compare the young star KLF binned to projected radii from $1''$ to $12''$ derived here to ones published in \cite{Bartko2010} and \cite{Do2013} in \autoref{fig:klf_1_12}. To compare the three studies we have normalized the curves with the area under the curve up to magnitude bin $K_{mag}=13$, up to which the sample should be complete in all studies. The overall agreement between this work and \cite{Bartko2010} is good. While the star sample within $12''$ is very comparable between this work and \cite{Bartko2009}, the way the completeness is derived is fundamentally different. Furthermore, the number of bright stars in the brightest bin has decreased, which is due to refined magnitude measurements which are typically lower than the ones reported in \cite{Bartko2010}. 
The discrepancy of the KLF reported by \cite{Do2013} and \cite{Bartko2010} consequently remains.

\section{K-Band Luminosity function of the central region}\label{appendix:sstars_klf}
\begin{figure}
    \centering
    \includegraphics[width=0.485\textwidth]{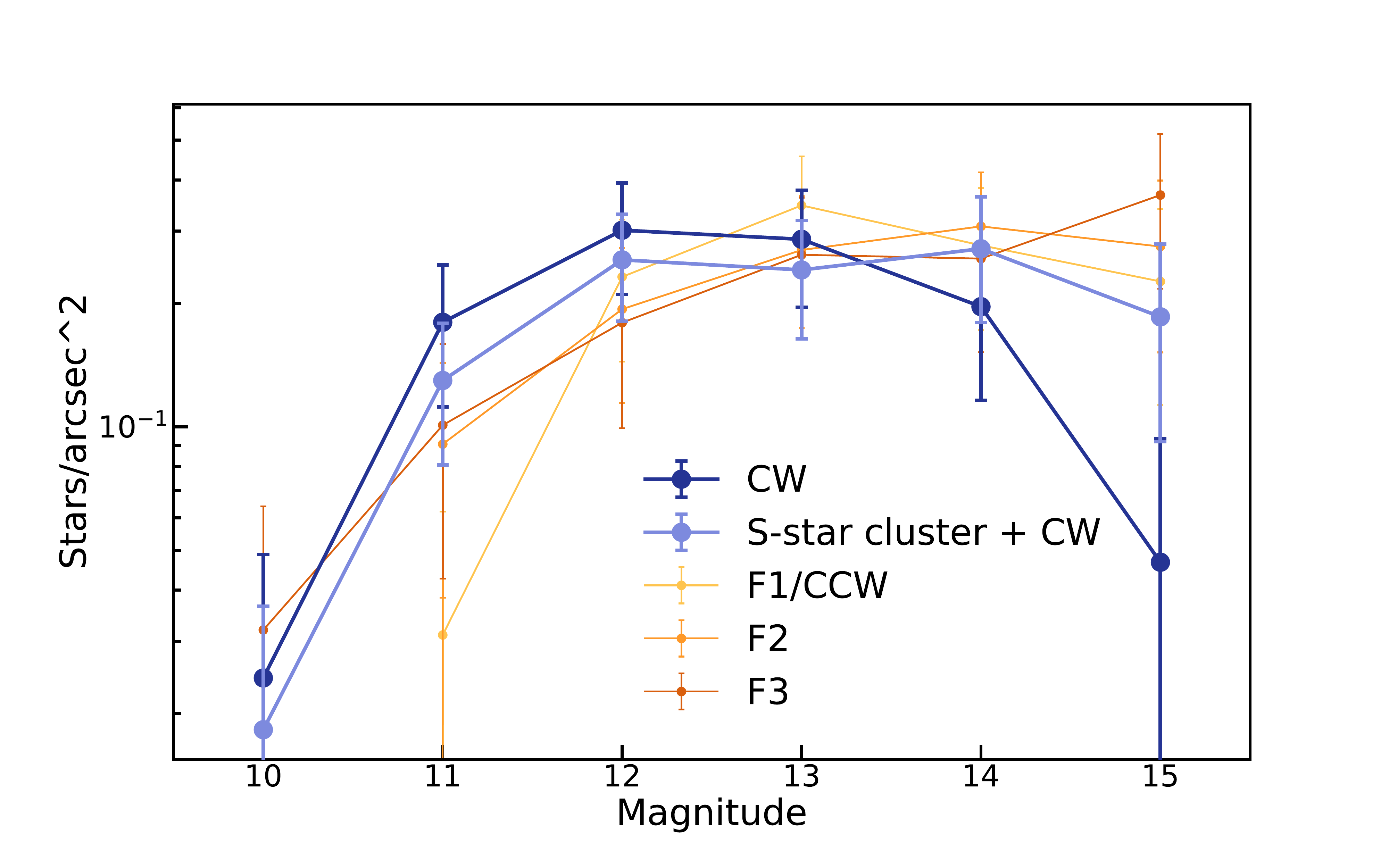}
    \caption{Similar to \autoref{fig:klf_disks}, but including the combined KLF of the S-star and the CW disk. As in \autoref{fig:klf_disks} and \autoref{fig:klf_1_12} the KLFs have been normalized to make them comparable, see text for details.}
    \label{fig:klf_stars_cw}
\end{figure}

The scenario proposed by \cite{Chen2014} connects the dynamically distinct S-star cluster to the young stars in the clockwise disk. There is no commonly accepted definition of S-stars, and typically the term also refers to central old stars. Here we define as an S-stars as young stars within $0.8''$. Since the central region is populated by stars typically fainter than $K_{\rm{mag}}=14$, including these stars in the K-band luminosity function of the CW disk alters the KLF. \autoref{fig:klf_stars_cw} displays the derived KLF, and compares it to the other KLFs reported in \autoref{fig:klf_disks}. If the S-stars and CW disk indeed form a common structure, their KLF is less top heavy. We caution however that in such a scenario the KLF is not a direct tracer of the initial mass function. In the \cite{Chen2014} scenario large stars are rapidly destroyed and thus do not appear in the star count. Further, this raises the question of including the other dynamical features and the other young stars not associated with any feature. 

\section{Anistropy in the central arcsecond}\label{appendix:anisotropy}
\begin{figure}
    \centering
    \includegraphics[width=0.485\textwidth]{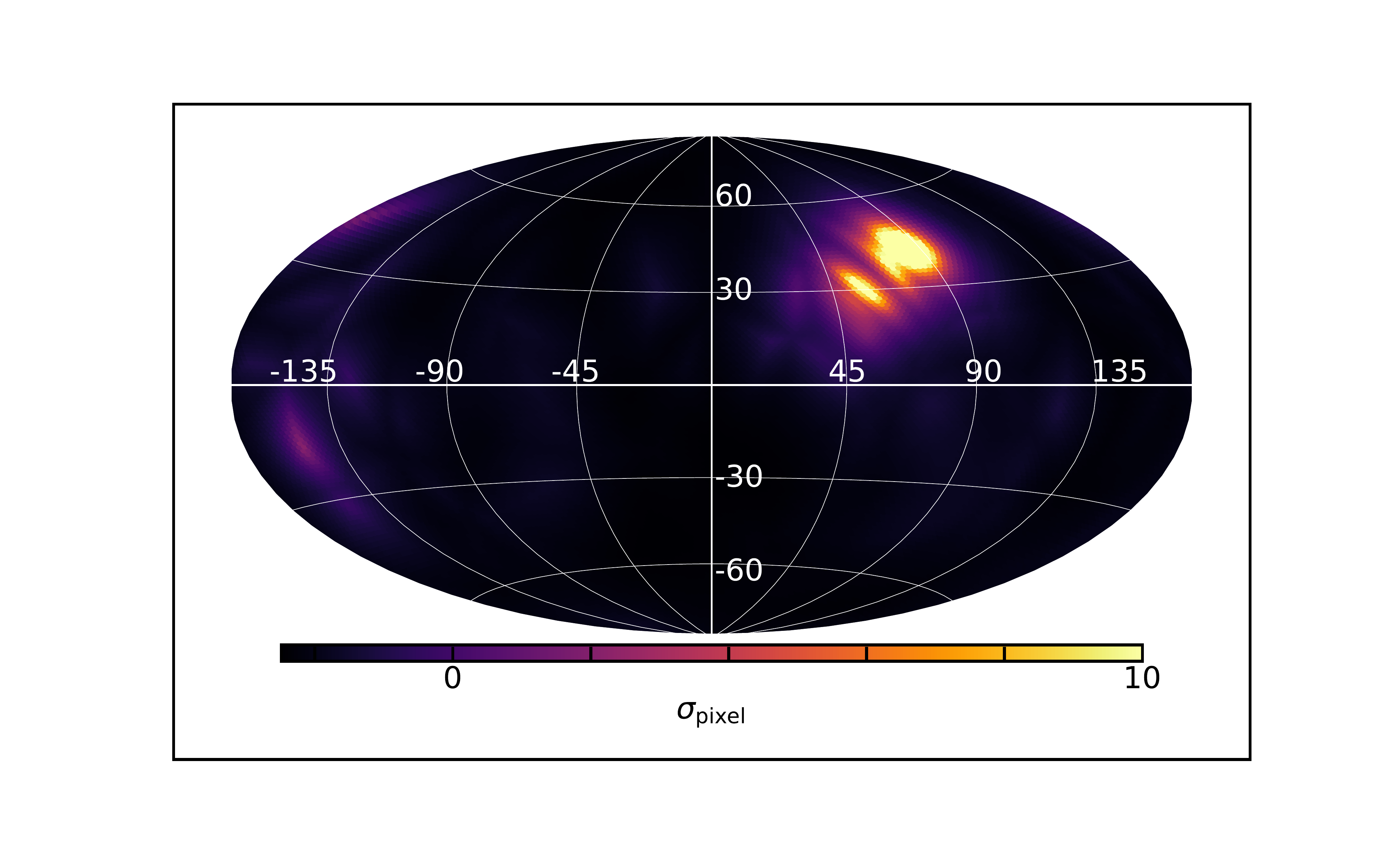}
    \caption{Like figure \autoref{fig:significance48} for the central arcsecond, plotted in linear scale. The color scale is truncated at $10\sigma_{\rm{pixle}}$ which corresponds to a p-value of $\sim99\%$.}
    \label{fig:basel_diss}
\end{figure}

\cite{Basel2020} report an apparent overdensity in the inclination of the orbits of the early and late type stars in the central arcsecond. \autoref{fig:basel_diss} shows the significance of the overdensity of the angular momentum distribution of the young stars studied in this paper. No significant overdensity is discernible other than the inner onset of the clockwise disk is apparent. This result entails two caveats: compared to \cite{Basel2020} we do not include late type stars, and the significance is calculated based on the comparison against an isotropic cluster, which also entails constraints on all orbital elements.

\section{Tables: Young stars with orbits}\label{app:youngst_stars_with_orbits}
{\setlength\tabcolsep{0.3pt} 
\begin{longtable}{lcc|cccccccccccccc}
  MPE & Pau. &    UCLA &     $\rm{a}$ & $\sigma\rm{a}$ &     $\rm{e}$ & $\sigma\rm{e}$ &       $\rm{i}$ & $\sigma\rm{i}$ &   $\Omega$ & $\sigma\Omega$ &  $ \omega$ & $\sigma\omega$ & $\rm{P}$ & $\sigma\rm{P}$ & $\rm{t_0}$ & $\sigma \rm{t_0}$\\
   & &  & [as] & [as] &  &  & [$\degree$] & [$\degree$] & [$\degree$] & [$\degree$] & [$\degree$] & [$\degree$] & [yr] & [yr] & [yr] & [yr]\\
\hline
  S18 &         &         &  0.28 &    0.01 &  0.24 &    0.01 &  107.18 &     0.6 &   51.23 &        0.37 &     3.5 &        5.45 &    54.12 &     3.41 &  1989.75 &      0.74 \\
  S13 &      E3 &   S0-20 &  0.26 &   $>$0.01 &  0.43 &   $>$0.01 &   22.96 &    0.12 &   74.05 &        0.55 &  245.86 &        0.77 &    49.51 &     0.15 &  2004.89 &      0.01 \\
   S9 &      E9 &    S0-5 &  0.27 &   $>$0.01 &  0.64 &   $>$0.01 &   82.63 &    0.11 &  156.18 &        0.05 &  153.22 &        0.45 &    50.93 &     0.52 &  1977.35 &       0.4 \\
   S4 &      E6 &    S0-3 &  0.35 &   $>$0.01 &   0.4 &   $>$0.01 &    80.3 &    0.05 &   258.8 &        0.04 &  293.66 &         0.9 &    75.26 &     0.71 &  1959.39 &      0.73 \\
 S175 &         &         &  0.37 &    0.01 &  0.98 &    0.01 &   87.83 &    0.19 &  325.99 &        0.27 &   68.14 &        0.14 &    81.76 &     2.21 &  2009.51 &     $>$0.01 \\
  S14 &      E2 &         &   0.3 &   $>$0.01 &  0.99 &   $>$0.01 &  105.27 &    0.89 &  222.92 &        0.67 &   330.5 &        0.54 &    60.76 &     0.53 &  2000.32 &      0.03 \\
  S60 &         &         &  0.41 &    0.01 &   0.7 &    0.01 &  124.87 &    0.55 &  168.31 &        1.65 &    25.8 &        1.04 &    95.07 &     5.16 &   2024.8 &      0.27 \\
  S12 &      E5 &   S0-19 &   0.3 &   $>$0.01 &  0.89 &   $>$0.01 &   33.59 &    0.32 &  231.09 &        0.96 &  316.94 &        0.83 &    59.06 &     0.13 &  1995.58 &      0.03 \\
  S31 &      E7 &    S0-8 &  0.44 &   $>$0.01 &  0.56 &   $>$0.01 &  109.57 &    0.04 &  137.39 &        0.05 &  311.38 &        0.31 &   104.92 &     0.34 &  2018.16 &      0.01 \\
   S8 &     E10 &    S0-4 &   0.4 &   $>$0.01 &  0.81 &   $>$0.01 &   74.19 &    0.24 &  315.15 &        0.14 &  347.42 &        0.29 &    92.08 &     0.35 &  1983.95 &      0.19 \\
  S29 &         &         &  0.39 &   $>$0.01 &  0.97 &   $>$0.01 &  144.11 &    1.37 &    6.26 &        1.99 &  205.08 &        1.85 &    89.85 &      0.7 &  2021.43 &      0.02 \\
   S1 &      E4 &    S0-1 &  0.61 &    0.01 &  0.57 &    0.01 &  119.19 &    0.09 &  342.32 &        0.13 &  122.98 &        0.52 &   174.78 &     3.63 &  2001.79 &      0.07 \\
  S19 &         &         &  0.59 &    0.02 &  0.78 &    0.02 &   71.76 &    0.08 &  344.94 &        0.15 &  158.67 &         0.4 &   165.66 &    10.38 &  2005.64 &      0.03 \\
  S33 &         &         &  0.75 &    0.01 &  0.54 &    0.01 &   64.35 &    0.86 &   99.28 &        2.04 &  299.96 &        1.88 &   236.01 &     6.22 &  1904.31 &      8.04 \\
  S42 &         &         &  0.38 &    0.01 &  0.75 &    0.01 &   39.85 &    0.18 &  319.29 &        1.88 &   42.63 &        1.14 &    84.42 &     2.25 &  2022.59 &      0.08 \\
  S67 &     E15 &    S1-3 &  1.16 &    0.01 &  0.27 &    0.01 &  135.66 &    0.48 &    95.9 &        3.04 &  208.75 &        0.81 &   453.13 &     6.77 &  1685.57 &     10.92 \\
  S71 &         &         &  0.98 &    0.01 &   0.9 &    0.01 &   74.18 &    0.13 &   34.84 &        0.29 &  338.96 &        1.19 &    354.7 &     3.65 &  1686.51 &      3.86 \\
  S66 &     E17 &    S1-2 &  1.37 &    0.02 &  0.14 &    0.02 &  131.64 &     0.5 &   92.87 &        2.12 &  121.65 &        8.01 &   583.93 &    14.64 &  1774.34 &     21.17 \\
  S96 &     E20 &   S1-11 &  1.56 &    0.03 &  0.13 &    0.03 &  126.14 &    0.46 &  115.64 &        0.36 &   234.1 &         1.9 &   710.53 &    20.48 &  1623.76 &      8.99 \\
  S91 &         &         &  1.91 &    0.07 &  0.34 &    0.07 &  115.63 &    0.24 &  110.07 &        0.18 &  344.37 &        0.65 &   957.92 &    52.64 &  1075.96 &     52.77 \\
  S83 &     E16 &   S0-15 &  1.18 &   $>$0.01 &  0.17 &   $>$0.01 &  129.95 &    0.05 &   97.04 &         0.2 &  188.27 &        0.02 &   464.74 &     0.88 &  2022.31 &      0.49 \\
  R14 &     E22 &   S1-14 &  2.78 &    0.08 &  0.44 &    0.08 &  118.86 &    0.25 &  113.28 &        0.76 &   158.8 &        1.17 &  1659.39 &     73.8 &  3510.03 &     74.77 \\
  S97 &     E23 &   S1-16 &  2.41 &    0.38 &  0.37 &    0.38 &   113.6 &    1.03 &  112.67 &        0.99 &   24.61 &        9.26 &  1360.81 &   325.27 &  2125.26 &      19.9 \\
  R44 &         &   S2-21 &  4.97 &    0.48 &  0.39 &    0.48 &  127.85 &    1.11 &    85.3 &        1.72 &  218.21 &         3.1 &  3964.44 &   576.82 &  1934.91 &     13.95 \\
  S87 &     E21 &   S1-12 &  4.14 &    0.03 &  0.36 &    0.03 &  115.27 &     0.2 &  106.64 &        0.84 &  305.55 &        5.76 &  3059.27 &    29.16 &  -877.69 &     40.06 \\
   S2 &      E1 &         &  0.12 &   $>$0.01 &  0.88 &   $>$0.01 &  134.68 &    0.03 &  228.17 &        0.03 &   66.26 &        0.03 &    16.05 &    $>$0.01 &  2018.38 &     $>$0.01 \\
  R34 &         &         &  1.85 &    0.02 &  0.61 &    0.02 &  136.94 &    0.36 &     333 &        1.37 &   58.88 &        0.88 &   902.73 &    13.67 &  1506.63 &      9.97 \\
  S22 &         &         &  1.06 &    0.02 &  0.39 &    0.02 &   107.7 &    0.12 &  291.01 &        0.14 &   84.69 &        1.98 &   398.26 &    10.27 &  1991.82 &      0.85 \\
   S6 &     E11 &    S0-7 &  0.65 &    0.01 &  0.85 &    0.01 &    87.5 &    0.07 &   84.46 &        0.13 &  117.36 &        0.51 &    190.3 &     3.14 &  2111.16 &      1.85 \\
  R85 &     E56 &  I.34W &  6.59 &     1.7 &  0.34 &     1.7 &  128.27 &    2.87 &  110.77 &        1.03 &  181.05 &        2.38 &  6055.99 &  2343.35 &  2019.38 &     31.46 \\
  R70 &     E54 &   S4-36 &  3.48 &    0.01 &  0.35 &    0.01 &  147.27 &    0.18 &  115.55 &        0.98 &   43.19 &         1.8 &  2326.65 &    12.68 &  3667.56 &     16.75 \\
   R1 &     E29 &    S2-7 &  2.63 &    0.06 &  0.53 &    0.06 &  125.09 &    0.32 &  117.28 &        1.52 &  243.87 &        1.17 &  1523.25 &    50.46 &  4005.64 &     45.51 \\
   S5 &      E8 &   S0-26 &  0.53 &    0.01 &  0.72 &    0.01 &  115.96 &    0.37 &  128.58 &        0.81 &  271.99 &         0.4 &   140.03 &      3.4 &  1954.49 &      1.41 \\
  S72 &     E18 &    S1-8 &   2.2 &    0.05 &  0.33 &    0.05 &  119.18 &    0.36 &  316.04 &        0.64 &  205.09 &        1.97 &  1184.02 &    41.41 &   1055.3 &      41.4 \\
  R39 &     E40 &    S3-5 &  3.22 &    1.46 &  0.01 &    1.46 &  122.23 &    8.07 &  107.05 &        0.05 &  359.99 &        4.71 &  2067.22 &  1403.78 &  2054.44 &      27.1 \\
  R30 &     E32 &   S2-15 &  6.36 &    0.14 &  0.61 &    0.14 &  113.02 &    0.38 &   94.51 &        1.49 &   12.33 &        1.74 &  5739.49 &   184.24 &   2242.8 &      7.55 \\
\caption[Young stars with orbital solutions]{Stars with determined orbital solutions. Pau. abbreviates \cite{Paumard2006}}
\label{tab:orbit_stars}
\end{longtable}
}

\section{Tables: Young stars without orbits}\label{app:youngStasPhaseSpace}
\begin{figure*}
    \centering
    \includegraphics[width=0.95\textwidth, trim={4cm 0 8cm 0}]{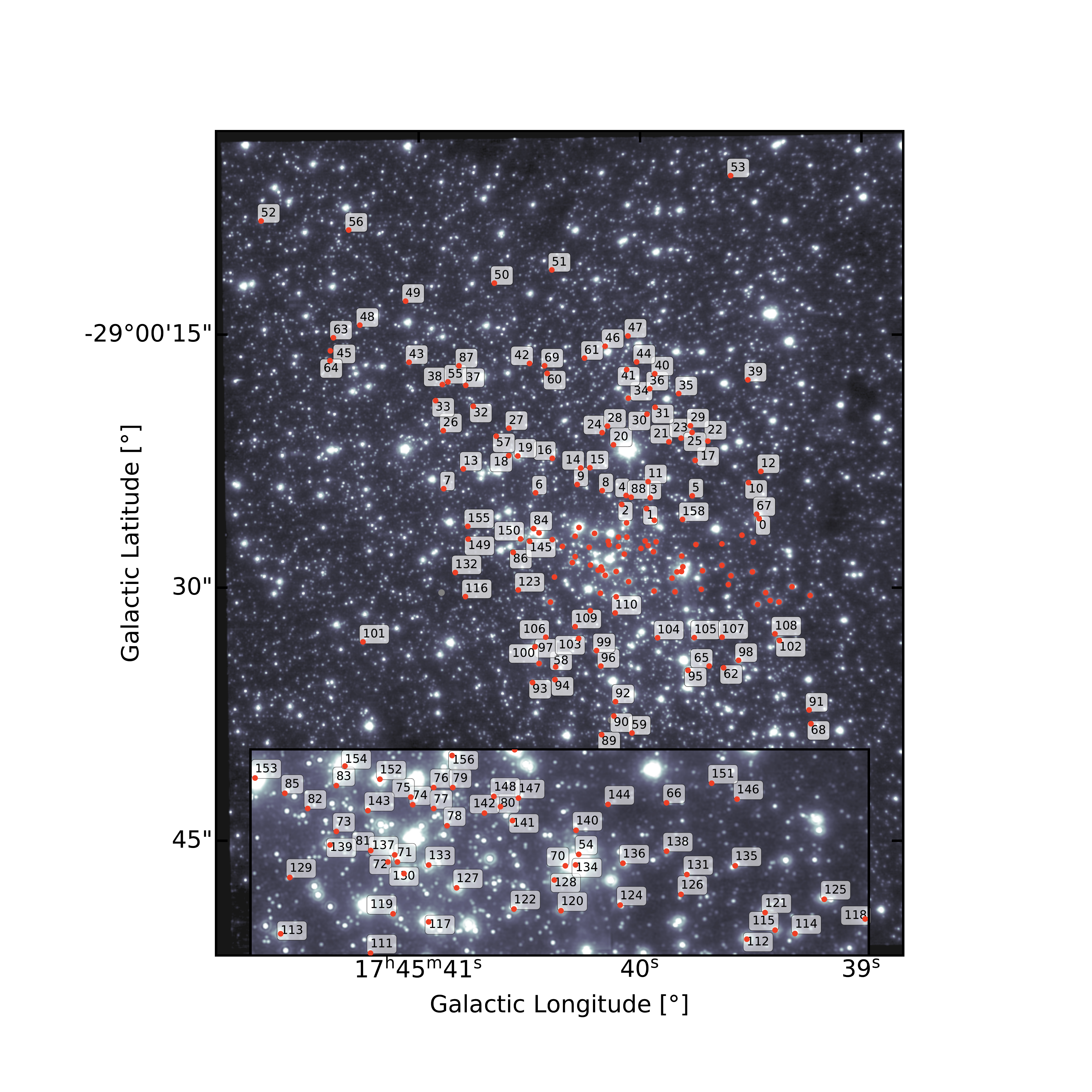}
    \caption[Lookup map of young stars]{Lookup map of young stars without accelerations: Young stars used in this study. The number indicated next to star corresponds to the row index in column \# in \autoref{tab:youngStasPhaseSpace}.}
    \label{fig:lookup}
\end{figure*}

\autoref{fig:lookup} indicates the young stars projected location and the reference number given in table \autoref{tab:youngStasPhaseSpace}. 

{\fontsize{10}{12}\selectfont
{\setlength\tabcolsep{0.01pt} 
\begin{longtable}{ccccc|ccccccccccccc}
{} &  Tri. & Pau. &  UCLA & Alt.  &  $K_{mag}$ & R & Sou. &  $R.A.$ &  $\sigma R.A.$ &  $Dec.$ &  $\sigma Dec.$ &  $v_{R.A.}$ &  $\sigma v_{R.A.}$ &  $v_{Dec.}$ &  $\sigma v_{Dec.}$ &  $v_{z}$ &  $\sigma v_{z}$ \\
\# & ID & ID & ID & names & &  $[\mathrm{''}]$ & vLSR & $[\mathrm{''}]$ & $[\mathrm{mas}]$& $[\mathrm{''}]$ & $[\mathrm{mas}]$& $[\mathrm{km/s}]$ & $[\mathrm{km/s}]$& $[\mathrm{km/s}]$& $[\mathrm{km/s}]$& $[\mathrm{km/s}]$ & $[\mathrm{km/s}]$ \\
\hline
0   &                41 &                   &   S7-216 &             I.6W &         10.9 &                                 7.9 &                          t.w. &   -7.73607 &               0.13 &    1.39504 &               0.18 &             87.5 &                    1.7 &            202.2 &                     2.8 &           71.0 &                  36.9 \\
1   &               163 &               E35 &    S2-16 &          I.29 &         12.3 &                                 2.3 &                      Pau. &   -1.00111 &               0.09 &    2.06472 &               0.07 &           -349.2 &                    0.6 &            -48.9 &                     0.4 &          -98.1 &                  68.5 \\
 & & & & NE1 & & && & & & & & & & & & \\
2   &               283 &               E36 &    S2-19 &                   &         12.7 &                                 2.3 &                      Pau. &    0.44190 &               0.08 &    2.29963 &               0.09 &           -319.9 &                    0.4 &             26.0 &                     0.5 &           41.2 &                  18.8 \\
3   &               571 &                   &    S3-17 &              &         13.5 &                                 3.2 &                          MPE &   -1.41736 &               0.64 &    2.84290 &               0.33 &            256.3 &                    9.3 &             60.6 &                     4.9 &          -67.4 &                  50.7 \\
\end{longtable}
}
}

\section{Tables: Properties of stars consistent with belonging to angular momentum overdensities}
\begin{longtable}{lllllllr}
Delta Evidence &  MPE &      Mag & Orbit & Paumard &          R &    UCLA &  index \\
\hline
       1.23236 &  S31 &    15.85 &   yes &      E7 &   0.140883 &    S0-8 &    8.0 \\
      0.284251 &  S67 &    12.35 &   yes &     E15 &   0.785569 &    S1-3 &   15.0 \\
      0.257495 &  S66 &    14.80 &   yes &     E17 &    0.91267 &    S1-2 &   17.0 \\
      0.111609 &  S96 &    10.75 &   yes &     E20 &    1.01648 &   S1-11 &   18.0 \\
      0.151909 &  S91 &    12.72 &   yes &         &    1.24012 &         &   19.0 \\
      0.156153 &  S83 &    13.73 &   yes &     E16 &   0.970887 &   S0-15 &   20.0 \\
      0.122042 &  R14 &  13.2434 &   yes &     E22 &    1.49567 &   S1-14 &   21.0 \\
\end{longtable}

\begin{longtable}{rlllllll}
 index &  MPE &     UCLA & Paumard &      Mag &          R & Delta Evidence & Orbit \\
\hline
     7 &  S12 &    S0-19 &      E5 &    15.48 &  0.0306114 &       0.869801 &   yes \\
     0 &      &   S7-216 &         &  10.8521 &    7.86087 &        1.48128 &    no \\
     3 &      &    S3-17 &         &    13.49 &    3.17676 &        1.42991 &    no \\
     6 &      &    S6-89 &         &    11.95 &    6.22556 &        1.91016 &    no \\
     9 &      &          &     E59 &    13.14 &    4.55295 &        1.25763 &    no \\
    11 &      &   S3-331 &     E52 &    13.60 &    3.84197 &        1.38002 &    no \\
    13 &      &    S10-7 &     E85 &    12.59 &    10.6752 &       0.850875 &    no \\
\end{longtable}

\begin{longtable}{rlllllll}
 index &  MPE &    UCLA & Paumard &      Mag &         R & Delta Evidence & Orbit \\
\hline
     6 &  S60 &         &         &    16.76 &  0.113085 &       0.629584 &   yes \\
    30 &  R70 &   S4-36 &     E54 &    12.82 &   2.08898 &       0.656347 &   yes \\
    10 &      &  S7-236 &         &   12.666 &   7.94508 &         1.3941 &    no \\
    30 &      &   S7-10 &     E73 &    11.62 &   7.70832 &        1.05606 &    no \\
    36 &      &   S9-23 &     E77 &    13.57 &   9.23117 &        1.30484 &    no \\
    40 &      &   S10-5 &     E83 &    11.97 &   10.1532 &        1.78867 &    no \\
    65 &      &         &         &  15.3536 &   8.67658 &        1.53632 &    no \\
    84 &      &  S5-237 &         &    13.25 &   5.59141 &        1.15228 &    no \\
    85 &      &  S4-169 &     E57 &    13.53 &   4.43282 &        1.95868 &    no \\
    86 &      &   S6-82 &     E72 &    13.56 &   6.73303 &        1.50384 &    no \\
    96 &      &   S7-16 &         &  12.6886 &   7.42579 &        1.71009 &    no \\
    99 &      &   S6-63 &     E69 &    11.36 &   6.57862 &       0.921401 &    no \\
   103 &      &         &     E65 &    12.13 &    6.3003 &        1.06481 &    no \\
   105 &      &         &         &  12.9989 &   6.79444 &         1.0242 &    no \\
   106 &      &         &         &  12.2293 &   7.36999 &        1.69583 &    no \\
   143 &      &   S2-22 &         &    13.03 &   2.33073 &        1.99339 &    no \\
   145 &      &  S5-231 &     E64 &    12.01 &   5.81382 &        1.60709 &    no \\
   150 &      &   S6-81 &     E67 &    11.08 &    6.3722 &        1.47119 &    no \\
   153 &      &         &     E63 &    9.328 &     5.297 &       0.938662 &    no \\
\caption[Stars consistent with belonging to the inner warp feature]{Stars consistent with belonging to the inner warp feature.}
\label{tab:inner_warp}
\end{longtable}
\begin{longtable}{lll|cccc}
 MPE &    UCLA & Paumard &      Mag &          R & $\Delta$ Evidence & Orbit \\
\hline
  S9 &    S0-5 &      E9 &    15.15 &  0.091365 &        1.38003 &   yes \\
 S60 &         &         &    16.76 &  0.113116 &       0.589293 &   yes \\
     &  S5-235 &         &    13.21 &   5.33721 &        1.95023 &    no \\
     &         &         &    14.78 &   6.96797 &       0.958125 &    no \\
     &         &         &    13.71 &   7.06065 &        1.15084 &    no \\
     &    S8-4 &     E75 &    11.12 &   8.54245 &       0.321072 &    no \\
     &   S9-23 &     E77 &    13.57 &    9.2312 &       0.983455 &    no \\
\end{longtable}

\begin{longtable}{lll|cccc}
 MPE &    UCLA & Paumard &      Mag &          R & $\Delta$ Evidence & Orbit \\
\hline
  S4 &     S0-3 &      E6 &    14.58 &    0.091998 &        1.90794 &   yes \\
 S14 &          &      E2 &    15.63 &  0.00271191 &      0.0173749 &   yes \\
  S2 &          &      E1 &    14.16 &   0.0107373 &        1.44597 &   yes \\
     &          &         &  12.4966 &     11.3631 &        1.85805 &    no \\
     &   IRS10W &         &  10.9925 &     8.29025 &        1.99354 &    no \\
     &          &     E90 &    11.57 &     12.7704 &       0.561536 &    no \\
     &          &         &  13.6698 &     9.79173 &       0.463197 &    no \\
\end{longtable}

\end{document}